\documentclass[cernpreprint,texlive=2016,UKenglish,texmf]{atlasdoc}
 
\usepackage[subfigure=true]{atlaspackage}
\usepackage{atlasbiblatex}
 
\usepackage{atlasphysics}
 
\graphicspath{{logos/}{figures/}{figures/general/}{figures/llqq}{figures/lvqq}{figures/vvqq}}
 
\usepackage{ANA-HDBS-2018-10-PAPER-defs}
\usepackage{xcolor,colortbl}

\AtlasTitle{Search for heavy diboson resonances in semileptonic final states in $pp$ collisions at $\sqrt{s}=13$~TeV with the ATLAS detector}
 
\AtlasAbstract{
This paper reports on a search for heavy resonances decaying into $WW$, $ZZ$ or $WZ$ using proton--proton collision data at a centre-of-mass energy of $\sqrt{s}=13$~TeV. The data, corresponding to an integrated luminosity of \intLumi\,\ifb, were recorded with the ATLAS detector from 2015 to 2018 at the Large Hadron Collider. The search is performed for final states in which one $W$ or $Z$ boson decays leptonically, and the other $W$ boson or $Z$ boson decays hadronically. The data are found to be described well by expected backgrounds.
Upper bounds on the production cross sections of heavy scalar, vector or tensor resonances are derived in the mass range 300--5000~GeV within the context of Standard Model extensions with warped extra dimensions or including a heavy vector triplet. Production through gluon--gluon fusion, Drell--Yan or vector-boson fusion are considered, depending on the assumed model.
}
 
\author{The ATLAS Collaboration}
 
\AtlasRefCode{HDBS-2018-10}

\AtlasJournal{EPJC}
\AtlasJournalRef{Eur. Phys. J. C 80 (2020) 1165}
\AtlasDOI{10.1140/epjc/s10052-020-08554-y}
\PreprintIdNumber{CERN-EP-2020-049}
\hypersetup{pdftitle={ATLAS document},pdfauthor={The ATLAS Collaboration}}
\begin{document}
 
\maketitle
 
\tableofcontents

\newpage
% The next lines are included from the .//tex/introduction.tex input file
\section{Introduction}
\label{sec:introduction}
 
Many extensions to the Standard Model~(SM) predict the existence of heavy resonances that decay into pairs of vector bosons ($WW$, $WZ$, and $ZZ$, collectively referred to as $VV$ with $V=W,Z$). These theoretically well-motivated extensions include the two-Higgs-doublet model~\cite{Branco:2011iw}, composite Higgs models~\cite{Dugan:1984hq,Agashe:2004rs}, technicolour~\cite{Eichten:2007sx,Catterall:2011zf,Andersen:2011yj} models,  and warped extra dimensions~\cite{Randall:1999ee,Davoudiasl:1999tf}. The Large Hadron Collider~(LHC), as the world's highest-energy proton--proton ($pp$) collider, is a unique facility for the search for these heavy resonances. Indeed, both the ATLAS and CMS collaborations have reported searches for diboson resonances in various production modes and in a variety of decay final states of the vector bosons~\cite{HIGG-2013-20,EXOT-2014-18,CMS-B2G-16-007,CMS-B2G-18-006,EXOT-2016-01,EXOT-2016-19,EXOT-2017-31,HDBS-2018-31}.
 
\begin{figure}[!h]
\centering
\subfigure[gluon--gluon fusion]{\includegraphics[width=0.325\textwidth]{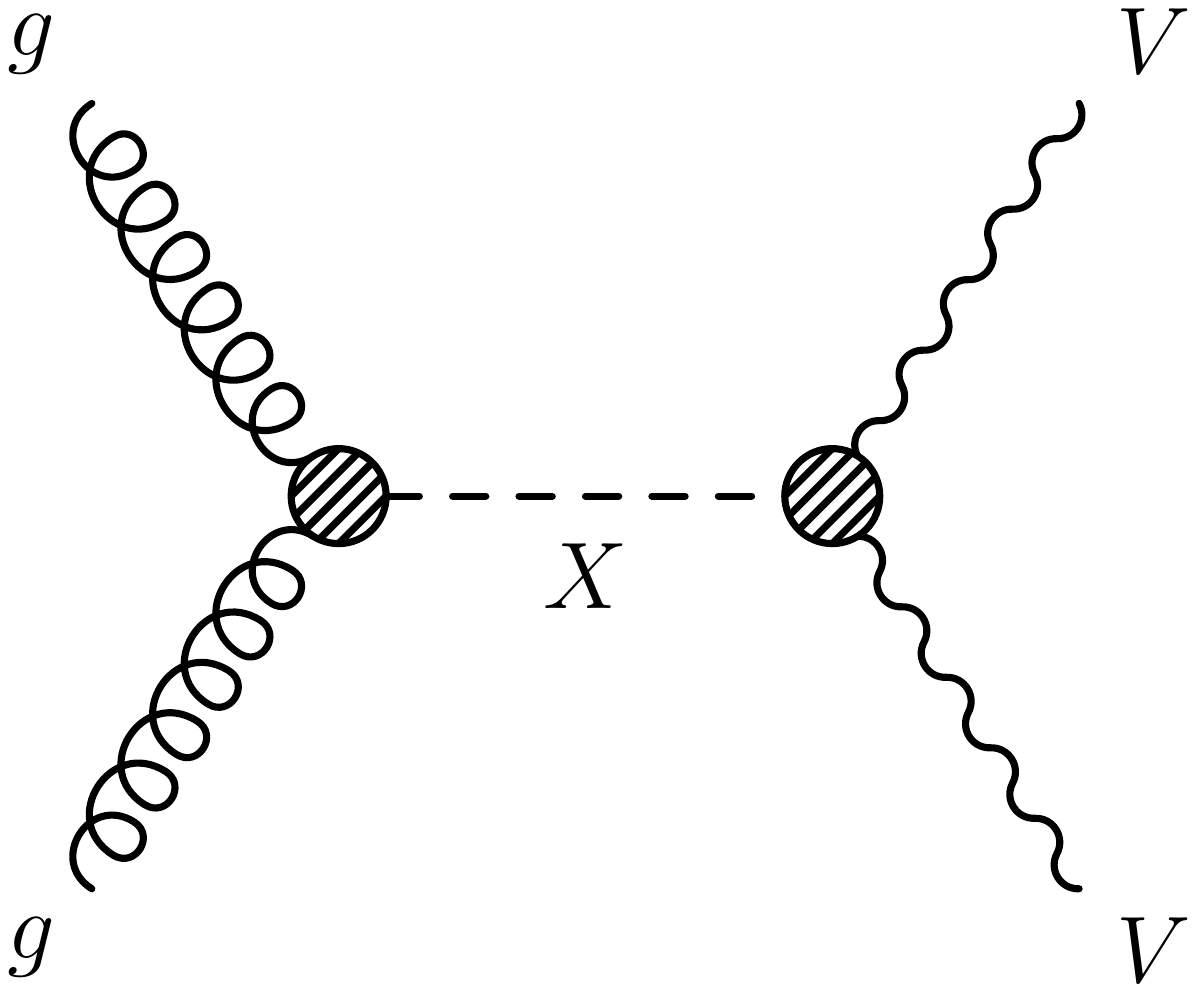}}
\subfigure[Drell--Yan]{\includegraphics[width=0.325\textwidth]{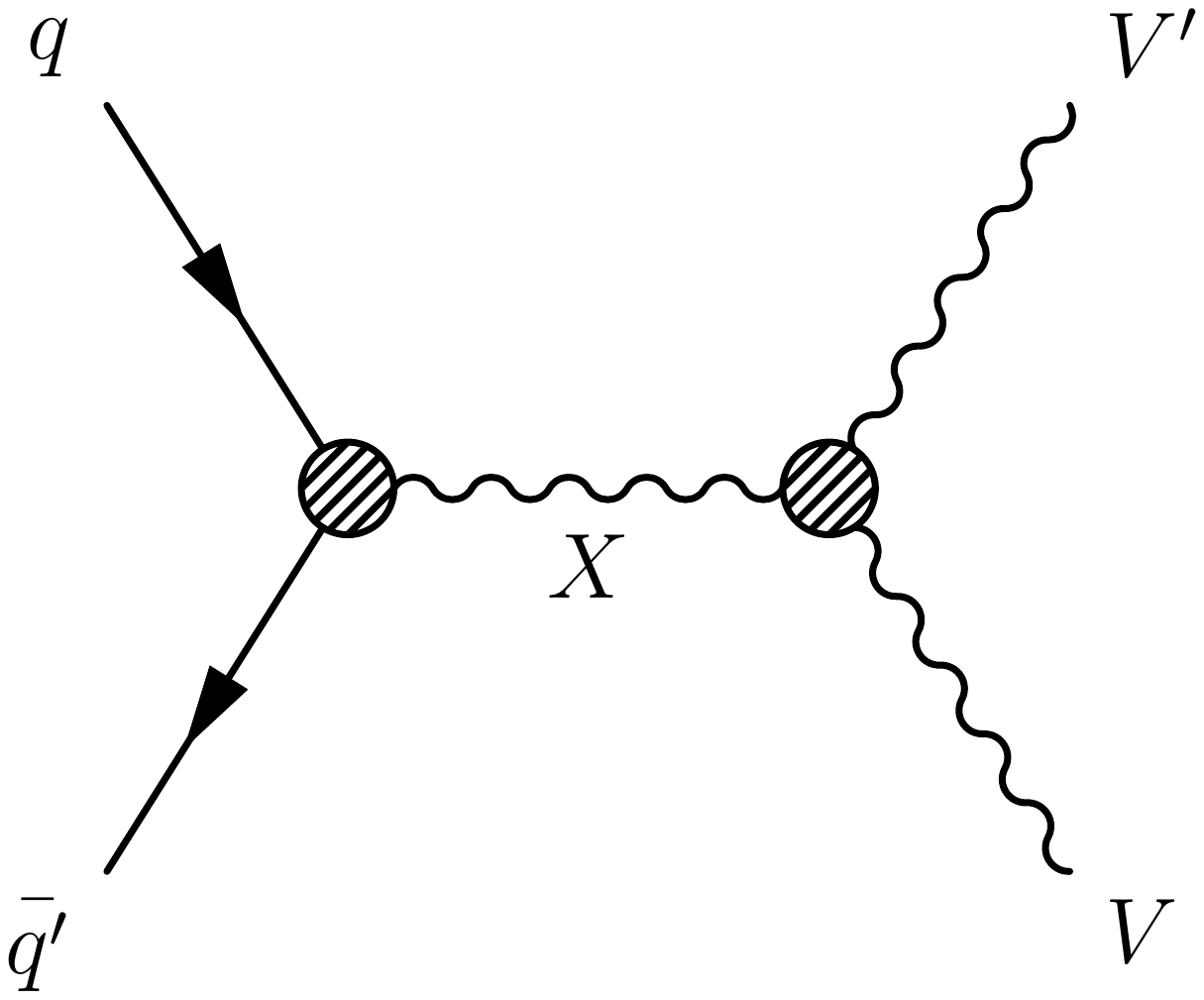}}
\subfigure[vector-boson fusion]{\includegraphics[width=0.325\textwidth]{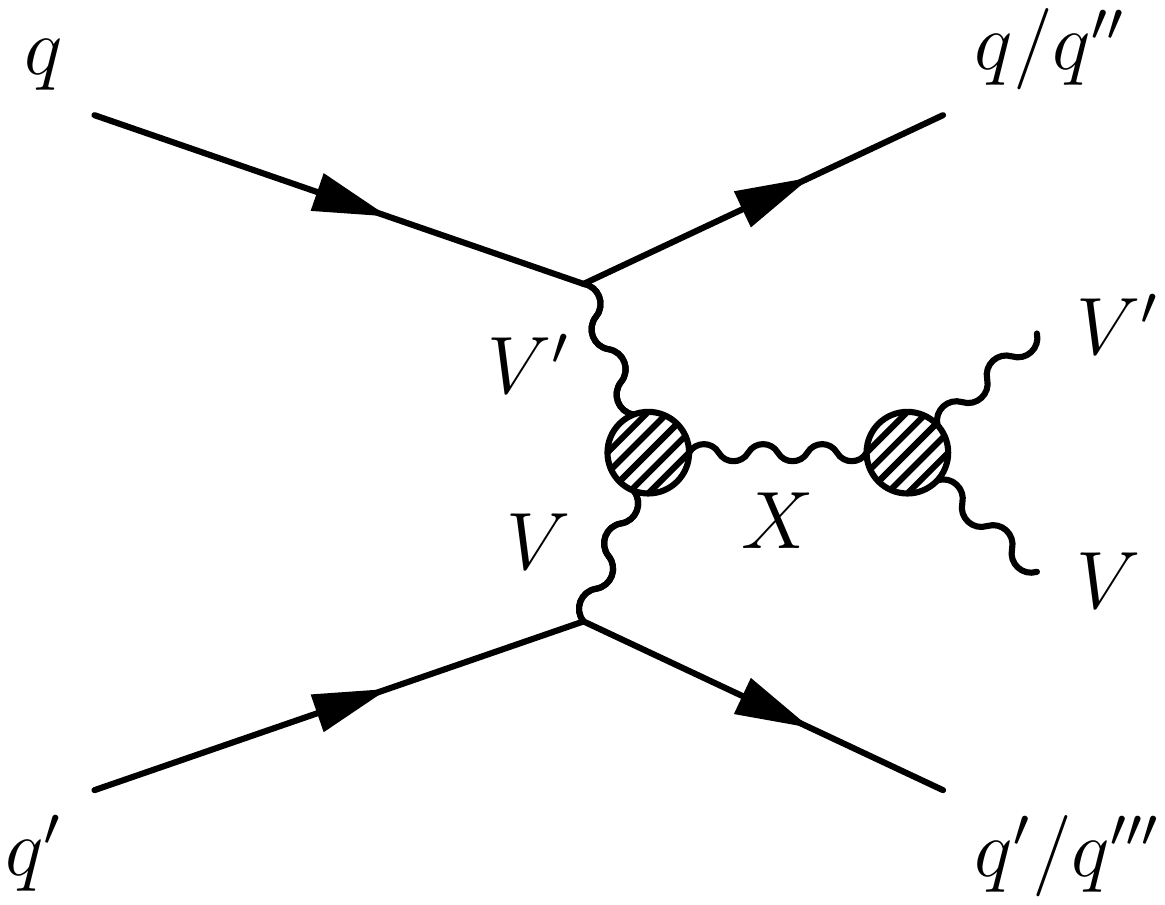}}
\caption{\label{fig:diagram} Representative Feynman diagrams for the production of heavy resonances $X$ with their decays into a pair of vector bosons. The hashed circles represent direct or effective couplings.}
\end{figure}

Depending on the assumed model, the predicted diboson resonances can be produced through gluon--gluon fusion~(ggF), Drell--Yan~(DY), or vector-boson fusion~(VBF) processes. Representative Feynman diagrams of these processes are shown in Figure~\ref{fig:diagram}.
 
This paper reports on a search for heavy resonances $X$ in the mass range 300~GeV to 5~TeV in the $X\to VV$ diboson decay in $pp$ collisions at $\sqrt{s}=13$~TeV. Three types of diboson resonances are considered in the search. The first is a neutral scalar resonance, the radion ($R$)~\cite{Goldberger:1999uk,Goldberger:1999un} which appears in some Randall--Sundrum (RS) models and which can decay into $WW$ or $ZZ$. The second is the heavier versions of the SM $W$ and $Z$ bosons, $\Wprime$ and $\Zprime$ bosons, as parameterised  in the Heavy Vector Triplet (HVT) framework~\cite{Pappadopulo:2014qza}, which can decay through $\Wprime\to WZ$ and $\Zprime\to WW$.  The third diboson resonance is a spin-2 graviton ($\Gkk$) of the first Kaluza--Klein~(KK) excitation in a bulk RS model~\cite{Han:1998sg,Randall:1999ee,Agashe:2007zd} and decays into $WW$ or $ZZ$.
 
Semileptonic $VV$ final states in which one vector boson decays leptonically ($V_\ell$: $\Wlv$, $\Zll$ or $\Zvv$) while the other decays hadronically ($V_h$: $\Vqq$) are considered, leading to three distinct channels: $ZV\to\vvqq$ (\zlep), $WV\to\lvqq$ (\olep), and $ZV\to\llqq$ (\tlep). Here $\ell$ denotes either an electron ($e$) or a muon ($\mu$). The hadronic $\Vqq$ decays are reconstructed either as two separate small-radius jets (small-$R$ jet, or $j$) or as one large-radius jet (large-$R$ jet, or $J$) depending on the transverse momentum (\pt) of the boson. The reconstructed transverse mass (\mT) of the $VV$ system for the \zlep channel and $VV$ invariant mass ($m_{VV}$) for the \olep and \tlep channels are used for signal--background discrimination via maximum-likelihood fits to their observed distributions.
 
Compared with previous searches in these final states~\cite{EXOT-2016-28,EXOT-2016-29}, the current search is performed with a data set approximately four times larger. Several improvements are made which include utilising a multivariate technique to identify and distinguish production processes, using tracking information in the large-$R$ jet reconstruction, and introducing $b$-quark jet tagging for large-$R$ jets.

% End of text imported from the .//tex/introduction.tex input file
 
% The next lines are included from the .//tex/detector.tex input file
\section{Detector and data sample}
\label{sec:detector}
 
\newcommand{\AtlasCoordFootnote}{
ATLAS uses a right-handed coordinate system with its origin at the nominal interaction point (IP) in the centre of the detector and the \(z\)-axis along the beam pipe. The \(x\)-axis points from the IP to the centre of the LHC ring, and the \(y\)-axis points upwards. Cylindrical coordinates \((r,\phi)\) are used in the transverse plane, \(\phi\) being the azimuthal angle around the \(z\)-axis. The pseudorapidity is defined in terms of the polar angle \(\theta\) as \(\eta = -\ln \tan(\theta/2)\). Angular distance is measured in units of \(\Delta R \equiv \sqrt{(\Delta\eta)^{2} + (\Delta\phi)^{2}}\).}
 
The ATLAS experiment~\cite{PERF-2007-01,PIX-2018-001} at the LHC is a multipurpose particle detector with a forward--backward symmetric cylindrical geometry and a near \(4\pi\) coverage in solid angle.\footnote{\AtlasCoordFootnote} It consists of an inner tracking detector surrounded by a thin superconducting solenoid providing a \SI{2}{\tesla} axial magnetic field,  electromagnetic and hadron calorimeters, and a muon spectrometer.
 
The inner tracking detector~(ID) covers the pseudorapidity range \(|\eta| < 2.5\). It consists of silicon pixel, silicon microstrip, and transition radiation tracking detectors. Lead/liquid-argon (LAr) electromagnetic calorimeters~(ECAL) provide electromagnetic (EM) energy measurements with high granularity. A steel/scintillator-tile hadron calorimeter~(HCAL) covers the central pseudorapidity range (\(|\eta| < 1.7\)). The endcap and forward regions are instrumented with LAr calorimeters for EM and hadronic energy measurements up to \(|\eta| = 4.9\). The muon spectrometer~(MS) surrounds the calorimeters and is based on three large air-core toroidal superconducting magnets with eight coils each. The field integral of the toroids ranges between \num{2.0} and \SI{6.0}{\tesla\metre} across most of the detector. The muon spectrometer includes a system of precision tracking chambers and fast detectors for triggering. A two-level trigger system \cite{TRIG-2016-01} selects events to be recorded at a reduced rate. The first level is a hardware implementation aiming to reduce the rate to around \SI{100}{\kilo\hertz}, while the software-based high-level trigger provides the remaining rate reduction to approximately \SI{1}{\kilo\hertz}.
 
This search uses the $pp$ collision data at $\sqrt{s}=13$~\TeV\ recorded by the ATLAS detector during the data-taking between 2015 and 2018 with a total integrated luminosity of $139.0\pm 2.4$~\fbi~\cite{ATLAS-CONF-2019-021}.
 
A combination of multiple single-lepton and missing transverse momentum (\met) triggers with varying thresholds, as well as lepton quality and isolation requirements is used~\cite{TRIG-2016-01,TRIG-2018-05}.
During data-taking, as the instantaneous luminosity increased, the thresholds for unprescaled single-lepton triggers with tight isolations were increased in stages: the electron transverse energy ($\et$) threshold was increased from 24 to 26~\GeV, and the muon transverse momentum ($\pt$) threshold was increased from 20 to 26~\GeV.  Similarly, the threshold of the \met triggers increased from 70 to 110~\GeV. Lepton triggers with tight isolations were complemented by those with looser isolations but higher $\et $ or $\pt$ thresholds.
The search uses the \met triggers in the \zlep channel and single-lepton triggers in the \tlep channel. The trigger efficiencies are greater than 90\% for signal events targeted by these two channels, independent of the resonance mass.
For the \olep channel,  the single-electron triggers were used in the electron case, but the single-muon triggers were used only for $\pt(\mu\nu)<150$~\GeV\ in the muon case. For $\pt(\mu\nu)>150$~\GeV, since the calculation of \met at the trigger level does not account for the presence of minimum ionising muons, the \met triggers were used instead. Using the unprescaled \met triggers miminises the impact of the efficiency loss from the limited geometric coverage of the muon triggers. The trigger efficiency for the \olep channel increases from approximately 80\% at 300~\GeV\ to be above 90\% at a resonance mass of 500~\GeV.

Events were retained for analysis if they were recorded with all detector systems operating normally and pass data-quality requirements~\cite{ATLAS-CONF-2015-029}.
Collision vertices are formed from tracks with $\pt>500$~\MeV.
The vertex candidate with the highest $\sum\pt^2$ of its associated tracks is selected as the primary vertex. All events are required to contain a primary vertex with at least two associated tracks.

% End of text imported from the .//tex/detector.tex input file
 
% The next lines are included from the .//tex/simulation.tex input file
\section{Simulation of signal and background processes}
\label{sec:simulation}
 
Monte Carlo~(MC) simulations were used for background modellings and estimations, evaluations of signal efficiencies, optimisations of event selections, and estimations of systematic uncertainties. Generated signal and background events were processed through the full ATLAS detector simulation program~\cite{SOFT-2010-01} based on $\GEANT4$~\cite{Agostinelli:2002hh}. Multiple overlaid $pp$ collisions~(pile-up) were simulated with the soft QCD processes of \textsc{Pythia}~8.186~\cite{Sjostrand:2007gs} using the A3 set of tuned parameters~\cite{ATL-PHYS-PUB-2016-017} and the \texttt{NNPDF23lo} parton distribution function~(PDF) set~\cite{Ball:2012cx}. All simulated events are processed with the same trigger and reconstruction algorithm as the data. Scale factors were used to correct differences between the data and simulations.

\subsection{Signal models and simulation}
 
Three types of resonances corresponding to different spins are considered in the search. The first one is a scalar neutral radion, introduced in some bulk RS models to stabilise the radius of the compactified extra dimension $r_\mathrm{c}$~\cite{Goldberger:1999uk,Goldberger:1999un}. The coupling of the RS radion field to SM fields is inversely proportional to $\Lambda_R = \mathrm{e}^{-k\pi r_\mathrm{c}} \sqrt{6M^3_5/k}$~\cite{Oliveira:2014kla,Barger:2011qn,Csaki:2007ns}, where $M_5$ is the five-dimensional Planck mass, and $k$ is the curvature factor.
The RS radion events were simulated with $k\pi r_\mathrm{c}=35$ and $\Lambda_R=3$~\TeV~\cite{Barger:2011qn}. The RS radion couples to SM fermions with a strength proportional to the fermion mass and to SM vector bosons with a strength proportional to the square of the boson mass, similarly to a heavy Higgs boson. However, the RS radion has a much narrower width due to its overall weaker couplings to SM particles. For example, the intrinsic width of a 3~\TeV\ RS radion is approximately 3\% of its mass, assuming $\Lambda_R=3$~\TeV. RS radions can be produced through both the ggF and VBF processes at the LHC as shown in Figure~\ref{fig:diagram}.

The second type considered comprises two heavy vector bosons described in the HVT framework~\cite{Pappadopulo:2014qza}: an electrically charged $\Wprime$ boson and an electrically neutral $\Zprime$ boson produced through the DY and VBF processes. The new heavy vectors couple to the Higgs and the SM gauge bosons via a combination of parameters $g_V c_H$ and to the fermions via the combination $g^2/g_V~c_F$. The parameter $g_V$ represents the typical strength of the vector boson interaction, while the parameters $c_H$ and $c_F$ are expected to be of the order of unity in most models.
Benchmark Model A~\cite{Pappadopulo:2014qza}  ($g_V =1$) is representative of a model of weakly coupled vector resonances in an extension of the SM gauge group where the HVT bosons have comparable decay branching ratios into SM fermions and vector bosons. Model B~\cite{Pappadopulo:2014qza} with $g_V = 3$, is representative of a composite model scenario where the HVT boson couplings to fermions are suppressed. In \textit{Model C}, $g_V=c_H=1$ and the HVT boson coupling to fermions was set to zero, so that only VBF production is possible. The $\Wprime\to WZ$ and $\Zprime\to WW$ decays were considered in this search.
 
The third benchmark resonance searched for is a spin-2 bulk RS graviton $\Gkk$ which appears as the first KK excitation of the gravitational field in a bulk RS graviton model~\cite{Han:1998sg,Randall:1999ee,Agashe:2007zd}. The $\Gkk$ couplings to light fermions are suppressed and therefore decays into final states involving heavy quarks, Higgs or vector bosons are favoured.  The strength of the coupling depends on $\kMPl$, where $k$ corresponds to the curvature of the warped extra dimension and $\MPl$ is the effective four-dimensional Planck scale. The value of $\kMPl$ is typically of ${\cal O}(1)$, and this and the $\Gkk$ mass are the only two free parameters. The $\Gkk$ has a mass-dependent width, which is 3.7\% of its mass at 500~\GeV\ and 6.4\% at 5~\TeV\ for $\kMPl=1$. It can be produced through  the ggF and VBF processes and decays into $WW$ and $ZZ$ with sizeable branching ratios. The $\Gkk$ samples were generated with $\kMPl=1$.

Signal events for the HVT and bulk RS graviton (radion) models were generated with \MGMCatNLO v2.2.2~(v2.6.1)~\cite{Alwall:2014hca}
at leading order (LO) using the \texttt{NNPDF23lo} PDF set.
For the production of resonances in the HVT model, both the DY and VBF mechanisms were simulated,
and the RS radion and \Gkk\ resonances were produced via both the ggF and VBF mechanisms.
For all signal models and production mechanisms, the generated events were interfaced
to \PYTHIAV{8.186} (8.230 for the RS radion model)~\cite{Sjostrand:2014zea} for parton showering, hadronisation, and the underlying event.
This interface relied on the A14 set of tuned parameters~\cite{ATL-PHYS-PUB-2014-021}
for events generated with \MGMCatNLO at LO.

As examples, Table~\ref{tabl:signal_parameters} shows the theoretical cross-sections, the diboson decay branching ratios, and the total widths of the resonances for two different mass values.
 
% The next lines are included from the .//tex/signaltable.tex input file
 
\begin{table}[h]
\centering
\scriptsize
\caption{List of benchmark signal models. Predictions of cross-section $\sigma$, branching ratio $\mathcal{B}$ into $WW$, $WZ$, or $ZZ$, and intrinsic width divided by the resonance mass $\Gamma/m$, for the given hypothetical new particle at $m=800\,\GeV$ and $3\,\TeV$ are summarised.}
\label{tabl:signal_parameters}
\begin{tabular}{l|c|c|c|ccc|ccc}
\hline \hline
\multicolumn{3}{c|}{\multirow{2}{*}{Model}}                   & \multirow{2}{*}{Spin} & \multicolumn{3}{c|}{$m = 800\,\GeV$}                   & \multicolumn{3}{c}{$m = 3\,\TeV$} \\  \cline{5-10}
\multicolumn{3}{c|}{}                                         &                     & $\sigma$ [pb]          & $\mathcal{B}$  &  $\Gamma/m$            & $\sigma$ [fb]              & $\mathcal{B}$  & $\Gamma/m$   \\
\hline\hline
\multicolumn{2}{c|}{RS radion ($k\pi r_\mathrm{c}=35$,}     & $R \to WW$        & \multirow{2}{*}{0} & 0.54 (ggF)                 & 0.43          & \multirow{2}{*}{$2.6 \times 10^{-3}$} & 1.38 (ggF)             & 0.44 & \multirow{2}{*}{0.032} \\
\multicolumn{2}{c|}{$\Lambda_R=3\,\TeV$)}       & $R \to ZZ$        &                    & $1.1\times 10^{-3}$ (VBF)  & 0.21          &                      & $5.5\times 10^{-3}$ (VBF)               & 0.22 &  \\
\hline
\multirow{6}{*}{HVT} & \multirow{2}{*}{Model A} & $\Wprime \to WZ$  & \multirow{6}{*}{1} & 53   & 0.024 & \multirow{2}{*}{0.026}  & 79     & \multirow{2}{*}{0.020} & \multirow{2}{*}{0.025} \\
&                          & $\Zprime \to WW$  &                    & 26   & 0.023 &                         & 36                       &      & \\
\cline{2-3}\cline{5-10}
& \multirow{2}{*}{Model B} & $\Wprime \to WZ$  &                    & 1.6  & 0.43  & \multirow{2}{*}{0.040}    & 5.5    & \multirow{2}{*}{0.47} & \multirow{2}{*}{0.031} \\
&                          & $\Zprime \to WW$  &                    & 0.86 & 0.41  &                           & 2.5                     &      & \\
\cline{2-3}\cline{5-10}
& Model C                  & $\Wprime \to WZ$  &                    & $4.0 \times 10^{-3}$  & 0.50 & \multirow{2}{*}{$3.5\times 10^{-3}$}   & $1.6 \times 10^{-3}$  & \multirow{2}{*}{0.50} & \multirow{2}{*}{$3.3 \times 10^{-3}$} \\
& (VBF)                    & $\Zprime \to WW$  &                    & $2.7 \times 10^{-3}$  & 0.49 &                           & $1.0 \times 10^{-3}$  & & \\
\hline
\multicolumn{2}{c|}{Bulk RS $\Gkk$}             & $\Gkk \to WW$     & \multirow{2}{*}{2} & 1.9 (ggF)       & 0.28 & \multirow{2}{*}{0.051} & 0.47 (ggF) & 0.20 & \multirow{2}{*}{0.062} \\
\multicolumn{2}{c|}{($\kMPl = 1.0$)}            & $\Gkk \to ZZ$     &                    & 0.050 (VBF)             & 0.14 &                & $1.6\times 10^{-2}$ (VBF)               & 0.10 &\\
\hline \hline
\end{tabular}
\end{table}

% End of text imported from the .//tex/signaltable.tex input file
 
\subsection{Background process simulation}
 
Background processes include $W$ and $Z$ boson production in association with jets ($\Wjets$ and $\Zjets$, collectively denoted by $\Vjets$), top-quark production (both top-quark pair, \ttbar, and single-top-quark), non-resonant diboson production ($WW, WZ$ and $ZZ$), and multijet production. MC samples were produced to model these background processes with the exception of multijet production, for which data were used to estimate its contribution.
 
The production of $V+$jets was simulated with the
\textsc{Sherpa}~v2.2.1~\cite{Bothmann:2019yzt}
generator using the matrix elements (ME) with next-to-leading order~(NLO) accuracy for up to two jets, and with leading-order~(LO) accuracy
for up to four jets, calculated with the Comix~\cite{Gleisberg:2008fv}
and OpenLoops~\cite{Cascioli:2011va,Denner:2016kdg} libraries. They
were matched with the \textsc{Sherpa} parton shower~\cite{Schumann:2007mg} using the MEPS@NLO
prescription~\cite{Hoeche:2011fd,Hoeche:2012yf,Catani:2001cc,Hoeche:2009rj}
using the set of tuned parameters developed by the \textsc{Sherpa} authors.
The \texttt{NNPDF30nnlo} set of PDFs~\cite{Ball:2014uwa} was used and the samples
were normalised to a next-to-next-to-leading-order (NNLO)
prediction~\cite{Anastasiou:2003ds} with a flat $K$-factor.
Simulated $V$+jets events from \textsc{MadGraph5\_aMC@nlo}~v2.2.2~\cite{Alwall:2014hca} using LO-accurate
ME with up to four final-state partons were used to estimate the possible mismodelling of the \textsc{Sherpa} sample.
The ME calculation employed the \texttt{NNPDF30nlo} set of PDFs~\cite{Ball:2014uwa} or the
\texttt{NNPDF23lo} set of PDFs. Events were
interfaced to \textsc{Pythia}~8.186 for the modelling
of the parton shower, hadronisation, and  underlying event.
The A14 tune~\cite{ATL-PHYS-PUB-2014-021} of \textsc{Pythia} was used with the
\texttt{NNPDF23lo} PDF set.
The decays of bottom and charm hadrons were performed by EvtGen~v1.2.0.

Samples of \ttbar\ and single-top-quark events were generated with \textsc{Powheg-Box}~\cite{Frixione:2007nw,Nason:2004rx,Frixione:2007vw,Alioli:2010xd}~v2 at NLO with the \texttt{NNPDF30nlo} PDF set.
The parameter $h_{\mathrm{damp}}$, which regulates the high-\pt\ radiation in the \textsc{Powheg},
was set to 1.5~$m_{t}$ to obtain good data--MC agreement at high \pt~\cite{ATL-PHYS-PUB-2016-020}, where $m_{t} = 172.5~\GeV$ was the top-quark mass used in the simulation.
The parton shower, fragmentation, and underlying event were simulated using \textsc{Pythia}~8.230~\cite{Sjostrand:2014zea} with the \texttt{NNPDF23lo} PDF set and the A14 tune.
The decays of bottom and charm hadrons were performed by EvtGen~v1.6.0.

Diboson processes were simulated with \textsc{Sherpa}~v2.2.1 using the ME at NLO accuracy in QCD for up to one additional parton and at LO accuracy for up to three additional parton emissions, including off-shell effects and Higgs boson contributions. The \texttt{NNPDF30nnlo} PDF set was used.
The electroweak $VVjj$ samples were generated by \textsc{MadGraph5\_aMC@NLO}~2.4.3~\cite{Alwall:2014hca} and were used, together with the \textsc{Sherpa} diboson sample, for the VBF analysis. The \texttt{NNPDF30lo} PDF set was used. The parton showers and hadronisation were modelled with \textsc{Pythia}~8.186 using the A14 tune.
 
Theoretical cross-sections were used to normalise background contributions.
The cross-sections of single-top-quark $t$- and $s$-channel production were calculated with the Hathor v2.1 program~\cite{Kant:2014oha, Aliev:2010zk}, while the $Wt$-channel followed the prescriptions from Refs.~\cite{Kidonakis:2011wy, Kidonakis:2010ux}. Cross-sections for diboson production were calculated at NLO~\cite{Campbell:2011bn,Alioli:2010xd}.
The normalisations of $\Vjets$ and \ttbar contributions were estimated from data using the control regions as described in Section~\ref{sec:controlRegions}.

% End of text imported from the .//tex/simulation.tex input file
 
% The next lines are included from the .//tex/objects.tex input file
\section{Object reconstruction and identification}
\label{sec:objects}
 
Leptons, jets, and \met are basic building blocks for this search. Their identification requirements are summarised briefly in this section.
 
\subsection{Leptons}
\label{sec:leptons}
 
Electrons are reconstructed from energy clusters that are consistent with EM showers in the ECAL and are matched to tracks in the ID~\cite{EGAM-2018-01}. They are required to have transverse energy $\et>7$~\GeV\ and pseudorapidity $|\eta|<2.47$, excluding the ECAL barrel--endcap transition region: $1.37<|\eta|<1.52$. To reduce backgrounds from misidentification and non-prompt sources, electrons must meet a likelihood-based criterion~\cite{EGAM-2018-01}. The likelihood is used to classify electrons as having either \textit{Loose}, \textit{Medium}, or \textit{Tight} quality.
 
Muons are identified by matching MS tracks with those in the ID and are required to have transverse momentum $\pt>7$~\GeV\ and pseudorapidity $|\eta|<2.5$. An identification requirement based on information from the ID and MS systems is applied to reduce backgrounds from misreconstruction and muons originating from hadron decays in flight. Similarly to electrons, muons are classified as having either \textit{Loose}, \textit{Medium}, or \textit{Tight} quality~\cite{PERF-2015-10}.
 
Leptons are required to have associated tracks satisfying $|d_0/\sigma_{d_0}|<5~(3)$ and $|z_0\times\sin\theta|<0.5$ mm for electrons~(muons), where $d_0$ is the transverse impact parameter relative to the beam line, $\sigma_{d_0}$ is its uncertainty, and $z_0$ is the distance between the longitudinal position of the track along the beam line at the point where $d_0$ is measured and the longitudinal position of the primary vertex.
 
Leptons from $W$ and $Z$ boson decays are required to have $\pt>30$~\GeV. They are expected to be isolated from other energy deposits in the detector. Thus, isolation criteria based on the sum of track $\pt$, the sum of calorimeter $\et$, or both, in small cones around the lepton direction are used to further reduce backgrounds from non-isolated sources. Leptons of \textit{Loose} quality with $\pt<100$~\GeV\ are required to pass a \texttt{FixedCutLoose} isolation requirement and no isolation requirement is applied for $\pt>100$~\GeV\
so that the leptons from high-\pt $Z \to \ell\ell$ decays are not removed in the presence of nearby leptons.
Details can be found in Refs.~\cite{EGAM-2018-01,PERF-2015-10}.

\subsection{Jets}
\label{sec:jets}
 
Small-$R$ jets are reconstructed from calorimeter energy clusters using the anti-$k_t$ algorithm~\cite{Cacciari:2008gp,Cacciari:2011ma} with a radius parameter of $R=0.4$. Energy- and $\eta$-dependent correction factors derived from MC simulations are applied in order to correct jets back to the particle level~\cite{PERF-2016-04}. Jets are required to have $\pT>30$~\GeV\ and $|\eta|<4.5$. To suppress jets from pile-up interactions, a jet vertex tagger~\cite{ATLAS-CONF-2014-018} is applied to jets with $\pT<120$~\GeV\ and $|\eta|<2.5$, based on information about tracks associated with the primary vertex and pile-up vertices. For forward jets, the uncertainty on pileup modelling is taken into account.
 
A multivariate algorithm for the identification of small-$R$ jets containing $b$-hadrons ($b$-tagging)~\cite{FTAG-2018-01} is used. The algorithm is based on information such as track impact-parameter significances and positions of reconstructed secondary decay vertices. The identified jets, called $b$-jets, are restricted up to $|\eta|<2.5$ due to the ID coverage. The $b$-tagging algorithm has an efficiency of 85\% for $b$-hadrons in simulated $\ttbar$ events, a light-flavour jet rejection factor of 33 and a $c$-jet rejection of about 3~\cite{FTAG-2018-01}.
 
Large-$R$ jets are reconstructed from track-calo clusters~\cite{ATL-PHYS-PUB-2017-015} with the anti-$k_t$ algorithm, but with the radius parameter increased to $R=1.0$.
The track-calo clusters are formed by combining information from the calorimeter and the ID, utilising the excellent angular resolution of the ID and the improving energy resolution of the calorimeter at high energies. A trimming algorithm~\cite{Krohn:2009th} is applied to reduce the impact of pile-up and soft radiation overlapping with the jet. The constituents of each jet are reclustered with the $k_t$ algorithm~\cite{Ellis:1993tq} into smaller $R = 0.2$ subjets and those subjets are removed if $p_{\mathrm{T}}^{\mathrm{subjet}} / \ptJ < 0.05$, where $p_{\mathrm{T}}^{\mathrm{subjet}}$ and \ptJ\ are the transverse momenta of the subjet and the large-$R$ jet, respectively.
The large-$R$ jets are required to have $\pt>200$~\GeV, $|\eta|<2.0$, and a jet mass ($\mJ$) greater than 50~\GeV.

Variable-radius (VR) jets are used to identify $b$-jets from boosted hadronic $\Vqq$ decays that are reconstructed as large-$R$ jets. They are reconstructed from ID tracks associated with large-$R$ jets by using the anti-$k_t$ algorithm with a $\pT$-dependent radius $R$ parameter between 0.02 and 0.4 and a $\rho$-parameter of 30~\GeV~\cite{ATL-PHYS-PUB-2017-010}. They are required to have $\pt>10$~\GeV\ and $|\eta|<2.5$. The same $b$-tagging algorithm which is used for small-$R$ jets is applied to identify variable-radius jets from $b$-hadrons.

\subsection{Overlap removal}
An overlap-removal procedure is applied to the selected leptons and jets. If two electrons share the same track, or the separation between their two energy clusters satisfies $|\Delta\eta|<0.075$ and $|\Delta\phi|<0.125$, then the lower-\pt electron is discarded. Electrons that fall within $\Delta R=0.02$ of a selected muon are also discarded.
For nearby electrons and small-$R$ jets, the jet is removed if the separation between the electron and jet satisfies $\Delta R<$ 0.2; the electron is removed if the separation satisfies 0.2~$<\Delta R<$~0.4. For nearby muons and small-$R$ jets, the jet is removed if the separation between the muon and jet satisfies $\Delta R<$ 0.2 and if the jet has less than three tracks or the energy and momentum differences between the muon and the jet are small; otherwise the muon is removed if the separation satisfies $\Delta R<$ 0.4. To prevent double-counting of energy from an electron inside the large-$R$ jet, the large-$R$ jet is removed if the separation between the electron and the large-$R$ jet satisfies $\Delta R < 1.0$.

\subsection{Missing transverse quantities}
\label{sec:met}
 
The missing transverse momentum (\metvec) is calculated as the negative vectorial sum of the transverse momenta of calibrated electrons, muons, small-$R$ jets, and unassociated tracks. Large-$R$ jets are not included in the \metvec\ calculation to avoid double-counting of energy between the small-$R$ jets and large-$R$ jets. Energy depositions due to the underlying event and other soft radiation are taken into account by constructing a `soft term' from ID tracks associated with the primary vertex but not with any reconstructed object~\cite{PERF-2016-07}. Similarly, the track-based missing transverse momentum, $\mptvec$, is the negative vectorial sum of the transverse momenta of all good-quality inner-detector tracks that are associated with the primary vertex.

% End of text imported from the .//tex/objects.tex input file
 
% The next lines are included from the .//tex/selections.tex input file
\section{Event classification and selections}
\label{sec:selections}
 
\begin{figure}[!hbt]
\centering
\includegraphics[width=0.9\textwidth]{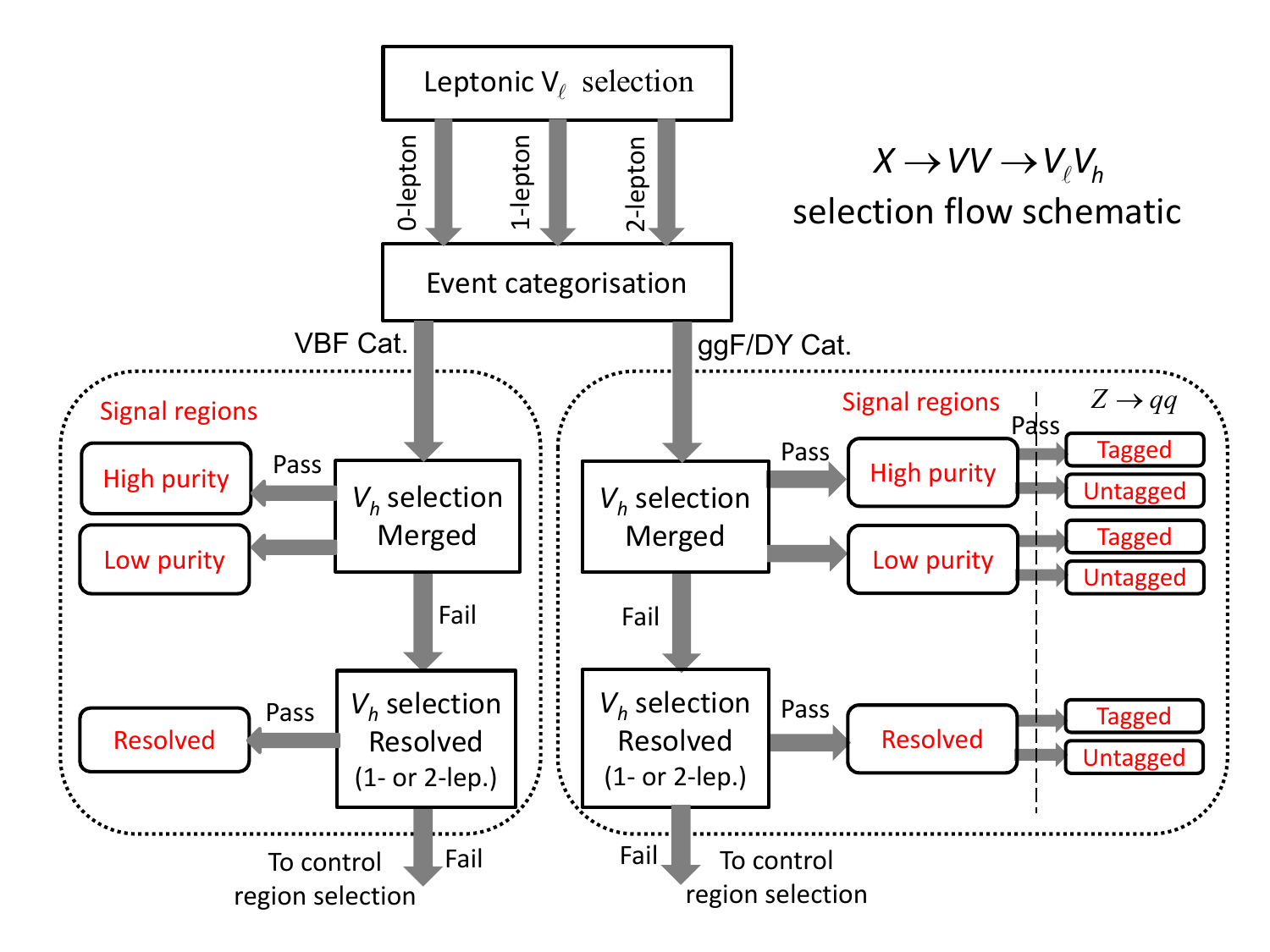}
\caption{
Illustration of the selection flow and signal regions of the $X\to VV\to\Vl\Vh$ search. The VBF category targets VBF production whereas the ggF/DY category is for the rest. Three signal regions (high purity, low purity and resolved) are selected for each category, based on the $\Vqq$ reconstruction. For the \zlep channel, no resolved selection is considered. For final states with hadronically decaying $Z$ bosons, the three signal regions in the ggF/DY category are each further split into tagged and untagged according to the $b$-tagging information about jets from $\Zqq$ decays. }
\label{fig:flow}
\end{figure}

The search begins with the selection of the leptonically decaying boson $\Vl$. Candidate events are first selected according to the number of \textit{Loose} leptons and assigned to \zlep ($\Vl=Z,\,Z\to\nu\nu$), \olep ($\Vl=W,\,W\to\ell\nu$) and \tlep ($\Vl=Z,\,Z\to\ell\ell$) channels. Other lepton multiplicities are excluded from the analysis. Although specific selections differ, the three channels follow the same analysis flow as illustrated in Figure~\ref{fig:flow}. For each channel, events are further classified into two exclusive VBF and ggF/DY categories as described in Section~\ref{sec:RNN}, targeting their corresponding production processes for heavy resonances.
 
The selection proceeds to identify the hadronically decaying boson $\Vh$. Depending on the $V_h$-boson momentum, the energy deposits of the two jets from the hadronically decaying $V$ bosons can be well separated or can largely overlap in the detector. Thus the $\Vqq$ decay, including $Z\to bb$, can be either reconstructed from two resolved small-$R$ jets ($\Vjj$) for low-energy bosons or identified as one merged large-$R$ jet ($\VJ$) for energetic bosons. The $\Vh$ candidates are identified  first through the merged $\VJ$ identification and then, if it fails, through the resolved $\Vjj$ reconstruction.
 
Selections specific to each channel are presented in Section~\ref{sec:selection}. Multiple signal regions (SRs) are defined in order to enhance search sensitivities, as described in Section~\ref{sec:regions}. Section~\ref{sec:discriminant} discusses the mass variables used as the final discriminants. The analysis flow is run twice, once for $\Vh=W$ and once for $\Vh=Z$, which involves selecting different ranges of $\mjj$ or $\mJ$.

\subsection{Categorisation of production processes}
\label{sec:RNN}
 
For the three production processes shown in Figure~\ref{fig:diagram}, the ggF and DY processes have the same final states while the VBF process possesses two additional jets, called VBF-tag jets. The kinematics of these jets differ from those from the $V$-boson hadronic decays. They are typically well separated in pseudorapidity and usually have large dijet invariant mass. These characteristics were used in previous searches~\cite{EXOT-2016-28,EXOT-2016-29} to separate VBF production from ggF/DY production. In this search, a recurrent neural network~(RNN) \cite{Sherstinsky:2018rnn, ATL-PHYS-PUB-2017-003} is used to classify the VBF and ggF/DY event topologies. It is built with the Keras~\cite{chollet2015keras} library using the Theano python library~\cite{2016arXiv160502688short} as a back end for mathematical computations. The RNN has 2 hidden layers with 25 recurrent cells to exploit the hidden correlation of the input sequence.

The RNN uses the four-momenta of small-$R$ jets as input. It is well suited for a variable-length input sequence such as the jet information.
The RNN allows to recover events with only one VBF-tag jet reconstructed ($\sim 30\%$ of signal events), and those events were not selected in previous searches where two VBF-tag jets were required~\cite{EXOT-2016-28,EXOT-2016-29}.
Although the RNN permits to deal with a large number of input jets, a maximum of two input jets is chosen to minimise the impact of systematic uncertainties associated with additional jets. Only a small increase (2--3\%) in the tagging efficiency of VBF events is observed if more than two jets are used as inputs.
 
For events with large-$R$ jets, small-$R$ jets with angular separations of $\Delta R<1$ from the leading large-$R$ jets are removed. If there is no large-$R$ jet in an event, the pair of small-$R$ jets with dijet invariant mass closest to the $V$-boson mass is removed. Up to two remaining small-$R$ jets with the highest $\pt$ are chosen as the input to the RNN. Events with no small-$R$ jets left are automatically classified as ggF/DY events.
 
The RNN score distributions depend on the assumed model of a heavy resonance, its mass and decay mode. The RNN trained with the 1~\TeV\ scalar resonance in the $X\to ZZ\to\llqq$ decay is applied for the three leptonic channels, the three resonance models, and all resonance masses.

\begin{figure}[!hbt]
\begin{center}
\subfigure[]{\includegraphics[width=0.48\textwidth]{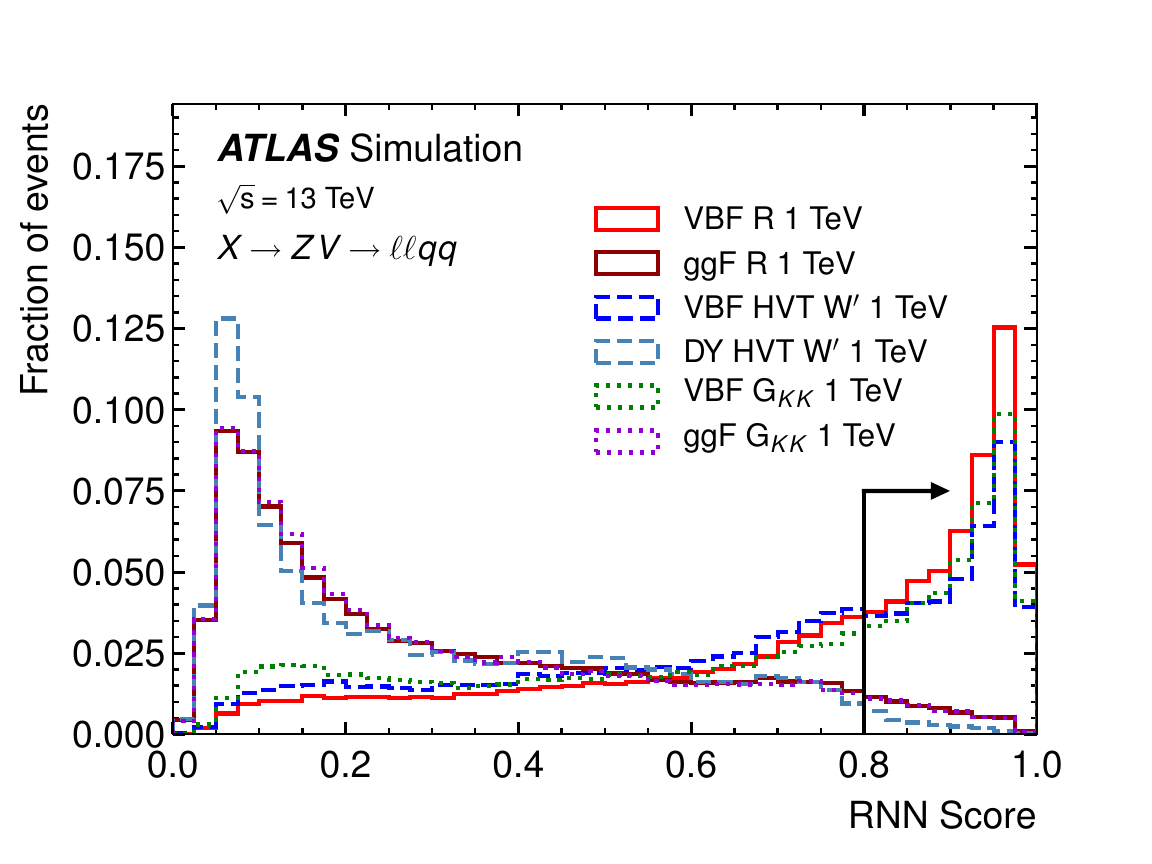}}
\subfigure[]{\includegraphics[width=0.48\textwidth]{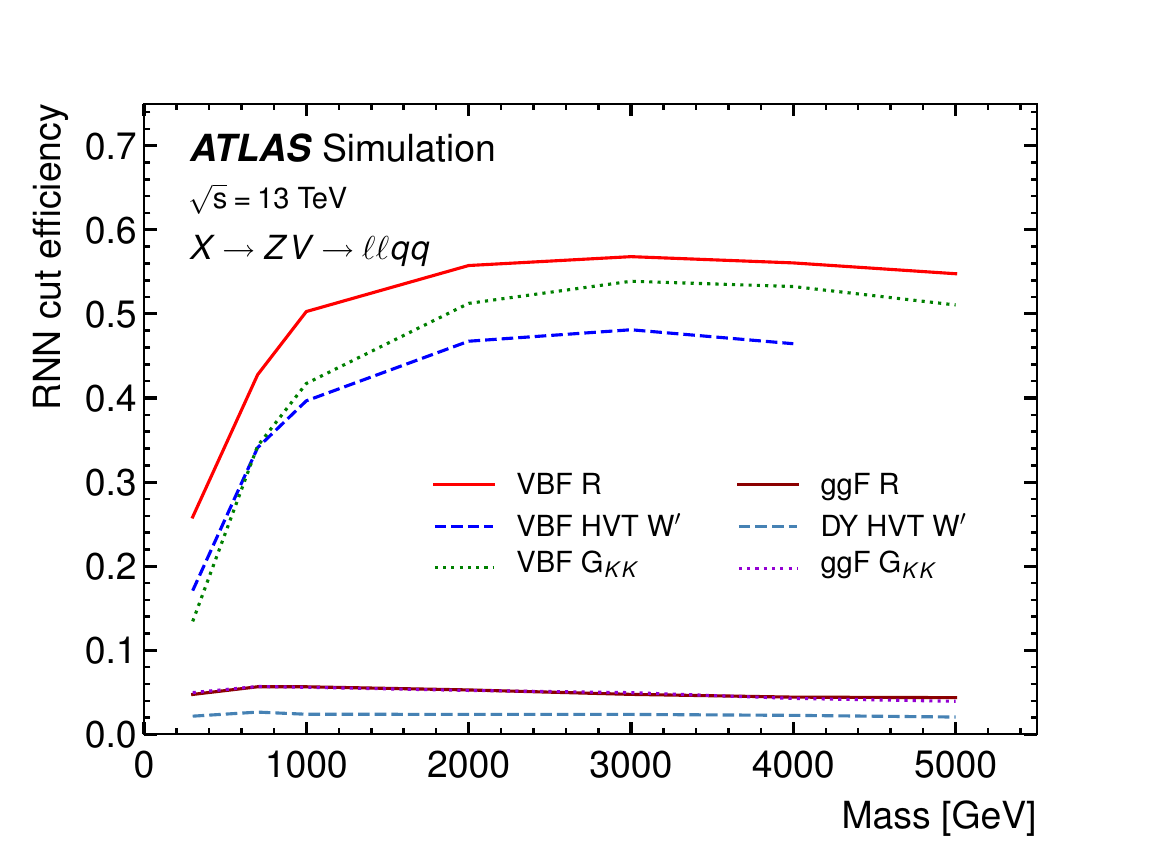}}
\end{center}
\caption{(a) RNN score distributions for the production of a 1~\TeV\ resonance in the signal models considered for this search;
(b) The fractions of signal events passing the VBF requirement on the RNN score as functions of the resonance mass for both VBF and ggF production.
}
\label{fig:RNN}
\end{figure}

Figure~\ref{fig:RNN}(a) compares the RNN score of simulated events from VBF and ggF/DY production of a 1~\TeV\ resonance in the signal models considered in this search. The RNN discrimination power increases with the resonance mass.
An event is classified as a VBF event if its RNN score is above 0.8 and otherwise as a ggF/DY event. The threshold is chosen to maximise the sensitivity to VBF signals.
Figure~\ref{fig:RNN}(b) shows the fractions of simulated signal events passing the RNN requirement as functions of the resonance mass for different signal models. The RNN correctly classifies VBF events more than 40\% of the time for a diboson resonance heavier than 1~\TeV\ with a ggF/DY contamination of about 2--5\%. It yields a relative increase in the VBF event selection efficiency from 10\% (5\%) at 0.5 TeV to 60\% (50\%) at 3 TeV for a scalar resonance (spin-1 or spin-2 resonance) compared to the previous cut-based selection~\cite{EXOT-2016-29,EXOT-2016-28}, with similar background rejections.

\subsection{Reconstruction and identification of the $\Vqq$ decay}
\label{sec:Vqq}

\begin{figure}[!hbt]
\begin{center}
\subfigure[]{\includegraphics[width=0.48\textwidth]{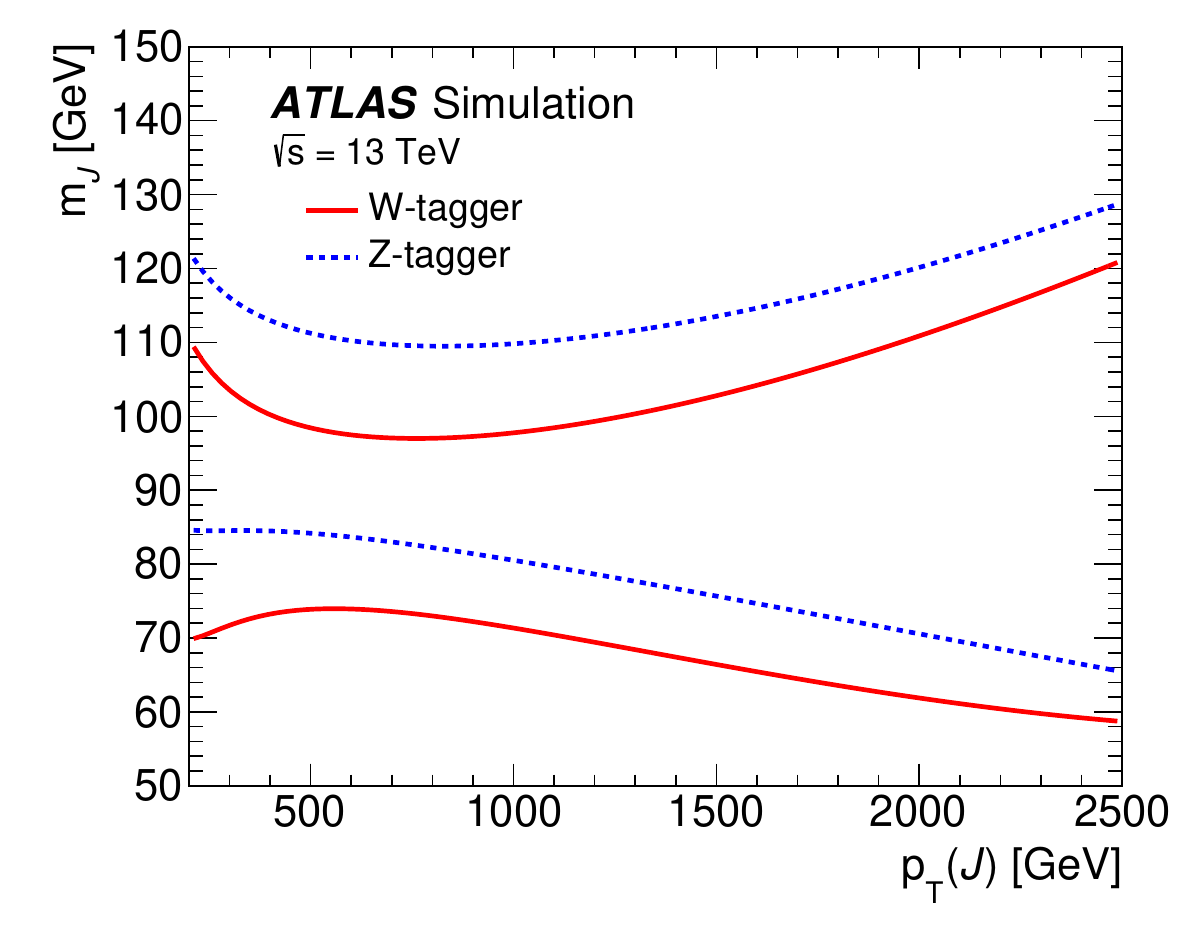}}
\subfigure[]{\includegraphics[width=0.48\textwidth]{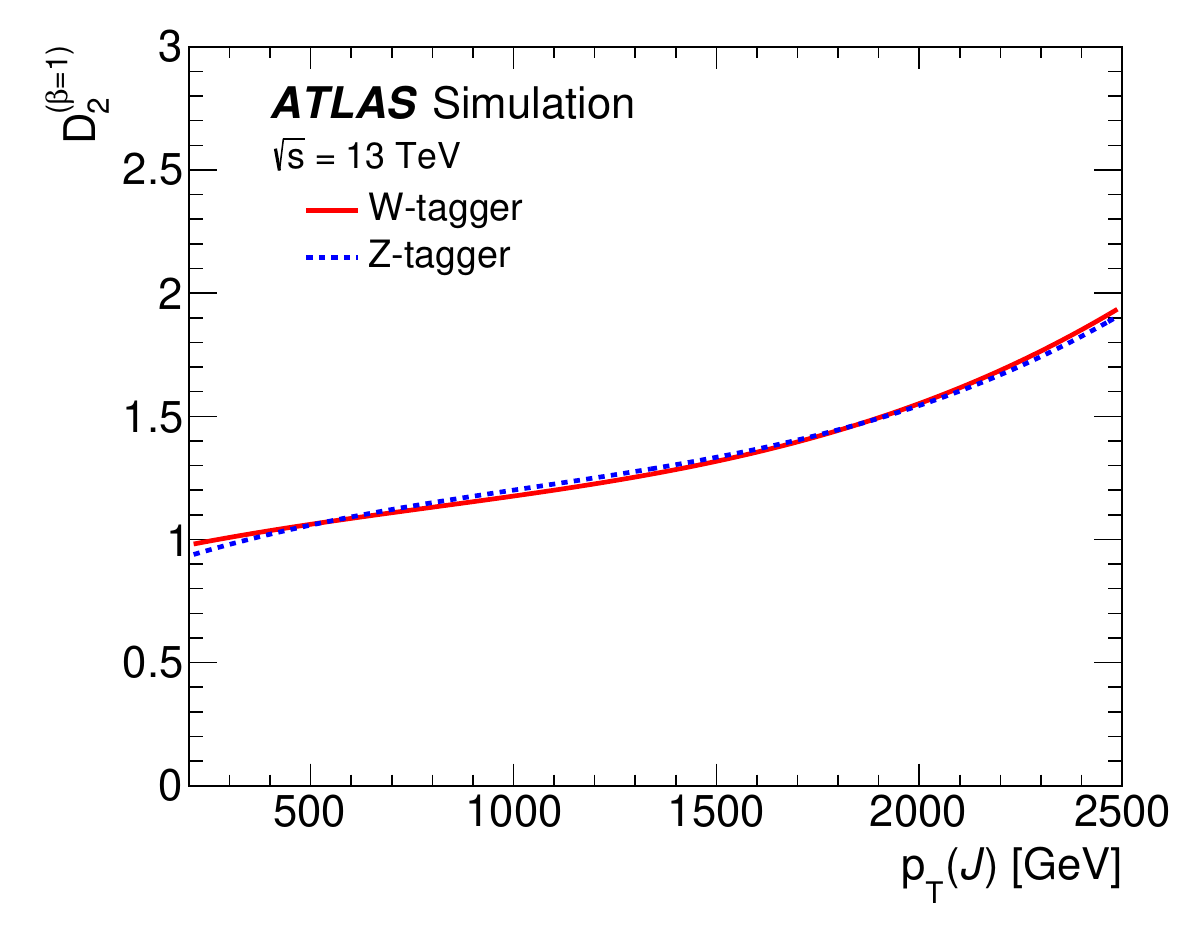}}
\end{center}
\caption{(a) The upper and lower bounds of $\mJ$ and (b) the upper~(lower) requirements on \DTwoBetaOne selecting the high~(low)-purity signal regions as functions of the large-$R$ jet $\pT$ for the $\VJ$ tagging for the $W$ boson and the $Z$ boson.}
\label{fig:VJmassD2}
\end{figure}

\begin{figure}[!hbt]
\begin{center}
\subfigure[]{\includegraphics[width=0.48\textwidth]{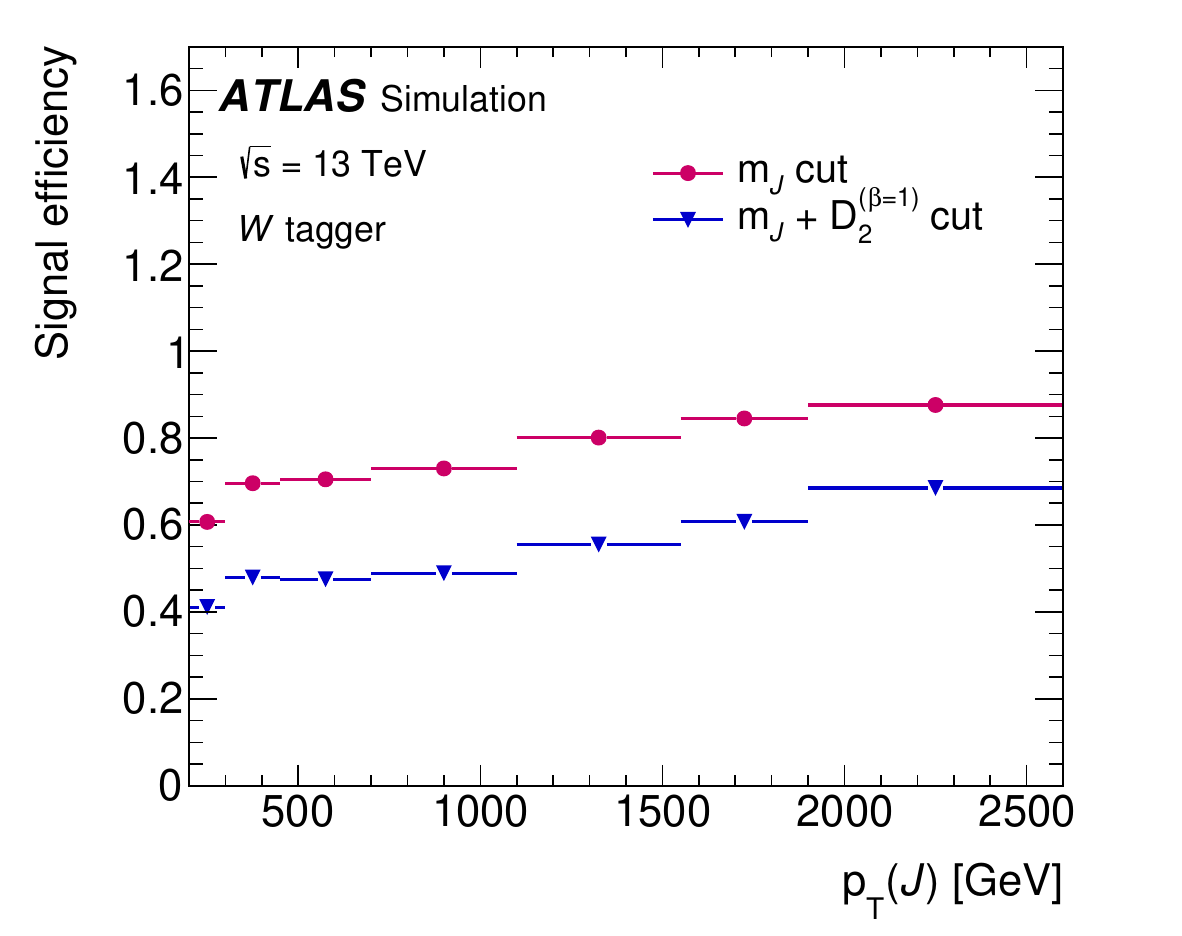}}
\subfigure[]{\includegraphics[width=0.48\textwidth]{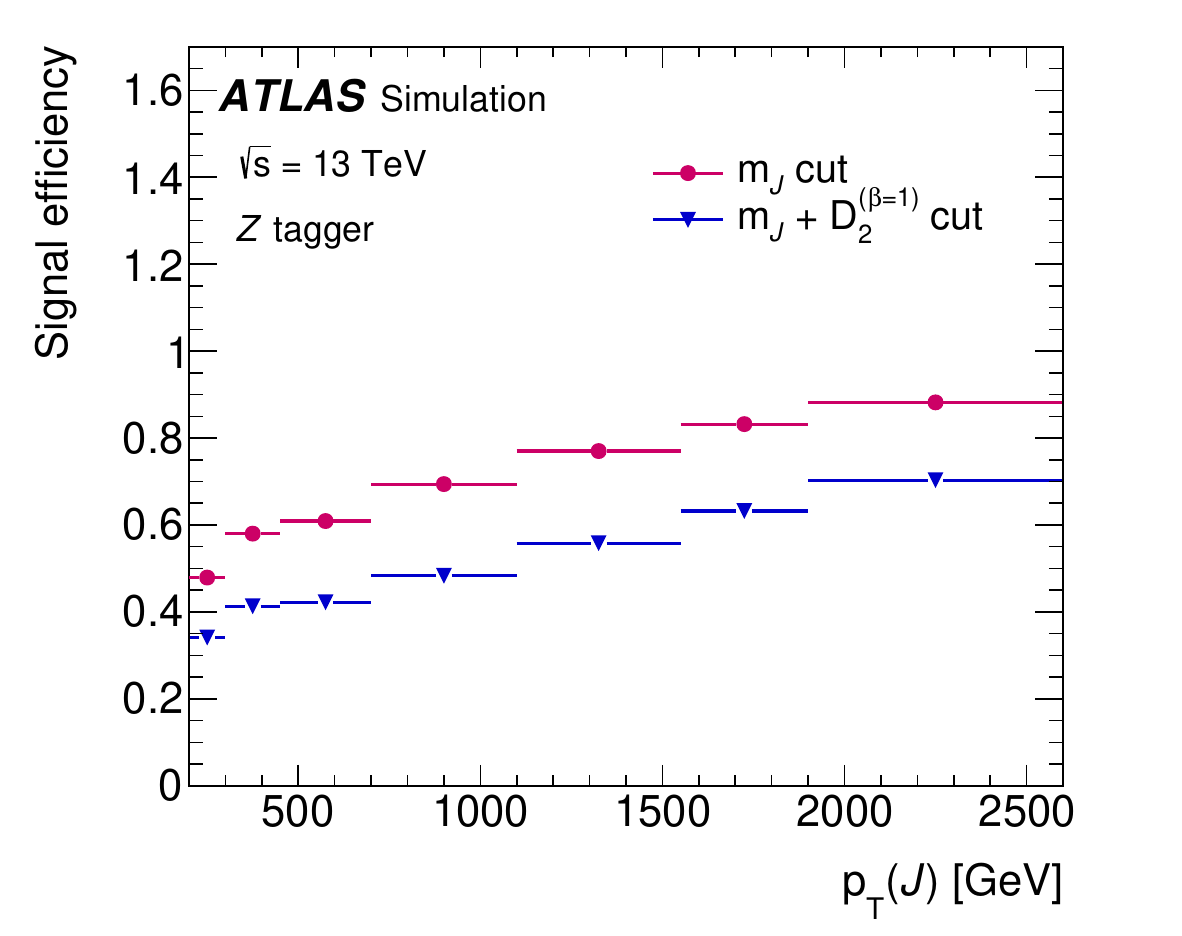}}
\end{center}
\caption{Efficiencies of the $\mJ$ and \DTwoBetaOne requirements as functions of the large-$R$ jet $\pT$ for the $\VJ$ tagging for (a) the $W$ boson and (b) the $Z$ boson.  }
\label{fig:VJmassD2Eff}
\end{figure}

The $\VJ$ candidates are identified from the highest-\pt large-$R$ jet in an event by requiring its mass $\mJ$ to be in a \pt-dependent window centred around the expected value of the $V$-boson mass from simulations~\cite{PERF-2015-03,ATL-PHYS-PUB-2015-033}, as shown in Figure~\ref{fig:VJmassD2}(a). The mass window depends on the jet mass resolution~\cite{ATL-PHYS-PUB-2017-015} and is approximately 30~\GeV\ wide at $\pt=500$~\GeV\ and increases to about 60~\GeV\ at $\pt=2.5$~\TeV.
A jet substructure variable \DTwoBetaOne is used to assess the quality of the $\VJ$ candidates. The variable \DTwoBetaOne is defined as the ratio of three-point to two-point energy correlation functions~\cite{Larkoski:2014gra,Larkoski:2015kga}  based on the energies and pairwise angular distances of particles within a large-$R$ jet. The variable is optimised to distinguish between jets originating from a single parton and those originating from $\Vqq$ decays.
A \pt-dependent upper~(lower) requirement on \DTwoBetaOne, shown in Figure~\ref{fig:VJmassD2}(b), is employed to select high~(low)-purity signal regions as described in Section~\ref{sec:regions}.
Efficiencies for the $\mJ$ requirement alone and for the combined $\mJ$ and \DTwoBetaOne requirements as functions of the large-$R$ jet $\pT$ are shown in Figure~\ref{fig:VJmassD2Eff}.
The efficiency  for tagging $\Vqq$ decay varies from approximately 40\% at low $\pt$ to 70\% at high $\pt$.
The background rejection factor of the $W$ ($Z$) tagger is estimated using the simulated $ W \to \ell \nu \left( Z \to \ell\ell \right)$+jets events, and is approximately 5~(6) at $\pt=200~$\GeV\ and 35~(30) at $\pt>700$~\GeV.
 
The $\Vjj$ candidates are reconstructed from two small-$R$ jets within $|\eta|<2.5$. The leading jet is required to have $\pt>60$~\GeV\ and the subleading jet is required to have $\pt>30\ (45)$~\GeV\ in the \tlep (\olep) channel. No resolved $\Vjj$ reconstruction is considered for the \zlep channel due to the large multijet background. The two highest-\pt small-$R$ jets in $|\eta|<2.5$
are chosen to form the $\Vjj$ candidate except for the $Z\to bb$ reconstruction, for which events are required to have exactly two $b$-tagged jets, and in which case they are used. The invariant mass of the two jets, $\mjj$, must be consistent with that of the $V$ boson by satisfying $62<\mjj<97$~\GeV\ for $\Wjj$ and $70<\mjj<105$~\GeV\ for $\Zjj$. Fixed mass windows are applied because the dijet mass resolution is largely independent of the dijet \pt for the resonance masses to which the resolved analysis is sensitive.

\subsection{Event selections for individual leptonic channels}
\label{sec:selection}

Event selections for all three leptonic channels consist of the selections for the leptonically and hadronically decaying $V$ bosons and an event-level selection designed to reduce backgrounds specific to each channel. The selection of hadronically decaying $V$ bosons is common to all three channels. It requires a $\Vqq$ candidate identified by either the merged or resolved technique as discussed above. The other selections are specific to individual leptonic channels and are described below. An overview of the selections is shown in Table~\ref{tab:selections}.
 
\renewcommand{\arraystretch}{1.2}
\begin{table}[!htb]
\scriptsize
\caption{Overview of the main $X\to VV\to \Vl \Vh$ selection criteria; the text gives more details. $\Rpt$ stands for $\min(\pt^{\Vl},\pt^{\Vh})/\mVV$.}
\begin{center}
\begin{tabular}{l|c|c|c}
\hline\hline
\multirow{2}{*}{Event selection}      &  \zlep              &  \olep                &   \tlep           \\
&  ($ZV\to\nu\nu \Vh$)       &  ($WV\to\ell\nu \Vh$) &  ($ZV\to\ell\ell \Vh$)  \\ \hline
\multirow{5}{*}{$\Vl$ selection}      &  No \textit{Loose} lepton  &  1 \textit{Tight} electron   &  2 \textit{Loose} leptons   \\
&  $\met>250$~\GeV            &  or 1 \textit{Medium} muon   &  with $p_\mathrm{T}^\ell > 30$~\GeV \\
&  $\mpt>50$~\GeV             &  with $p_\mathrm{T}^\ell > 30$~\GeV &  from the \\
&                            &  $\met>60$~\GeV   &   $\Zll$ candidate   \\
&                            &  $\ptVl>75$~\GeV  &    \\ \hline
 
\multirow{3}{*}{Event veto}          &  \multicolumn{3}{c}{No additional \textit{Loose} leptons} \\ \cline{2-4}
&             & Veto events with $b$-jets not associated & \\
&             & with the $V\to qq$ candidate &  \\ \hline
 
\multirow{2}{*}{Event categorisation} &  \multicolumn{3}{c}{$\ge 1$ large-$R$ jets or $\ge 2$ small-$R$ jets} \\
&  \multicolumn{3}{c}{VBF and ggF/DY classification according to RNN score}   \\ \hline
 
\multirow{6}{*}{$\Vh$ selection (Merged)}   &                          & $\met>100$~\GeV   &  \\
&                          & $\ptVl>200$~\GeV  &  \\ \cline{2-4}
& \multicolumn{3}{c}{$\ge 1$ large-$R$ jets}  \\
& \multicolumn{3}{c}{The leading jet passing \pt-dependent $\mJ$ requirement} \\ \cline{2-4}
&                          & $\Rpt>0.35$ (ggF/DY) & $\Rpt>0.35$ (ggF/DY)  \\
&                          & $\Rpt>0.25$ (VBF)   & $\Rpt>0.25$ (VBF)   \\ \hline
 
\multirow{6}{*}{$\Vh$ selection (Resolved)}       &  \cellcolor{gray}   &  \multicolumn{2}{c}{Failed merged selection} \\
&  \cellcolor{gray}   &  \multicolumn{2}{c}{$\ge 2$ small-$R$ jets with $|\eta|<2.5$} \\
&  \cellcolor{gray} Not   &  \multicolumn{2}{c}{$62<\mjj<97$~\GeV\ for $\Wjj$} \\
&  \cellcolor{gray} Performed  &  \multicolumn{2}{c}{$70<\mjj<105$~\GeV\ for $\Zjj$} \\ \cline{2-4}
&  \cellcolor{gray}   & $\Rpt>0.35$ (ggF/DY)  &  $\Rpt>0.35$ (ggF/DY)  \\
&  \cellcolor{gray}   & $\Rpt>0.25$ (VBF)     &  $\Rpt>0.35$ (VBF)  \\
 
\hline\hline
\end{tabular}
\end{center}
\label{tab:selections}
\end{table}

For the merged selection, since the leading large-$R$ jet is considered as the $\VJ$ candidate, any small-$R$ jet within an angular radius $R=1$ around it is removed. For the resolved selection, large-$R$ jets are ignored and no small-$R$ jets are removed.
 
An event veto based on $b$-tagging information is also applied. For signal events, $b$-jets can arise from the $Z\to bb$ decays.
For $\Vh = Z$, classification based on number of $b$-tagged jets in $Z \to qq$ candidates is applied, as described in Section~\ref{sec:regions}.
For $\Vh = W$, events are required to have at most one $b$-tagged jets in $W\to qq$ candidates, considering mis-identification rate for charm hadrons from $W \to sc$ decay.
In the merged selection, events are vetoed if there are more than two $b$-tagged variable-radius jets associated with the leading large-$R$ jets.

Unless specifically noted, the same selections are applied for the VBF and ggF/DY categories. The merged selection is applied first and the resolved selection is applied only to events failing the merged selection.

% The next lines are included from the .//tex/vvqq.tex input file
\subsubsection{\zlep: $ZV\to\vvqq$}
\label{sec:vvqq}
 
The \zlep channel targets the $ZV\to\vvqq$ final state from $R\to ZZ$,  $W'\to ZW$ and $\Gkk\to ZZ$ decays. Events in this final state have a large \met and a $\Vqq$ candidate.
Due to high \met trigger thresholds and the expected large multijet background from mismeasurement at low \met,
events are required to have $\met>250$~\GeV\ and no \textit{Loose} leptons,
to suppress background from multijet events and single $W$ bosons respectively.
 
Additional requirements are applied to further reduce the multijet background. These include $\mpt>50$~\GeV, and an azimuthal opening angle between $\metvec$ and $\mptvec$ satisfying $\Delta\phi(\metvec,\mptvec)<1$. Furthermore, the azimuthal angle between $\metvec$ and the nearest small-$R$ jet must satisfy $\min\Delta\phi(\metvec, j)>0.4$.  With these angular requirements along with the \met and \mpt\ requirements, the multijet background becomes negligible.
 
The high \met requirement is efficient only for signal events with very heavy resonances. Therefore, only the merged selection is considered for this channel.
% End of text imported from the .//tex/vvqq.tex input file
% The next lines are included from the .//tex/lvqq.tex input file
\subsubsection{\olep: $WV\to\lvqq$}
\label{sec:lvqq}

The \olep channel is designed for the $WV\to\lvqq$ final state from $R\to WW$,  $W'\to WZ$, $Z'\to WW$, and $\Gkk\to WW$. Events in this channel must have exactly one \textit{Tight} electron or \textit{Medium} muon, with $\pt > 30$~\GeV, and no other leptons satisfying the \textit{Loose} quality;  \met greater than 60~\GeV; and a transverse momentum of the lepton--$\metvec$ system (i.e.\ the reconstructed $\Vl$), $\ptVl$, greater than 75~\GeV.
 
For the merged selection, the \met and $\ptVl$ thresholds are raised to 100~\GeV\ and 200~\GeV, respectively. Considering that boson \pt\ is expected to be approximately $0.5 \mVV$, events are further required to have $\Rpt$, defined as $\min(\ptVl,\ptVh)/\mVV$, greater than $0.35\,(0.25)$ for the ggF/DY (VBF) category. Here $\ptVh$ is the \pt of the $\Vh$ candidate, i.e.\ the leading large-$R$ jet, and $\mVV$ is the invariant mass of the $VV$ system reconstructed from the $\ell\nu$ and the leading large-$R$ jet.\footnote{The unknown neutrino longitudinal momentum, $p_z$, is determined by fixing the invariant mass of the $\ell\nu$ system to the $W$-boson mass, resulting in a quadratic equation. The $p_z$ is chosen to be either the real component of the two complex solutions or the smaller of the two real solutions.} This requirement suppresses background significantly at large $\mVV$ while maintaining high efficiencies for signal events.
 
For the resolved selection, a set of azimuthal angular requirements are designed and applied to reduce large multijet backgrounds expected at low \met and $\ptVl$. They are $\Delta\phi(\ell,\metvec)<1.5$, $\Delta\phi(j_1, j_2)<1.5$, $\Delta\phi(\ell,j_{1/2})>1.0$ and $\Delta\phi(\metvec,j_{1/2})>1.0$. Here $j_{1/2}$ refers to both $j_1$ and $j_2$, which form the $\Vjj$ candidate. Similarly to the merged selection, a kinematic criterion of $\Rpt>0.35\,(0.25)$ is imposed for the ggF/DY (VBF) category, where $\ptVh$ is the \pt of the $\Vjj$ candidate and $\mVV$ is reconstructed from the $\ell\nu$ and dijet system.
 
Additional $b$-jets can originate from background \ttbar and single-top-quark events. To reduce this background, events are vetoed if there are one or more small-$R$ $b$-jets beyond those selected as the $V\to qq$ candidate.
% End of text imported from the .//tex/lvqq.tex input file
% The next lines are included from the .//tex/llqq.tex input file
\subsubsection{\tlep: $ZV\to\llqq$}
\label{sec:llqq}

The \tlep channel is intended for the $ZV\to\llqq$ final state from $R\to ZZ$, $W'\to ZW$ and $\Gkk\to ZZ$. The event selection begins with the identification of the $\Zll$ decay.
The $\Zll$ candidates are formed from two same-flavour leptons with $\pt > 30$~\GeV\ and satisfying the \textit{Loose} criteria defined in Section~\ref{sec:objects}. Muon pairs are required to have opposite charges. Because electrons are more susceptible to charge misidentification, no charge requirement is applied. The dilepton invariant mass, $\mll$, must be consistent with the $Z$ boson mass. A fixed $\mll$ window of $[83, 99]$~\GeV\ is applied to electron pairs, while the dilepton \pt is used to define a $\ptll$-dependent window of $[85.6\;\gev - 0.0117\times\ptll,\;  94.0\;\gev + 0.0185\times\ptll]$ that is required for the muon pairs because of the deteriorating muon momentum resolution at high \pt. The mass windows are chosen to maintain approximately 95\% selection efficiency for $\Zll$. Events with additional \textit{Loose} leptons are removed.

The selected $\Zll$ events are required to have $\Rpt>0.35\;(0.25)$ for the ggF/DY (VBF) category for the merged selection and  $\Rpt>0.35$ for both the ggF/DY and VBF categories for the resolved selection. Again, $\ptVl$ is the \pt of the leptonic $V$ candidate ($\ptVl=\ptll$ in this case), $\ptVh$ is the \pt of the $\Vh$ candidate, and $\mVV$ is the invariant mass of the $\Vl$ and $\Vh$ system.  This requirement exploits the kinematic feature of highly boosted boson decays expected from heavy resonances to reduce backgrounds.

% End of text imported from the .//tex/llqq.tex input file
 
\subsection{Signal region definitions}
\label{sec:regions}
 
Multiple signal regions are defined using the properties of the hadronically decaying $V$ boson as illustrated in Figure~\ref{fig:flow}.
Events passing the merged selection are assigned to either high-purity (HP) or low-purity (LP) signal regions according to the quality of their $\VJ$ candidates. Those with $\VJ$ candidates passing the \DTwoBetaOne requirement of the boson tagger~\cite{PERF-2015-03} are selected for the HP SR, otherwise for the LP SR. The combined $\mJ$ and \DTwoBetaOne\ efficiencies for the HR SR are shown in Figure~\ref{fig:VJmassD2Eff} as functions of $V$ $\pT$. Events passing the resolved selection form the resolved SR.
 
For $V_h=Z$, about 21\% of $\Zqq$ decays are $Z\to bb$, whereas jets from the dominant background source, $\Vjets$, have a smaller heavy-quark content. To exploit this difference, the HP, LP, and resolved SRs are each further split into tagged and untagged SRs in the ggF/DY category if the hadronically decaying boson is a $Z$ boson, i.e.\ $V=Z$.
The $b$-tagging is not applied in the VBF category due to the limited number of events.
Classification based on the $b$-tagging is not applied for $V_h~=~W$.
For the merged selection, the splitting is made by applying $b$-tagging to variable-radius jets associated to the leading large-$R$ jet. Events are tagged only if the two leading variable-radius jets are both $b$-tagged. For the resolved selection, events are tagged if the $Z\to jj$ is formed from two $b$-jets and untagged otherwise.

Classifications in terms of ggF/DY and VBF categories include: merged and resolved reconstruction of the $V_h\to qq$ decay, high and low purity for the merged reconstruction, tagged and untagged identification of the $Z\to qq$ decay, and different mass windows for the $\Wqq$ and $\Zqq$ decays. This results in 10 SRs for the \zlep channel and 15 SRs each for the \olep and \tlep channels, for a total of 40 SRs. Because of overlapping mass windows used to select the hadronic decays of the $W$ and $Z$ bosons, these SRs are not orthogonal. In terms of the diboson final states of resonance decays, there are 6 SRs for $X\to WW$, 15 for $X\to ZZ$, and 19 for $X\to WZ$. SRs in each diboson final state are orthogonal.
The $X\to WW$ and $X\to ZZ$ SRs are orthogonal by design, while the $X\to WZ$ SRs can overlap with either $X\to WW$ or $X\to ZZ$ SRs.
 
\subsection{Reconstruction of invariant and transverse resonance mass}
\label{sec:discriminant}
 
Either the invariant mass $\mVV$ or the transverse mass $m_{\mathrm{T}}$ of the selected $VV$ final states is used as the final discriminant to extract the signal. Heavy resonances would manifest themselves as resonant structures above the SM background in the invariant mass distributions or as broad enhancements in the transverse mass distributions.
 
For the \zlep channel of $X\to ZV\to\vvqq$, no resonance mass reconstruction is possible because of the two undetected neutrinos. Instead a transverse mass defined as:
$$\mT = \sqrt{(\ptJ+\met)^2-(\vec{p}_\textrm{T}^J+\metvec)^2}$$
is used as the discriminant for the merged selection.
For the \olep and \tlep channels, the $VV$ mass is calculated for both the merged  and the resolved reconstruction of the $\Vqq$ decay.
 
Muon momentum resolution deteriorates at high \pt, significantly impacting the $\Zmm$ mass resolution and consequently the resonance mass resolution in the \tlep channel.
This deterioration is particularly severe for very heavy resonances, especially in the merged selection.
To mitigate the impact, a scale of $m_Z/m_{\mu\mu}$ is applied to the four-momentum of the dimuon system, effectively fixing the dimuon mass to the $Z$ boson mass~\cite{Agashe:2014kda}.
This scaling improves the $m_{\mu\mu J}$ resolution by about 7\% in the merged analysis for a scalar resonance. The scaling is not applied for $W\to\mu\nu$ because of the undetected neutrino.
 
For the resolved selection, the $\mVV$ resolution is improved by $2\%$ through the scaling of the dijet four-momentum by a factor of $m_V/\mjj$, with $m_V$ being either the $Z$ or $W$ boson mass~\cite{Agashe:2014kda}.
No $m_V/\mJ$ scaling is applied to the merged selection as the improvement in the $\mVV$ resolution is found to be negligible.

\subsection{Signal efficiencies and mass resolutions}
\label{sec:signal}
 
\begin{figure}[!hbtp]
\centering
\subfigure[]{\includegraphics[width=0.48\textwidth]{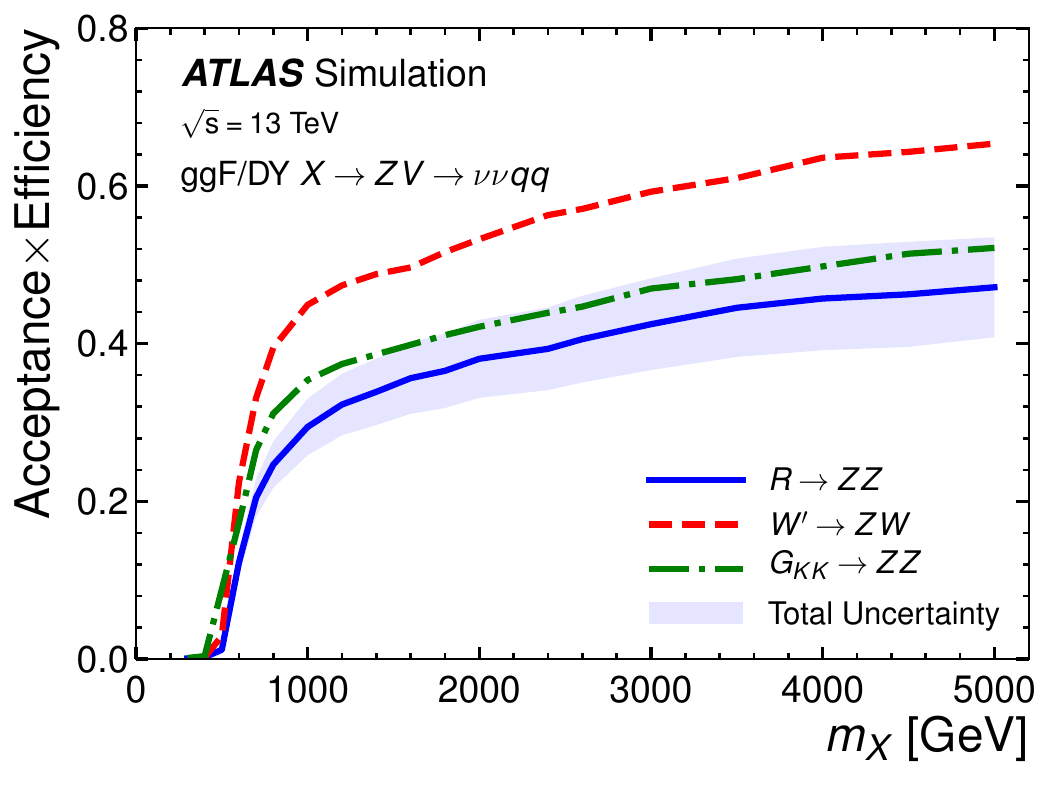}}
\subfigure[]{\includegraphics[width=0.48\textwidth]{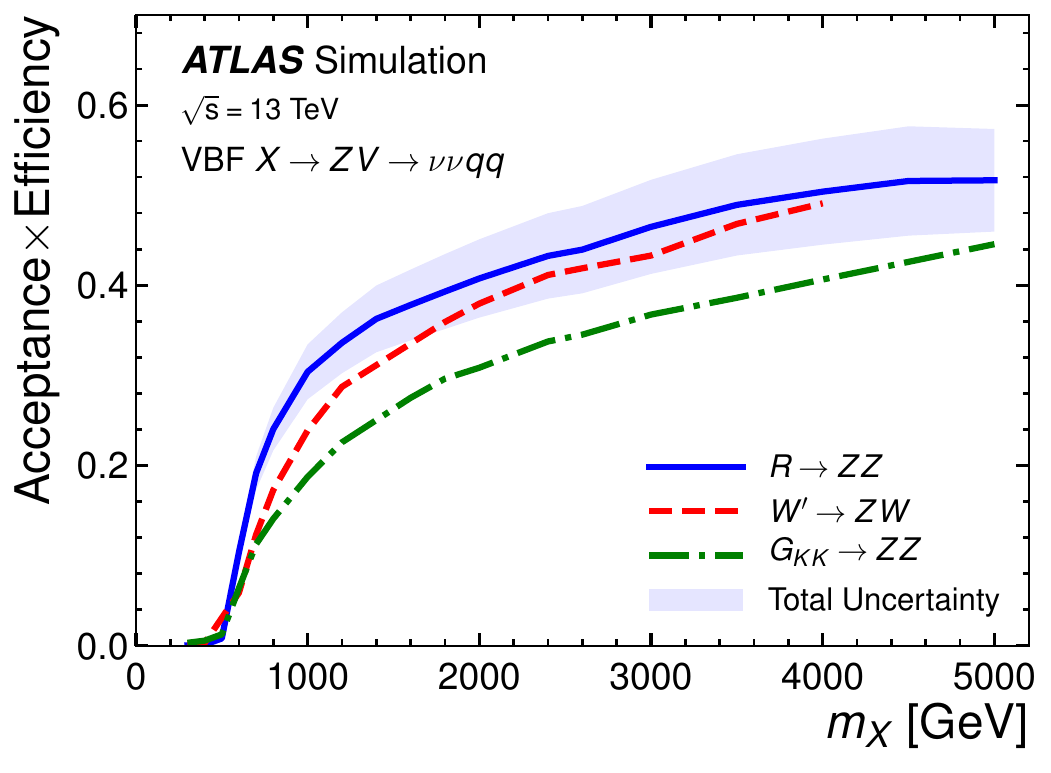}}
\caption{Selection acceptance times efficiency for the $X\to ZV\to\vvqq$ signal events from MC simulations as a function of the resonance mass for (a) ggF/DY and (b) VBF production, combining HP and LP signal regions.
The light shaded band represents the total statistical and systematic uncertainties for the RS radion model, and the total uncertainties
are similar for the other signal models.
}
\label{fig:0lep_acceptance}
\end{figure}

\begin{figure}[!hbt]
\centering
\subfigure[]{\includegraphics[width=0.48\textwidth]{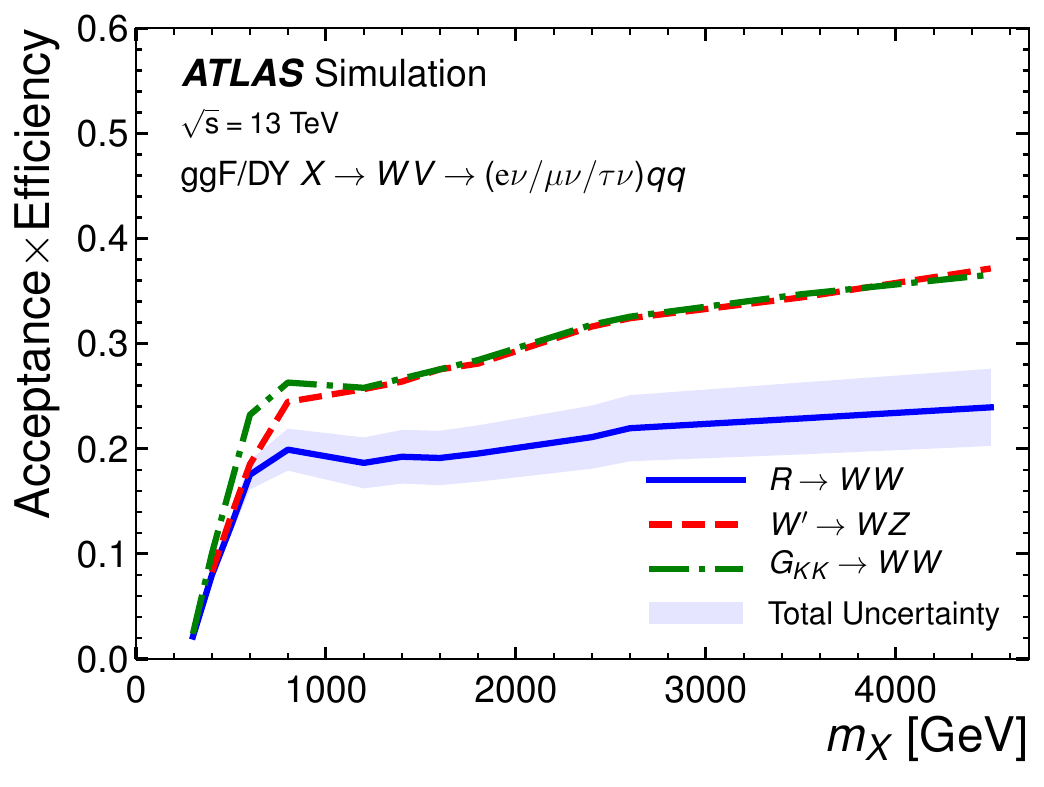}}
\subfigure[]{\includegraphics[width=0.48\textwidth]{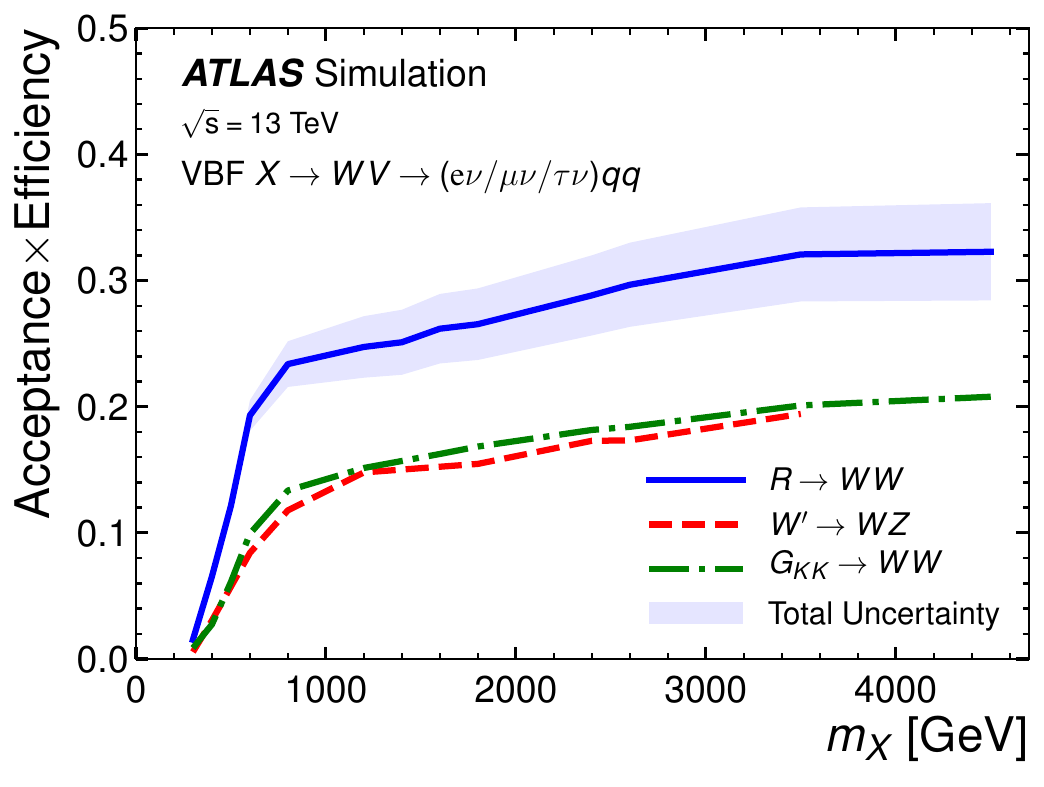}}
\caption{Selection acceptance times efficiency for the $X\to WV\to (e\nu/\mu\nu/\tau\nu)~qq$ signal events from MC simulations as a function of the
resonance mass for (a) ggF/DY and (b) VBF production, combining all SRs of both the resolved and merged analyses. Signal contributions from $W\to\tau\nu$ decays are included in the acceptance calculation.
The light shaded band represents the total statistical and systematic uncertainties for the RS radion model, and the total uncertainties  are similar for the other signal models. The `bump' structure around 800~\GeV\ is due to the decreasing contribution from the resolved analysis at higher masses.}
\label{fig:1lep_acceptance}
\end{figure}

\begin{figure}[!hbt]
\centering
\subfigure[]{\includegraphics[width=0.48\textwidth]{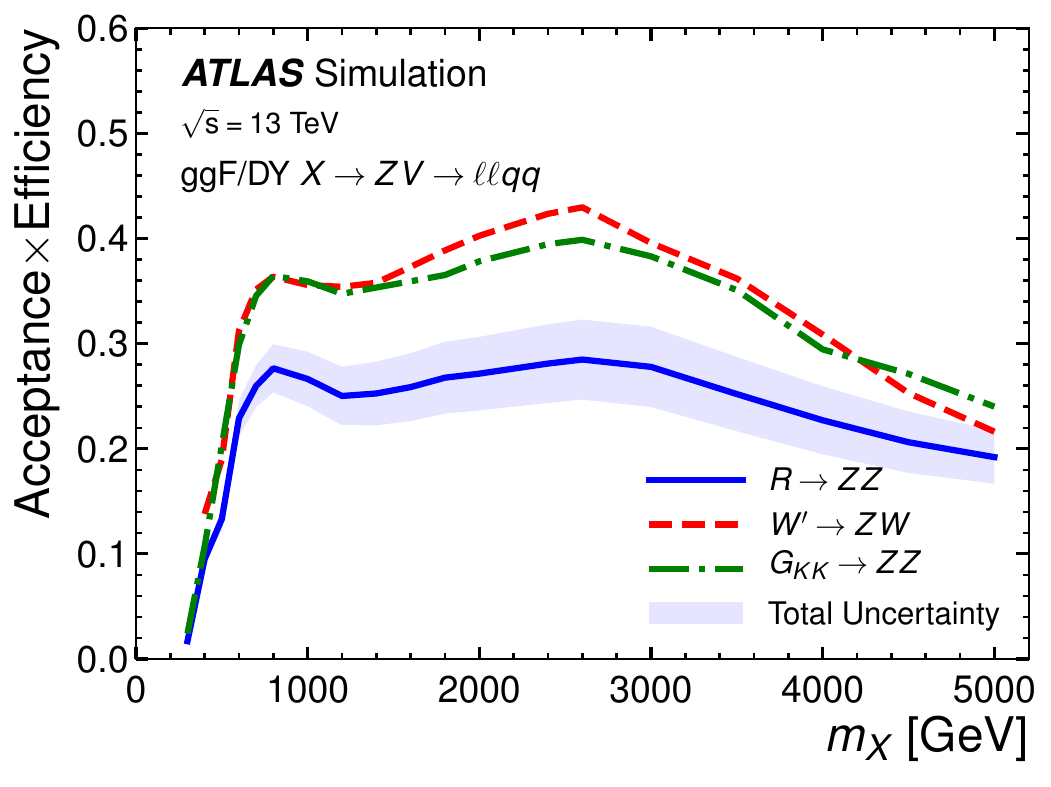}}
\subfigure[]{\includegraphics[width=0.48\textwidth]{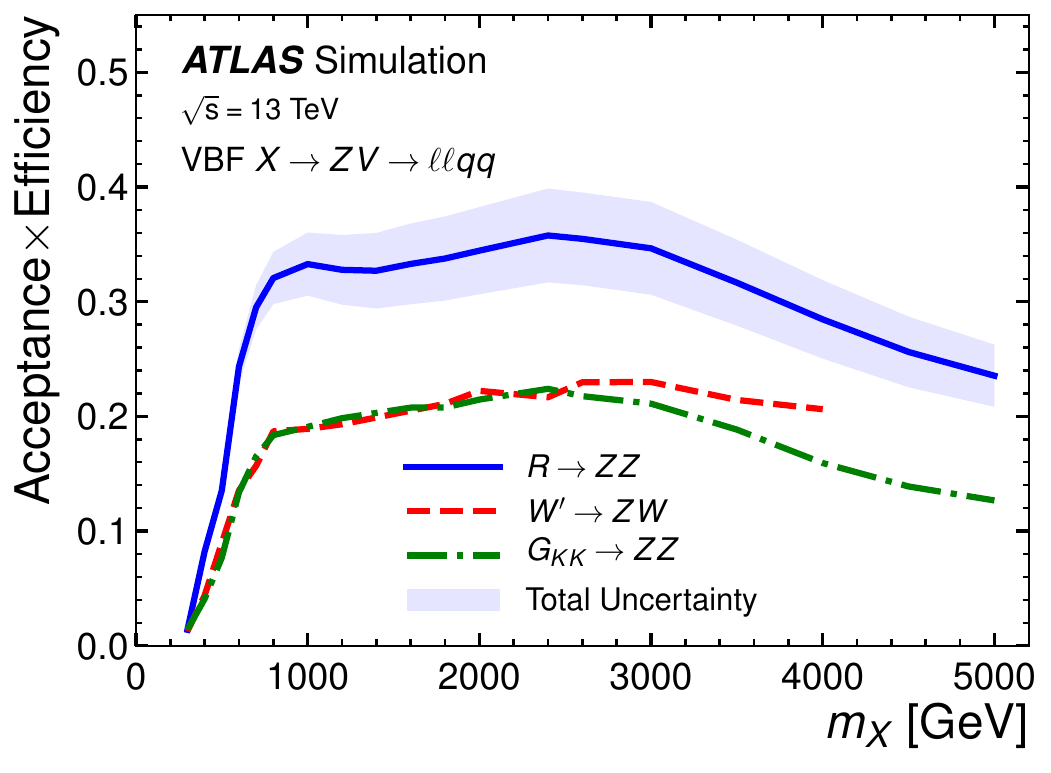}}
\caption{Selection acceptance times efficiency for the $X\to ZV\to\llqq$ signal events from MC simulations as a function of the
resonance mass for (a) ggF/DY and (b) VBF production, combining all SRs of both the resolved and merged analyses.
The light shaded band represents the total statistical and systematic uncertainties for the RS radion model, and the total uncertainties  are similar for the other signal models. The decreases in efficiencies for resonance masses above approximately 2.5~\TeV\ are due to the merging of electrons from the highly boosted $Z\to ee$ decays. The `bump' structure around 800~\GeV\ is due to the decreasing contribution from the resolved analysis at higher masses.}
\label{fig:2lep_acceptance}
\end{figure}

Signal selection efficiencies depend on the signal model, the production process, and the mass of heavy resonances.
Figures \ref{fig:0lep_acceptance}, \ref{fig:1lep_acceptance} and \ref{fig:2lep_acceptance} show the acceptance times efficiency ($A\times\epsilon$) of the signal events from MC simulations as a function of the resonance mass for (a) ggF/DY and (b) VBF production, combining all SRs of both the ggF/DY and VBF categories of both the resolved and merged analyses. The $A\times\epsilon$ curves are largely determined by the merged analyses. The resolved analyses contribute only in the low mass region, up to approximately 1~\TeV.

Large differences in $A\times\epsilon$ shown in the figures for different resonances are due to the different spins of these resonances. The spin-0 RS radions are produced with isotropic angular distributions for both ggF and VBF production. On the other hand, the spin-1 HVT resonances and spin-2 RS gravitons are produced more centrally (more forward) for ggF/DY (VBF) production. These different angular distributions lead to very different efficiencies of the \Rpt\ requirement. Moreover, the angular requirements between jets and \met in the \zlep channel are more efficient for DY production of HVT resonances than for ggF production of RS radions and RS gravitons due to the different colour factors for initial-state quarks and gluons.
 
Signal contributions from $W\to\tau\nu\to\ell\nu\nu\nu$ decays are included in the \olep channel, but not in the \tlep channels. Approximately 10--12\% of the signal events in SRs are from $W\to\tau\nu\to\ell\nu\nu\nu$ decays in the \olep channel. These events have mass distributions similar to those from $W\to\ell\nu$ decays.
In the \tlep channel, signal contributions from $Z\to\tau\tau\to 2\ell+4\nu$ decays are  suppressed by the small $\tau\tau\to2\ell+4\nu$ branching ratio and the $Z$ boson mass requirement. They are found to be negligible. The \zlep channel targeting the $X\to ZV\to\vvqq$ signal should also be sensitive to the $X\to WV\to\lvqq,\,\tau\nu qq$ signal due to either the inefficiency of the lepton veto or the lack of a $\tau$-veto. This additional `cross-channel' signal contribution is neglected in this search.

The resonance decays are fully reconstructed in the \olep and \tlep channels. In the \olep channel, the $\mlvjj$ distributions from the resolved $\Vjj$ reconstruction have widths of approximately 8\% of the resonance mass for narrow resonances, whose intrinsic widths are smaller than the detector mass resolution, with masses of 0.5--1~\TeV.\footnote{The width of the reconstructed mass distribution is defined as the standard deviation of a Gaussian function fit to the peak region.} The width of the $\mlvJ$ distribution from the merged $\VJ$ reconstruction varies from 7\% at 1~\TeV\ to 4\% at 5~\TeV. Similarly, in the \tlep channel the $\mlljj$ resolution is $\sim 6\%$ for resonance masses of 0.5--1~\TeV\ and the $\mllJ$ resolution varies from approximately 4\% at 1~\TeV\ to 2\% at 5~\TeV.

% End of text imported from the .//tex/selections.tex input file
 
% The next lines are included from the .//tex/backgrounds.tex input file
\section{Background estimations}
\label{sec:backgrounds}
 
Relevant background sources for the search are $\Vjets$, \ttbar and single top, non-resonant diboson, and multijet production.  Their relative importance depends on the final state. The largest contributions are from $\Zjets$ and $\Wjets$ in the \zlep channel and $\Wjets$ and \ttbar in the \olep channel. In the \tlep channel, the $\Zjets$ background dominates. The multijet background is negligible except in the resolved SRs of the \olep channel. In the tagged SRs, the \ttbar and single-top-quark contributions are enhanced and are in fact dominant in the \olep channel.
 
MC simulations are used to simulate kinematics of background events except for multijet events. Contributions from diboson and single-top-quark processes are normalised to their theoretical cross-sections, whereas the $\Vjets$ and \ttbar contributions are normalised using data through control regions. The multijet background is estimated from data. The definitions of control regions and the method of multijet estimation are described below.

\subsection{Control regions for $\Wjets$, $\Zjets$, and \ttbar}
\label{sec:controlRegions}
 
Control regions~(CR) are designed to constrain the normalisations of the $\Wjets$, $\Zjets$ and $\ttbar$ background contributions using data, eliminating the reliance on the theoretical cross-sections, which are often less reliable in the phase-space regions covered by this search. Events in CRs are selected from those failing the selections of the SRs, but are otherwise expected to have event topologies similar to those in SRs and small contaminations from potential signals.
 
CRs for the $\Wjets$ background are defined using events in the \olep channel by reversing the $\mJ$ or $\mjj$ requirements of the SR selections, but events are otherwise selected in the same way as those in the corresponding SRs. For the merged selection, $\mJ$ must fall outside of the mass windows of both $W$ and $Z$ boson tagging. For the resolved selection, $\mjj$ is required to be in the range from 50 to 150~\GeV, excluding the combined $W$ and $Z$ mass window of 62--105~\GeV. The $\Wjets$ events are expected to be the dominant contribution in these CRs, except in the $b$-tagged CRs, where the \ttbar contribution dominates.  CRs for the $\Zjets$ background are defined the same way, but using \tlep events. The $\Zjets$ events dominate in all $\Zjets$ CRs, even in the tagged CRs.
 
CRs for the \ttbar background are defined using \olep events, selected in the same way as the \olep SRs except for the requirement of an additional small-$R$ $b$-jet unassociated with the $V\to jj/J$ candidate instead of the $b$-jet veto. Moreover, $\mjj$ is required to be between 50 and 150~\GeV\ in the case of the resolved selection.

\subsection{Multijet background}

In the resolved SRs of the \olep channel, multijet production is a non-negligible background source. Multijet events can mimic signal events if there is a lepton from either jet misidentification or heavy-quark decay, along with a large \met from energy mismeasurements. The multijet contributions are difficult to model through MC simulations and are therefore estimated from data. A template method is used to estimate the multijet contributions. The method derives the shapes of the $\met$ distributions of the multijet contributions from multijet-enriched control regions~(MJCR), one for each signal and control region. MJCRs are designed to be orthogonal to both the SRs and CRs as defined above. For the muon channel, MJCRs are defined only for the single-muon trigger, i.e.\ events with $\pT(\mu\nu)<150$~\GeV, since the multijet contributions to the \MET-triggered events with $\pT(\mu\nu)>150$~\GeV\ are found to be negligible.
 
Events in MJCRs are selected by modifying the lepton requirements used for the SR and CR selections. Electron candidates are required to satisfy the \textit{Medium} quality criteria and not the \textit{Tight} quality criteria. Muon candidates must  pass a relaxed, but fail the tight, isolation requirement. All other selections remain unchanged. More than 80\% of the selected events in MJCRs are estimated to originate from multijet production. These MJCR samples are used to model the kinematics of multijet contributions in their corresponding CRs and SRs, after subtracting contributions from other sources.
The multijet scale factors, the ratios of the multijet contributions in the CRs to those in their MJCRs, are extracted through fits to the $\met$ distributions in CRs using the multijet $\met$ distribution shapes in MJCRs as templates. In the fits, contributions from other sources are constrained to their expectations from MC simulations within their uncertainties. These scale factors are then applied to their corresponding SRs to estimate multijet contributions.

% End of text imported from the .//tex/backgrounds.tex input file
 
% The next lines are included from the .//tex/systematics.tex input file
\section{Systematic uncertainties}
\label{sec:systematics}

Systematic uncertainties impact the search sensitivity through their effects on background estimations, signal selection efficiencies, and the distributions of the mass discriminants. The sources of these uncertainties can be classified broadly into two groups: (a) those experimental in nature related to the detector and reconstruction performance and (b) those of theoretical origins associated with the MC modelling of both the background and signal processes. The uncertainties and the methods used to evaluate them are discussed below. Unless explicitly stated, the uncertainties quoted are the uncertainties in the quantities themselves, not the impact on the search sensitivity.

\subsection{Experimental uncertainties}
Experimental uncertainties arise from the luminosity, triggers, and reconstruction and identification of leptons and jets, as well as the calculation of the \met. They also include uncertainties in the energy and momentum scales and resolutions of leptons and jets.

The uncertainty of the combined 2015--2018 integrated luminosity is 1.7\%. It is derived from the calibration of the luminosity scale using $x$--$y$ beam-separation scans, following a methodology similar to that detailed in Ref.~\cite{ATLAS-CONF-2019-021}, and using the LUCID-2 detector for the baseline luminosity measurement~\cite{Avoni:2018iuv}. A variation in the pile-up reweighting of MC events is included to cover the uncertainty in the ratio of the predicted and measured inelastic cross-sections~\cite{STDM-2015-05}.
 
Uncertainties in the efficiencies of lepton triggers are found to be negligible.
The modelling of the electron and muon reconstruction, identification and isolation efficiencies is studied with a tag-and-probe method using $\Zll$ events in data and simulation~\cite{EGAM-2018-01, PERF-2015-10}. Small corrections are applied to the simulation to better model the performance seen in data. These corrections have associated uncertainties of the order of 1\%. Uncertainties in the lepton energy (or momentum) scale and resolution, especially for muon momentum resolution (3\%), are also taken into account.

Uncertainties for the energy scale and resolution of the small-$R$ jets are determined using MC simulation and in situ techniques~\cite{PERF-2016-04}.
For central jets, the total relative uncertainty in the jet energy scale varies in the range 1--4\% for $\pt>20$~\GeV.
For forward jets, additional 2--4\% uncertainty depending on \pt\ is applied based on $\eta$-intercalibration study.
The uncertainty in the jet energy resolution ranges from $20\%$ for jets with a $\pt$ of $20$~\GeV\ to less than $5\%$ for jets with $\pt>200$~\GeV.

Uncertainties in the scale of the large-$R$ jet \pt\ are estimated by comparing the calorimeter- and track-based energy and mass measurements in data and simulation~\cite{JETM-2018-02}.
The precision of the relative jet energy scale is 1--2\% for $200\,\GeV <p_T< 2$~\TeV, while that of the mass scale is 2--10\%. The jet energy resolution uncertainty is estimated to be approximately 2\%.
The efficiency of the $W$/$Z$ boson tagging based on the $\mJ$ and \DTwoBetaOne requirements is estimated using data control samples, following the technique described in Ref.~\cite{ATLAS-PHYS-PUB-2020-008}.
The efficiency for large-$R$ jets from $W$/$Z$ boson decays is estimated using $t\bar{t}$ control samples for $\pt < 600$~\GeV.
The measurement is extrapolated to the higher \pt region with additional uncertainties estimated from simulations~\cite{ATL-PHYS-PUB-2020-017}.
The efficiency for background large-$R$ jets from gluons or light quarks is estimated using dijet and $\gamma$+jets samples.

Uncertainties in the efficiencies for tagging $b$-jets and for mis-tagging light-flavour jets  are determined from $\ttbar$ control samples~\cite{FTAG-2018-01,ATLAS-CONF-2018-006,ATLAS-CONF-2018-045}.  The total uncertainties are 1--10\%, 15--50\%, and 50--100\% for $b$-jets, $c$-jets, and light-flavour jets respectively.
 
Uncertainties in the \met trigger efficiencies have negligible impact on the search as the efficiencies for the selected signal events are high. The uncertainty in \met is calculated from those in the energy scales and resolutions of leptons and jets as well as  those in the energy deposits unassociated with any identified physics objects~\cite{PERF-2016-07}.

Multijet backgrounds are only important for the resolved analysis in the \olep channel and are estimated using data control regions. The dominant uncertainties are from the multijet \met and mass templates, obtained from MJCRs after subtracting \Wjets\ and \ttbar contributions. They are estimated by varying the \Wjets\ and \ttbar subtractions and are found to range from a few percent to up to 15\%.

\subsection{Theoretical uncertainties}
 
Theoretical uncertainties affect the normalisations of diboson and single-top-quark backgrounds, the shapes of mass distributions of background processes, and the signal acceptances. They arise from sources such as the choices of event generators, parton distribution functions~(PDFs), parton shower models, and underlying-event tunes. Modelling uncertainties in the shapes of the mass distributions are estimated by varying the renormalisation/factorisation scales, PDF set and $\alpha_\mathrm{s}$ values used in the nominal MC samples. Alternative generators are used to estimate the uncertainties due to the choices of generators, parton shower models and event tunes.
 
Background contributions from diboson and single-top-quark processes are estimated from MC simulations and are normalised to their theoretical cross-sections. For the diboson process, the cross-section uncertainty is estimated to be 10\%~\cite{Gleisberg:2008ta,Campbell:2011bn}.
An additional contribution from electroweak production, simulated with \textsc{Madgraph5\_aMC@NLO+Pythia8}, leads to an increase in the normalisation of the diboson background for the VBF process by a factor of 1.60 (1.85) in the resolved (merged) analyses. A uncertainty of 50\% is applied to the normalisation of the electroweak diboson contribution. The impact on the ggF/DY analysis is negligible. For the cross-section of single-top-quark processes, an uncertainty of 20\% is assumed~\cite{TOPQ-2015-16}.
 
Background contributions from $\Vjets$ and \ttbar are normalised using data control regions in the \olep and \tlep channels.
Their overall normalisations are free parameters in the likelihood fit (Section~\ref{sec:results}) and
thus only uncertainties in the shapes of discriminant variables are considered.
For $\Vjets$, the nominal \textsc{Sherpa} samples are compared with samples produced using \textsc{MadGraph5\_aMC@NLO}.
Moreover, the resummation scale and the CKKW~\cite{Hoeche:2009rj,Catani:2001cc} matching scale in the nominal samples are also varied.
The shape systematic uncertainty varies the background expectation in each bin and it is typically smaller than 10\%, with the \textsc{Sherpa--MadGraph} comparison reaching 25\% at the highest mass bin in the merged ggF/DY $WZ$ untagged signal regions for the \olep channel.
For \ttbar, the default \textsc{Powheg-Box} sample is compared with the alternative \textsc{MadGraph5\_aMC@NLO} sample interfaced with \textsc{Pythia}~8.230.
The difference is found to be approximately 4\% in the merged signal regions, twice the value in the resolved signal regions.
The difference between the \textsc{Pythia}~8.230 sample using the A14 tune and the alternative \textsc{Herwig} 7.04~\cite{Bahr:2008pv,Bellm:2015jjp} sample using the H7UE set of tuned parameters~\cite{Bellm:2015jjp} and the MMHT2014LO PDF set~\cite{Harland-Lang:2014zoa} is found to be between 2\% and 5\% in the various mass bins.
The changes resulting from varying the parameter values for the nominal generator are less than 5\%.
In the \zlep channel, there is no pure control region to evaluate the $\Vjets$ and \ttbar background, so the normalisation factors for the \zlep channel are assumed to be the same as for the \olep channel (\Wjets\ and \ttbar) and \tlep channel (\Zjets).
Systematic uncertainties in this normalisation are obtained by the data/prediction double ratio between the default and the alternative MC generator and is estimated to be between 10\% and 20\% for $\Vjets$ and up to 30\% for \ttbar.
The \ttbar background is negligible in the \tlep channel and therefore its uncertainty is not considered for this channel.

Uncertainties in the signal acceptances are estimated for the choice of PDF set and the modelling of  initial- and final-state radiation.
The PDF uncertainties are estimated by taking the acceptance difference due to applying internal PDF error sets and the difference due to choosing different PDF sets.
The uncertainty due to ISR/FSR modelling is studied by varying parameter values in the tunes used and applied to the HVT, the RS graviton, and the RS radion models.
These uncertainties, calculated for several resonant mass points, are retrieved for each model, production process and decay.
The PDF uncertainties are evaluated to be under 5\% for all models.
ISR/FSR uncertainties range from 2\% for the merged analysis of ggF HVT production to about 11\% for the resolved analysis of VBF HVT production.

\begin{table}[!ht]
\small
\caption{\label{tab:np_ranking}
Dominant relative uncertainties in the best-fit signal-strength parameter $\mu$ of hypothesised signal production of
ggF RS graviton with $m(\Gkk)=600$~\GeV\ and $m(\Gkk)=2$~\TeV. For this study, the RS graviton production cross-section is
assumed to be 100~fb at 600~\GeV\ and 2~fb at 2~\TeV, corresponding to approximately the expected median upper limits
at these two mass values.
Uncertainties with smaller contributions are not included.
}
\centering
\begin{tabular}{ l r@{}l c l r@{}l} \hline\hline
\multicolumn{3}{c}{$m(\Gkk)=600$~\GeV} &\hspace*{1.0cm}& \multicolumn{3}{c}{$m(\Gkk)=2$~\TeV}\\  \midrule
Uncertainty source    & \multicolumn{2}{c}{$\Delta\mu/\mu$ [\%]} && Uncertainty source    & \multicolumn{2}{c}{$\Delta\mu/\mu$ [\%]}     \\  \midrule
 
Total                                  & ~~~~~~~50&  &&  Total                             & ~~~~~~~59&   \\
Statistical                            & ~~~~~~~29&  &&  Statistical                       & ~~~~~~~48&   \\
Systematic                             & ~~~~~~~41&  &&  Systematic                        & ~~~~~~~34&   \\
\midrule
Large-$R$ jet                          & ~~~~~~~18&  &&  Large-$R$ jet                     & ~~~~~~~24&   \\
MC statistics                          & ~~~~~~~16&  &&  MC statistics                     & ~~~~~~~17&   \\
Background normalisations              & ~~~~~~~15&  &&  $W/Z$+jets modelling              & ~~~~~~~15&   \\
Diboson modelling                      & ~~~~~~~12&  &&  Flavour tagging                   & ~~~~~~~5&.5   \\
$W/Z$+jets modelling                   & ~~~~~~~11&  &&  \ttbar modelling                  & ~~~~~~~4&.2   \\
Small-$R$ jet                          & ~~~~~~~9&.7  &&  Diboson modelling                & ~~~~~~~3&.9   \\
\ttbar modelling                       & ~~~~~~~8&.1  &&  Single-$t$ modelling             & ~~~~~~~3&.3   \\ \hline\hline
 
\end{tabular}
\end{table}

\subsection{Impact of systematic uncertainties}

The effects of systematic uncertainties on the search are studied for hypothesised signals using the signal-strength parameter $\mu$, the ratio of the extracted cross-section (Section~\ref{sec:results}) to the injected hypothesised signal cross-section.
The expected relative uncertainties in the best-fit $\mu$ value from the leading sources of systematic uncertainties are shown in Table~\ref{tab:np_ranking} for the ggF production of an RS graviton with $m(\Gkk)=600$~\GeV\ and 2~\TeV.  Apart from the statistical uncertainties in the data, the uncertainties with the largest impact on the sensitivity of the searches are from the sizes of the MC samples, measurements of small-$R$ and large-$R$ jets, background normalisations and modellings. Uncertainties related to the jet measurements, such as jet energy scale and resolution, affect the search primarily through their impacts on the shapes of the discriminant mass distributions of both signal and background processes. Uncertainties on the normalisations of background contributions estimated using CRs arise from CR statistics as well as MC event generators used to extroplate from CRs to SRs. Background modelling uncertainties include uncertainties on their normalisations, if estimated from MC simulations, as well as on the shapes of the mass distributions. The normalisations are affected by the uncertainties on the theoretical cross sections and on the luminosity. The shapes are affected by, in addition to experimental sources, theoretical sources such as PDF, ISR/FSR, and MC generator etc.
For signals with higher mass, the data statistical uncertainty becomes dominant. The effects of systematic uncertainties for the other searches are similar.

% End of text imported from the .//tex/systematics.tex input file
 
% The next lines are included from the .//tex/results.tex input file
\section{Results and interpretations}
\label{sec:results}

\subsection{Statistical procedure}
The statistical analysis is based on the framework described in Refs.~\cite{HIGG-2012-17,Moneta:2010pm,Verkerke:2003ir}.
A profile-likelihood-ratio test statistic is used to test the compatibility of the background-only hypothesis and the observed data, and to test the signal-plus-background hypothesis for the production of a heavy resonance $X$, with its production cross-section in the $VV$ decay mode, $\sigma(pp\to X\to VV)$, as the parameter of interest.
Maximum-likelihood fits are made to the observed binned distributions of the final discriminants in SRs, \mT\ in \zlep, $\mlvJ$ or $\mlvjj$ in \olep and $\mllJ$ or $\mlljj$ in \tlep,  and to the numbers of observed events in CRs simultaneously.
The mass ranges fitted are 300--3000~\GeV\ for the resolved analysis and 500--6000~\GeV\ for the merged analysis.
The normalisations of the  $\Vjets$ and $\ttbar$ contributions are free parameters in these fits and are constrained by the data in both the CRs and SRs.
Systematic uncertainties, described in Section~\ref{sec:systematics}, and their correlations are incorporated as constraints into the likelihood calculations through nuisance parameters, where each is given a prior distribution based on individual studies or is allowed to float freely, constrained simultaneously by the SRs and CRs.

Two types of fits, referred to as the $WW+ZZ$ and $WZ$ fits below, are performed.
The $WW+ZZ$ fits include all 21 SRs of the $X\to WW$ and $X\to ZZ$ searches and the $WZ$ fits includes the 19 SRs of the $X\to WZ$ search, along with their respective CRs.
Separate fits are performed for the ggF/DY and VBF production modes and for different resonance mass hypotheses, but including SRs and CRs in both the ggF/DY and VBF categories.
The $WW+ZZ$ fits are used to search for the RS radion and RS graviton signals as both the $WW$ and $ZZ$ decay modes are expected from these resonances.
The fits are also used to search for HVT $Z'\to WW$ production.
In this case, the $X\to ZZ$ SRs effectively become additional CRs for the search.
The $WZ$ fits are used to search for HVT $W'\to WZ$ production.

\subsection{Data and background comparisons}
 
\begin{figure}[!htbp]
\centering
\subfigure[VBF, merged HP]{\includegraphics[trim=0 8mm 0 8mm, width=0.32\textwidth]{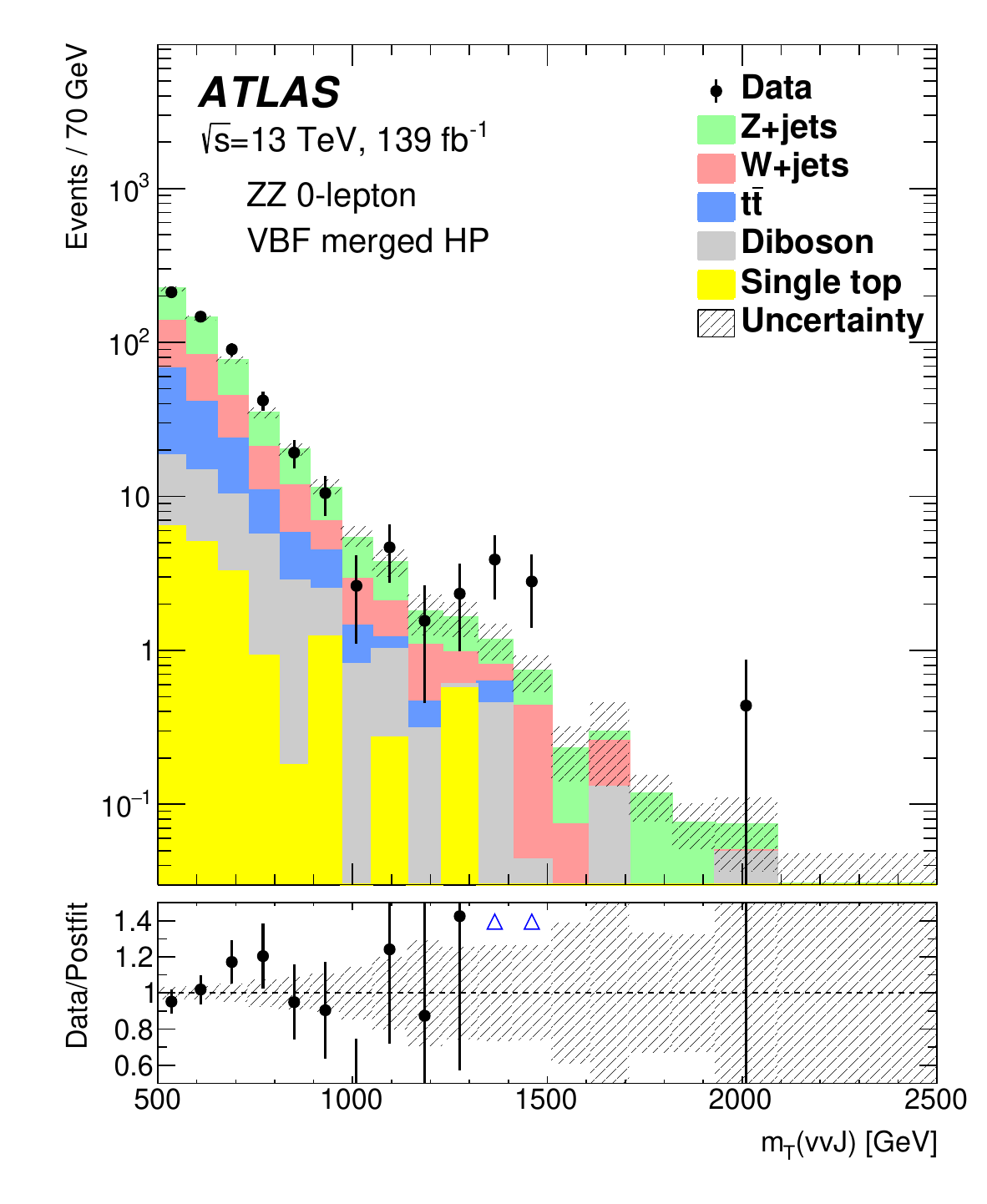}}
\subfigure[VBF, merged LP]{\includegraphics[trim=0 8mm 0 8mm, width=0.32\textwidth]{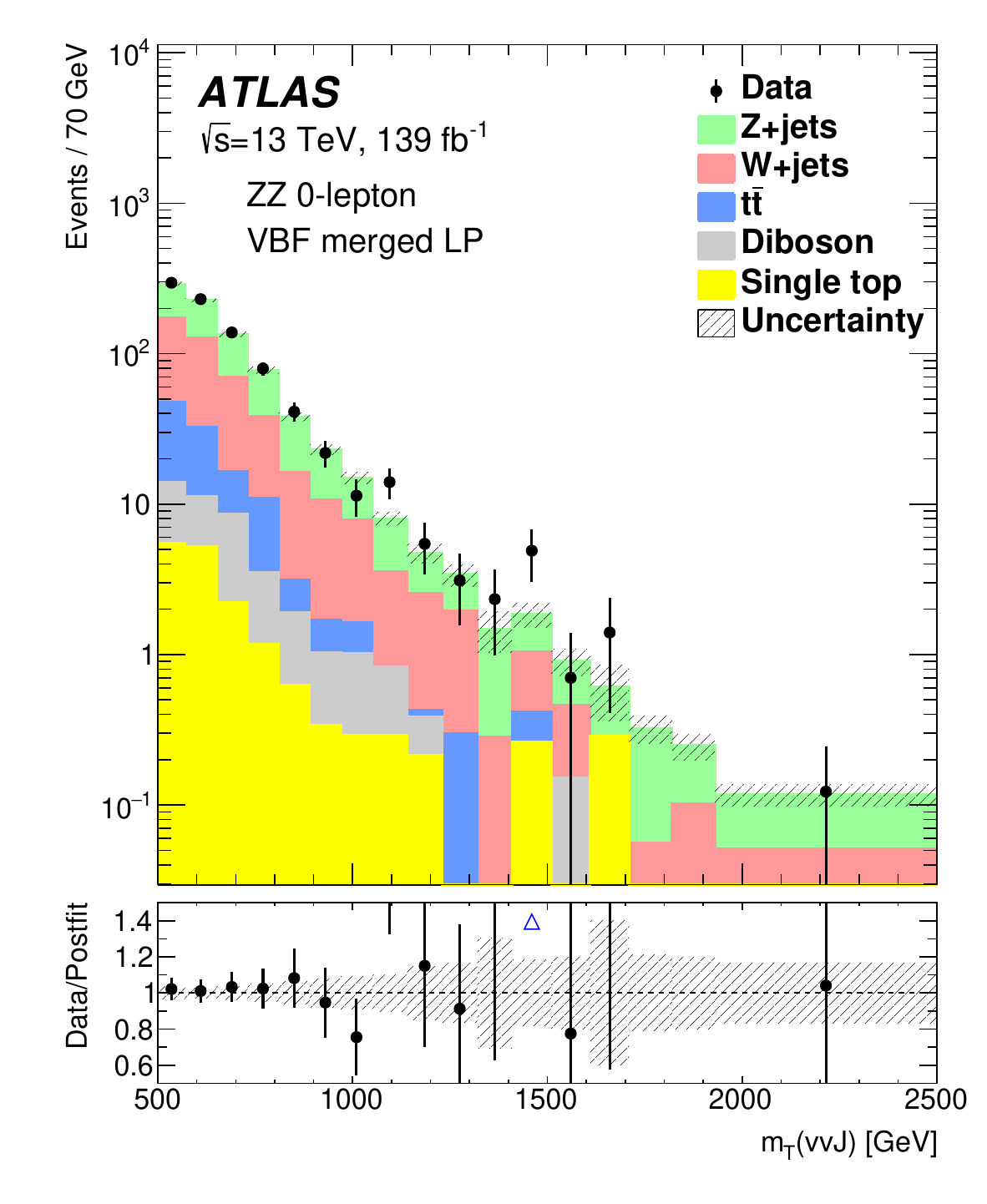}} \\
 
\subfigure[ggF/DY, merged HP tagged]  {\includegraphics[trim=0 8mm 0 8mm, width=0.32\textwidth]{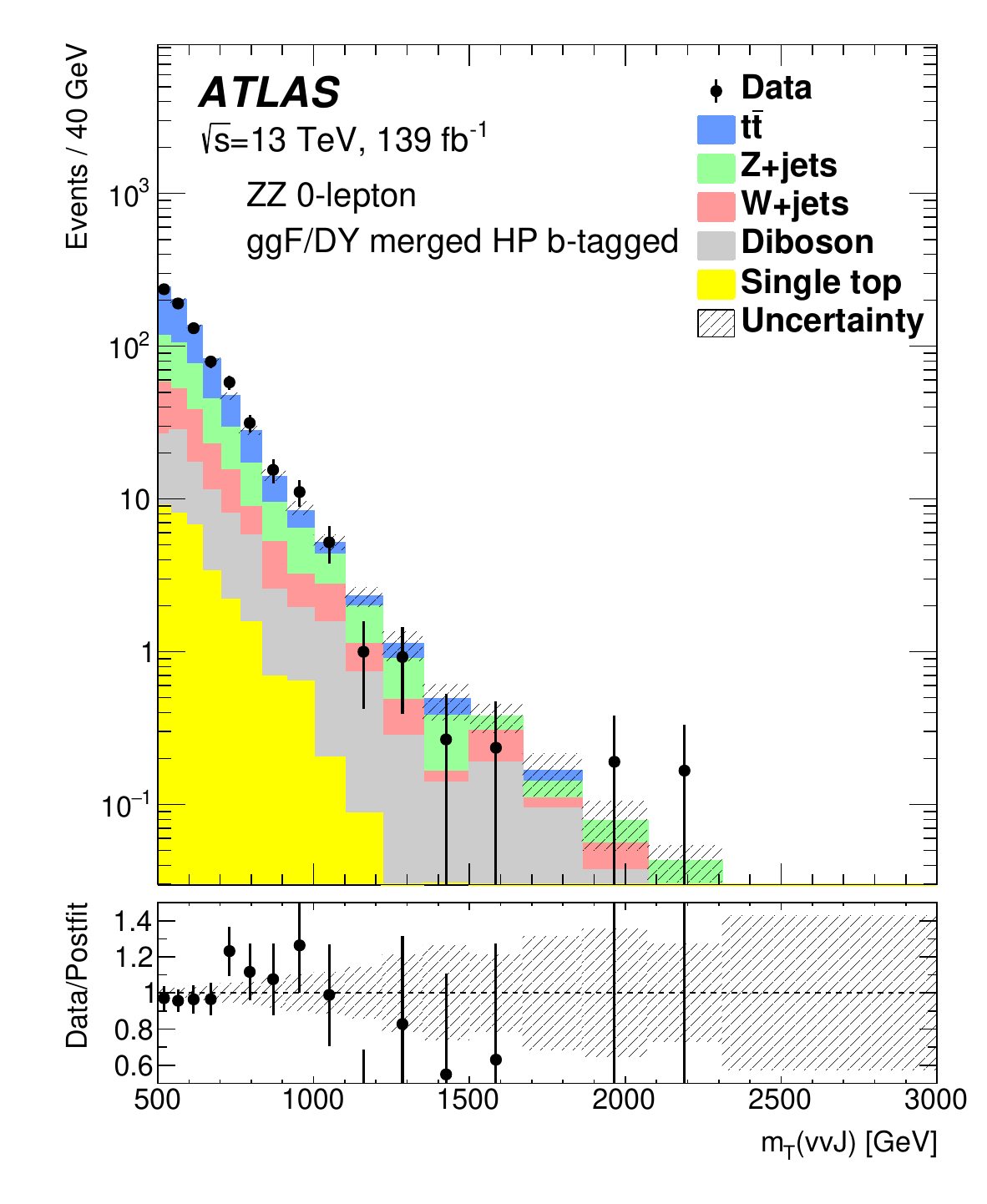}}
\subfigure[ggF/DY, merged HP untagged]{\includegraphics[trim=0 8mm 0 8mm, width=0.32\textwidth]{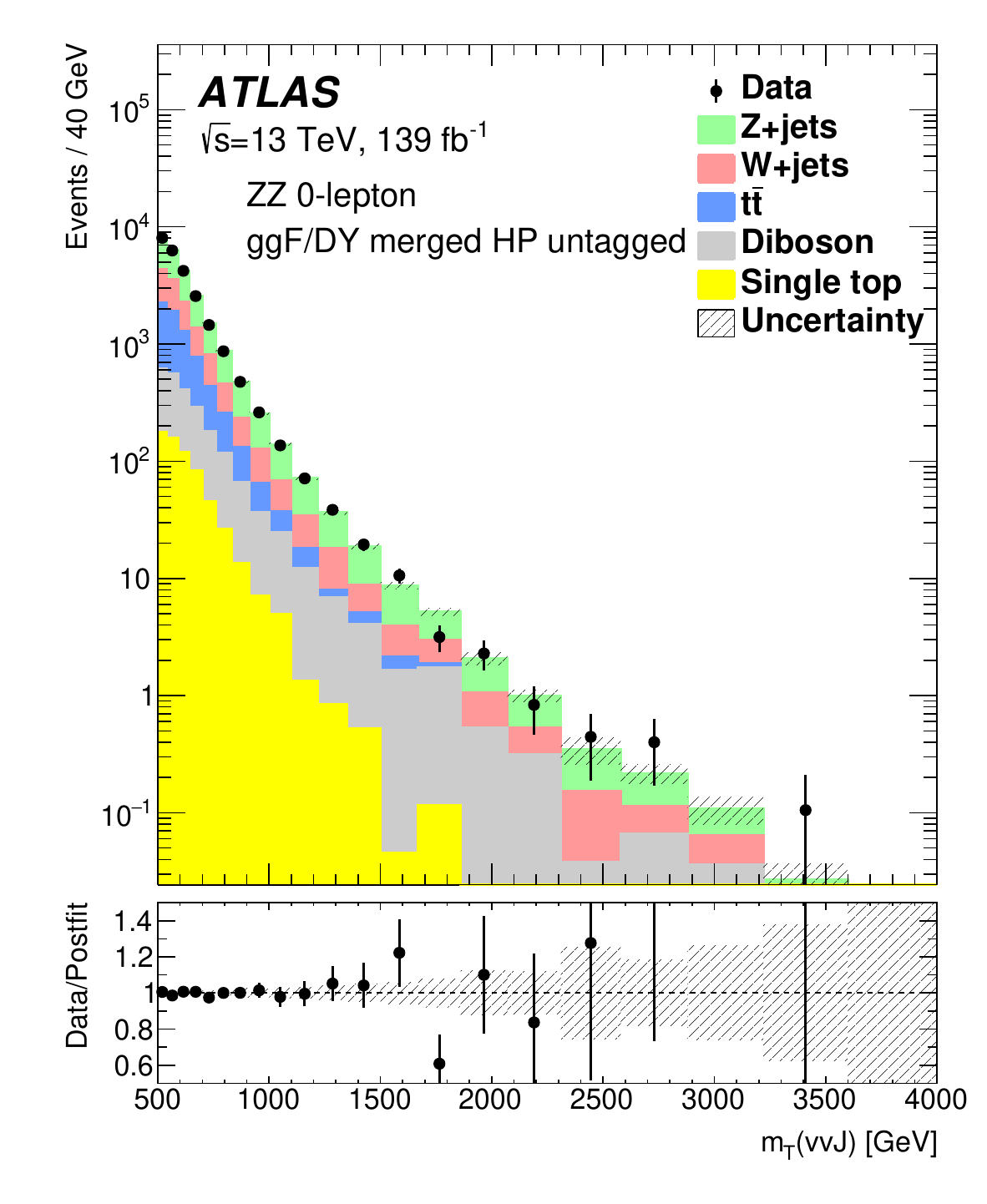}} \\
 
\subfigure[ggF/DY, merged LP tagged]  {\includegraphics[trim=0 8mm 0 8mm, width=0.32\textwidth]{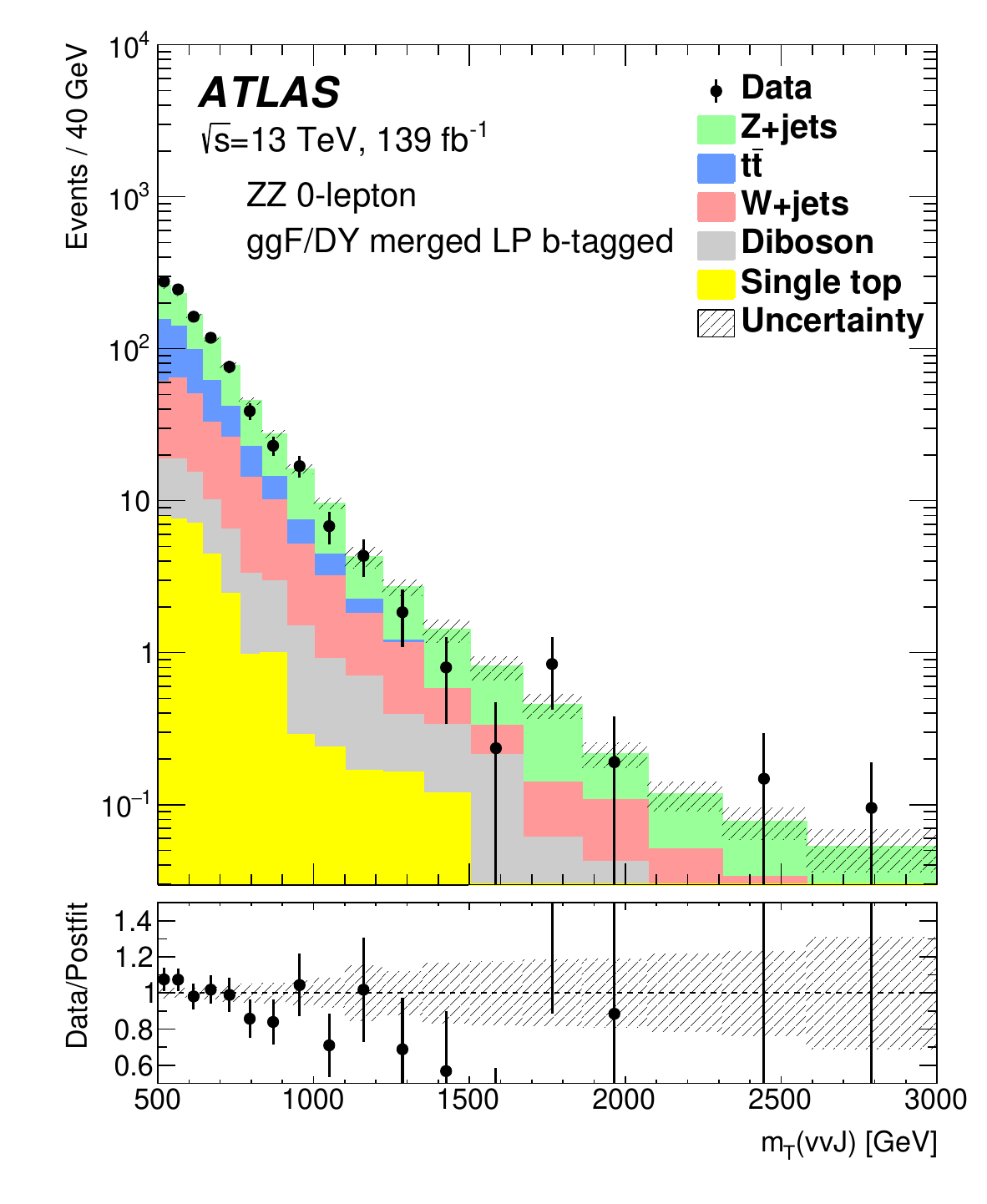}}
\subfigure[ggF/DY, merged LP untagged]{\includegraphics[trim=0 8mm 0 8mm, width=0.32\textwidth]{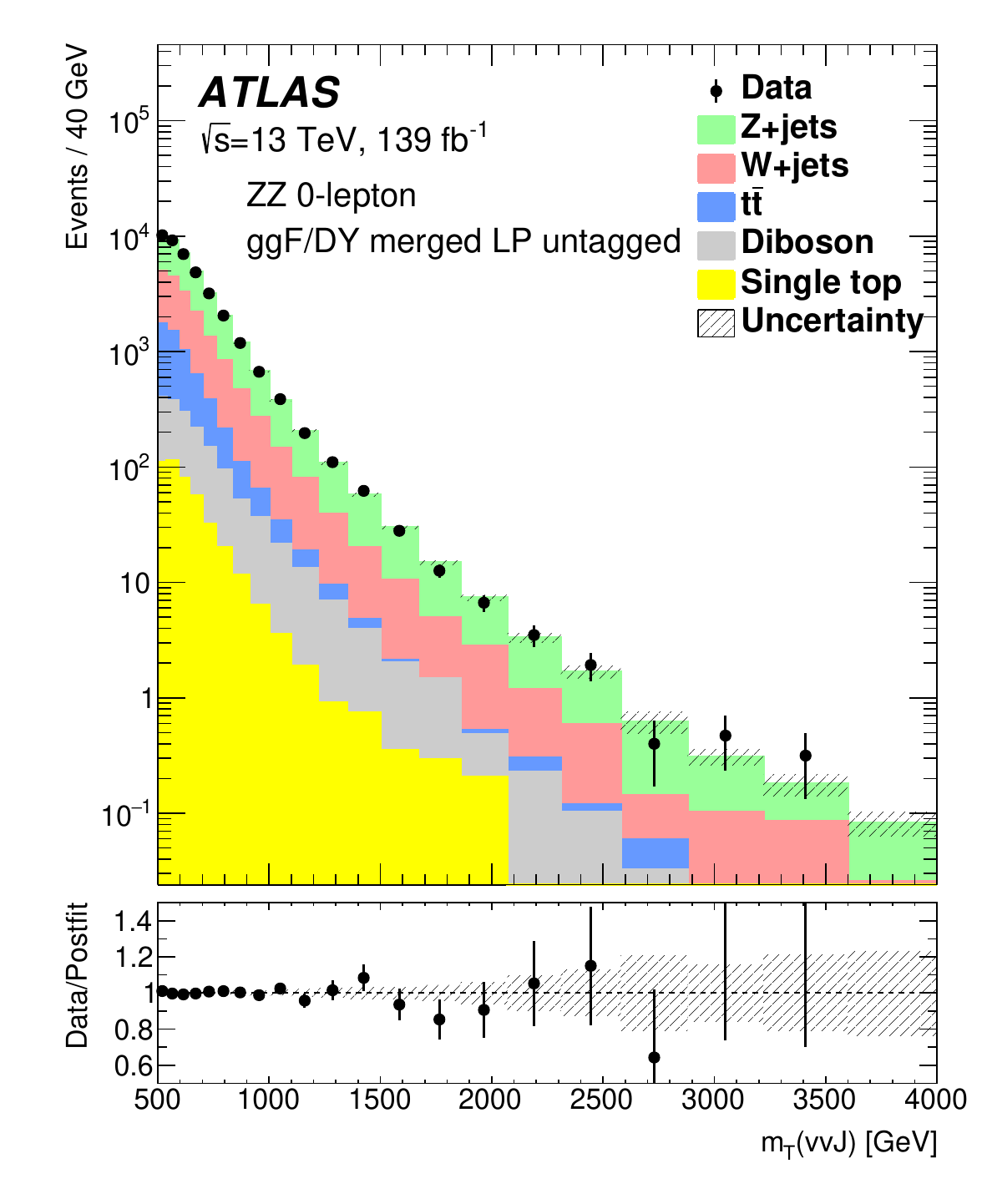}}
\caption{Comparisons of the observed data and the expected background distributions of \mT\ in the 6 $ZZ$ SRs of the \zlep channel.
The background predictions are obtained through a background-only simultaneous fit to the 6 $WW$ and 15 $ZZ$ SRs and their respective $\Vjets$ and \ttbar CRs (see text).
The bottom panes show the ratios of the observed data to the background predictions.
The blue triangles indicate bins where the ratio is non-zero and outside the vertical range of the plot.
The hatched bands represent the uncertainties in the total background predictions, combining statistical and systematic contributions.  }
\label{fig:mass0lep}
\end{figure}

\begin{figure}[!htbp]
\centering
\subfigure[VBF, merged HP]{\includegraphics[trim=0 8mm 0 8mm, width=0.32\textwidth]{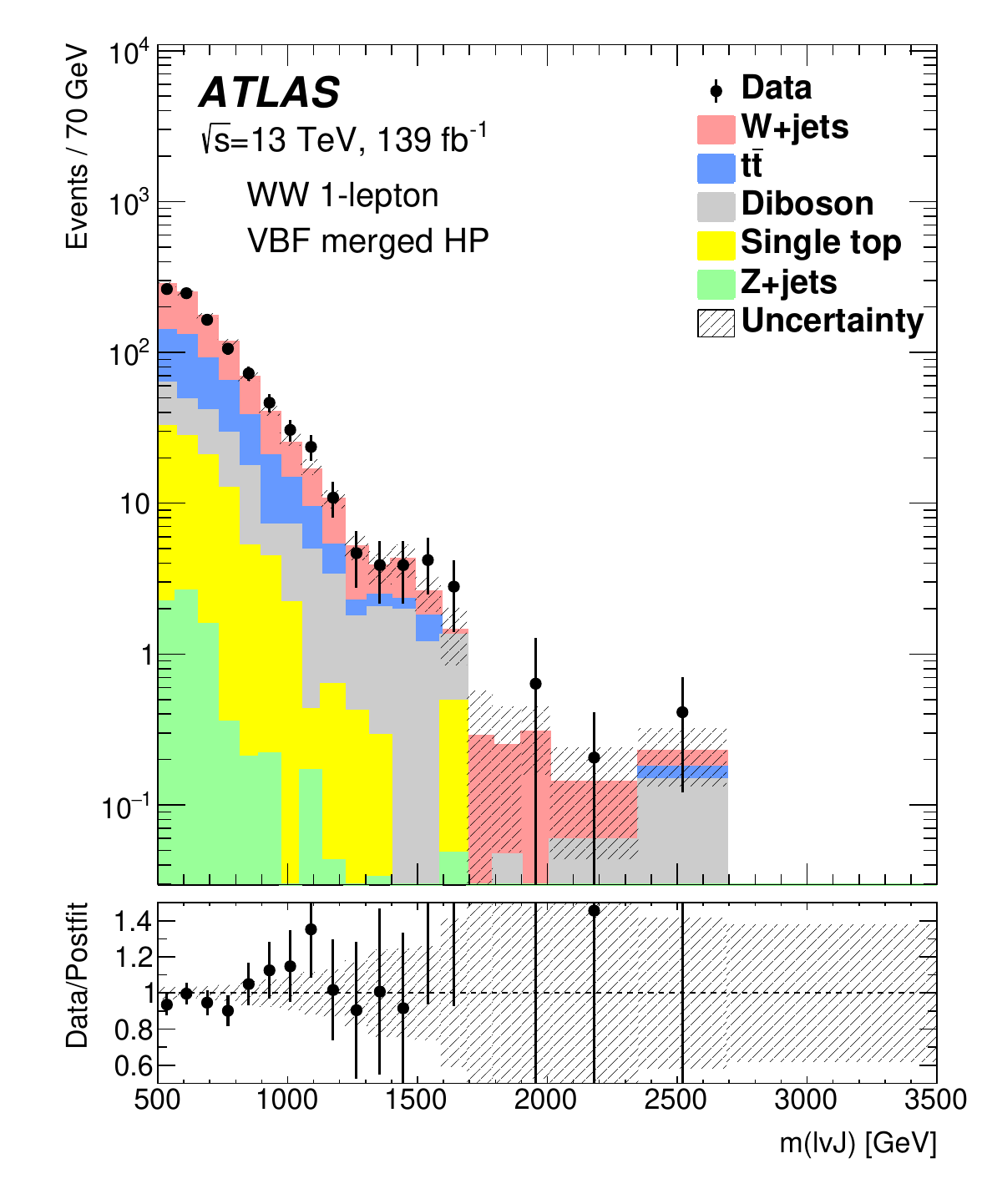}}
\subfigure[VBF, merged LP]{\includegraphics[trim=0 8mm 0 8mm, width=0.32\textwidth]{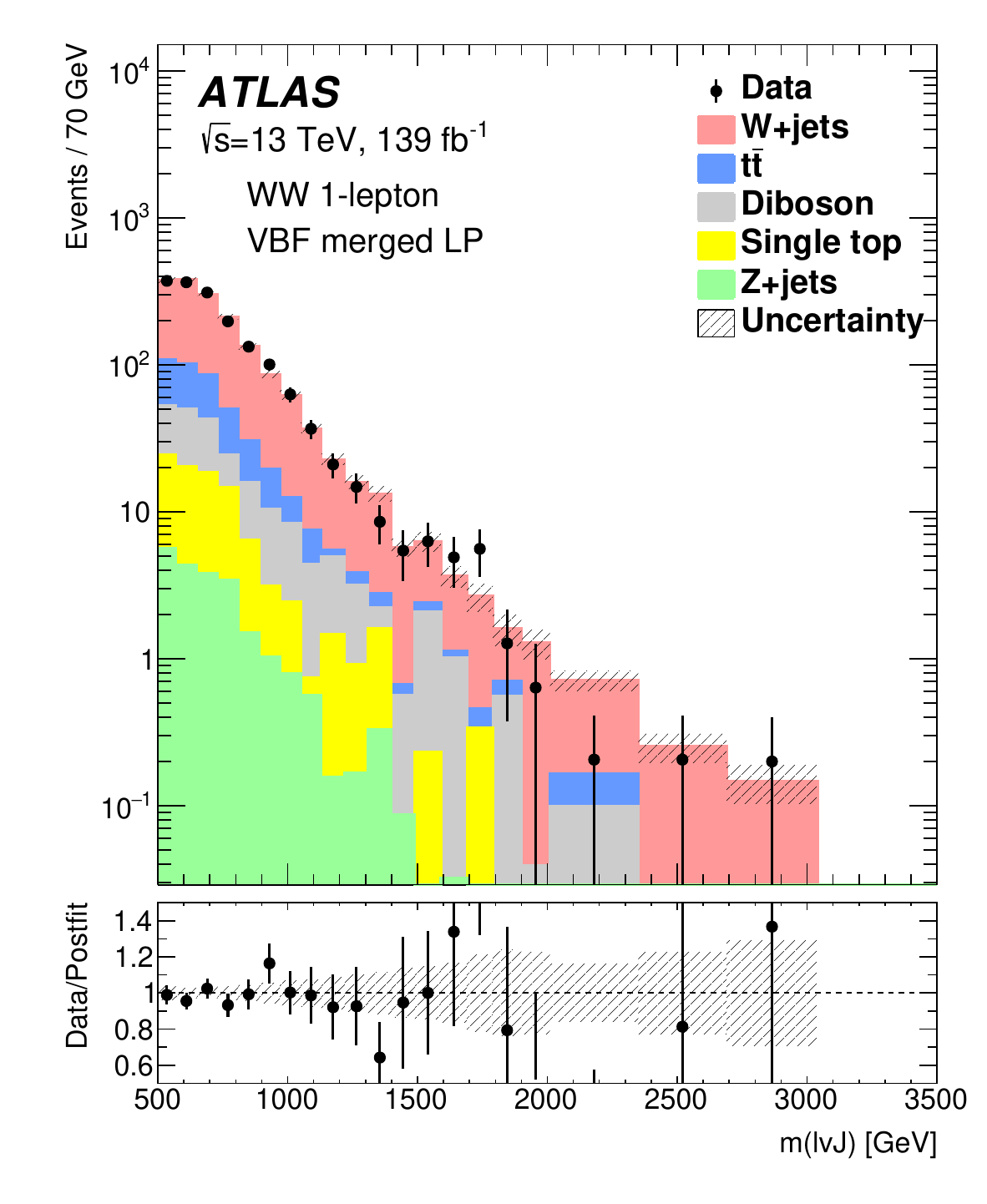}} \\
 
\subfigure[VBF, resolved] {\includegraphics[trim=0 8mm 0 8mm, width=0.32\textwidth]{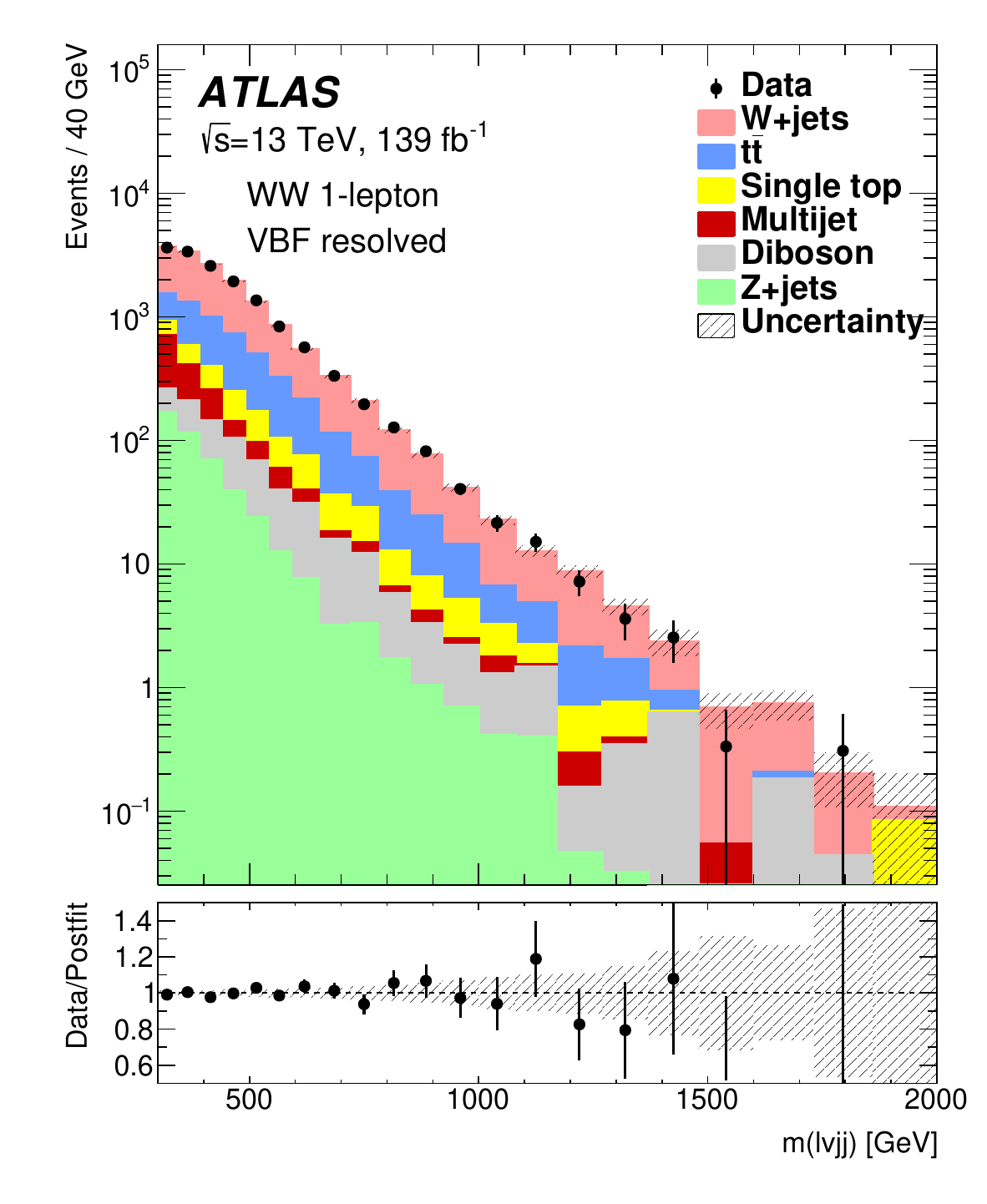}}
\subfigure[ggF/DY, merged HP]{\includegraphics[trim=0 8mm 0 8mm, width=0.32\textwidth]{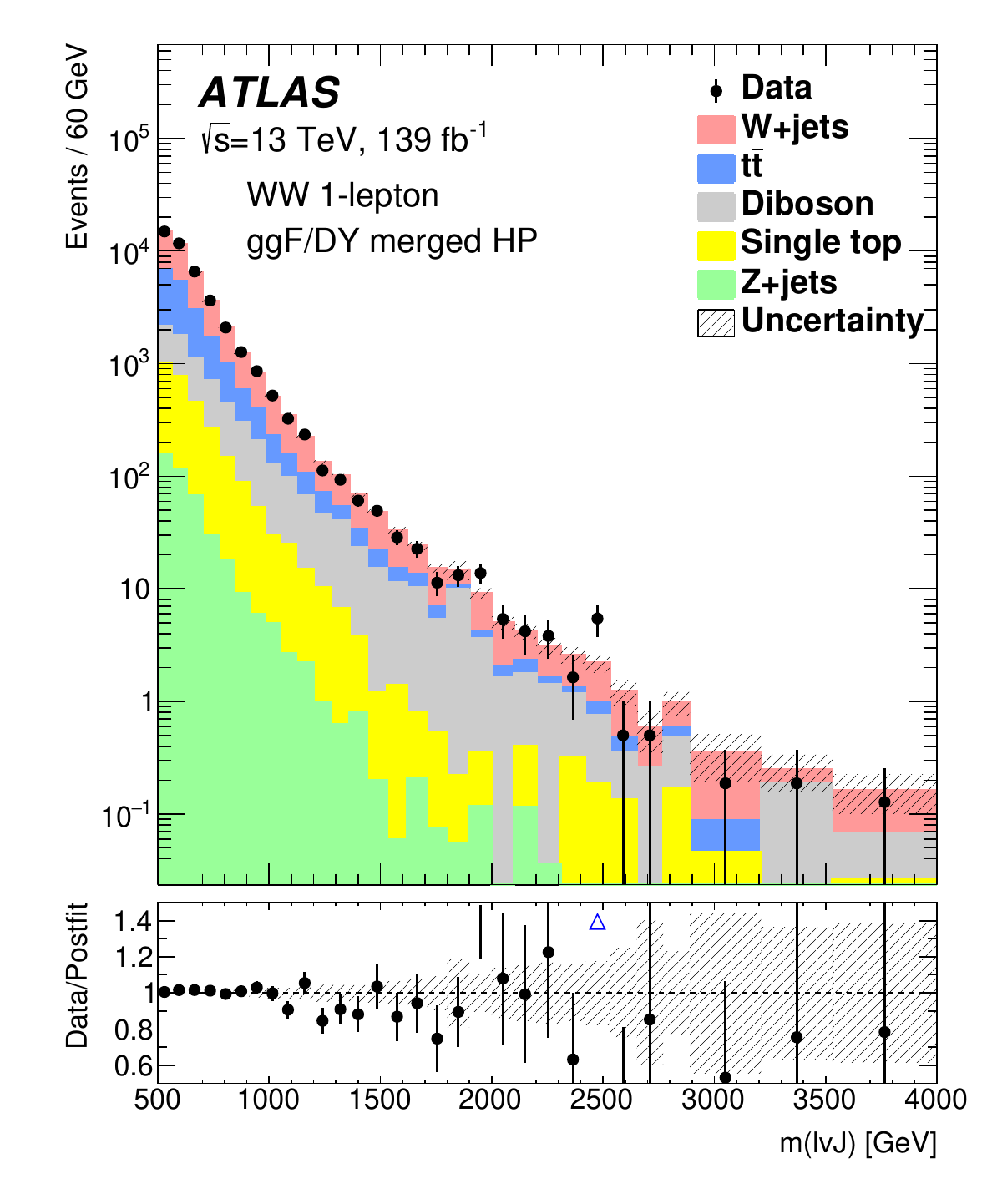}} \\
 
\subfigure[ggF/DY, merged LP]{\includegraphics[trim=0 8mm 0 8mm, width=0.32\textwidth]{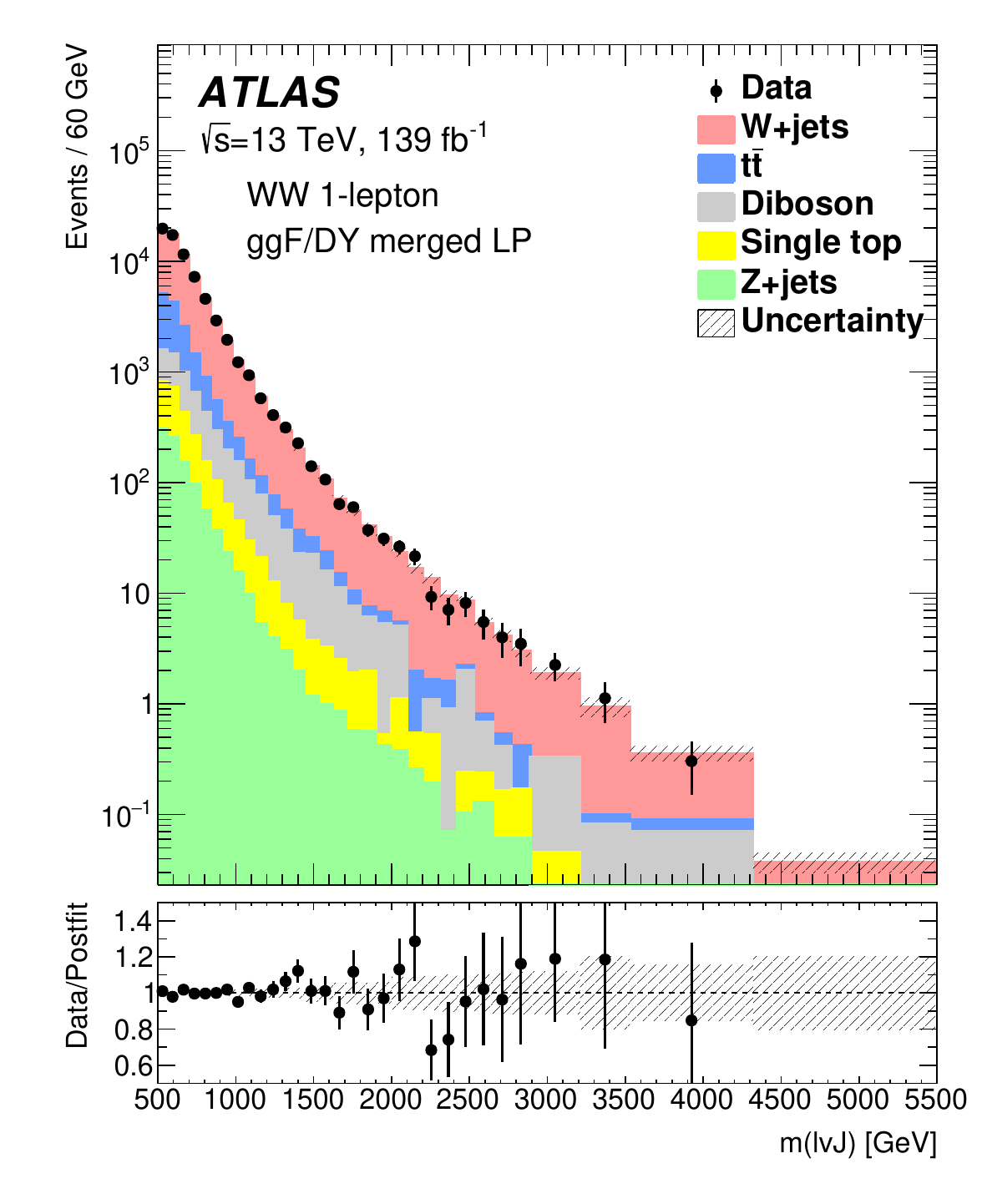}}
\subfigure[ggF/DY, resolved] {\includegraphics[trim=0 8mm 0 8mm, width=0.32\textwidth]{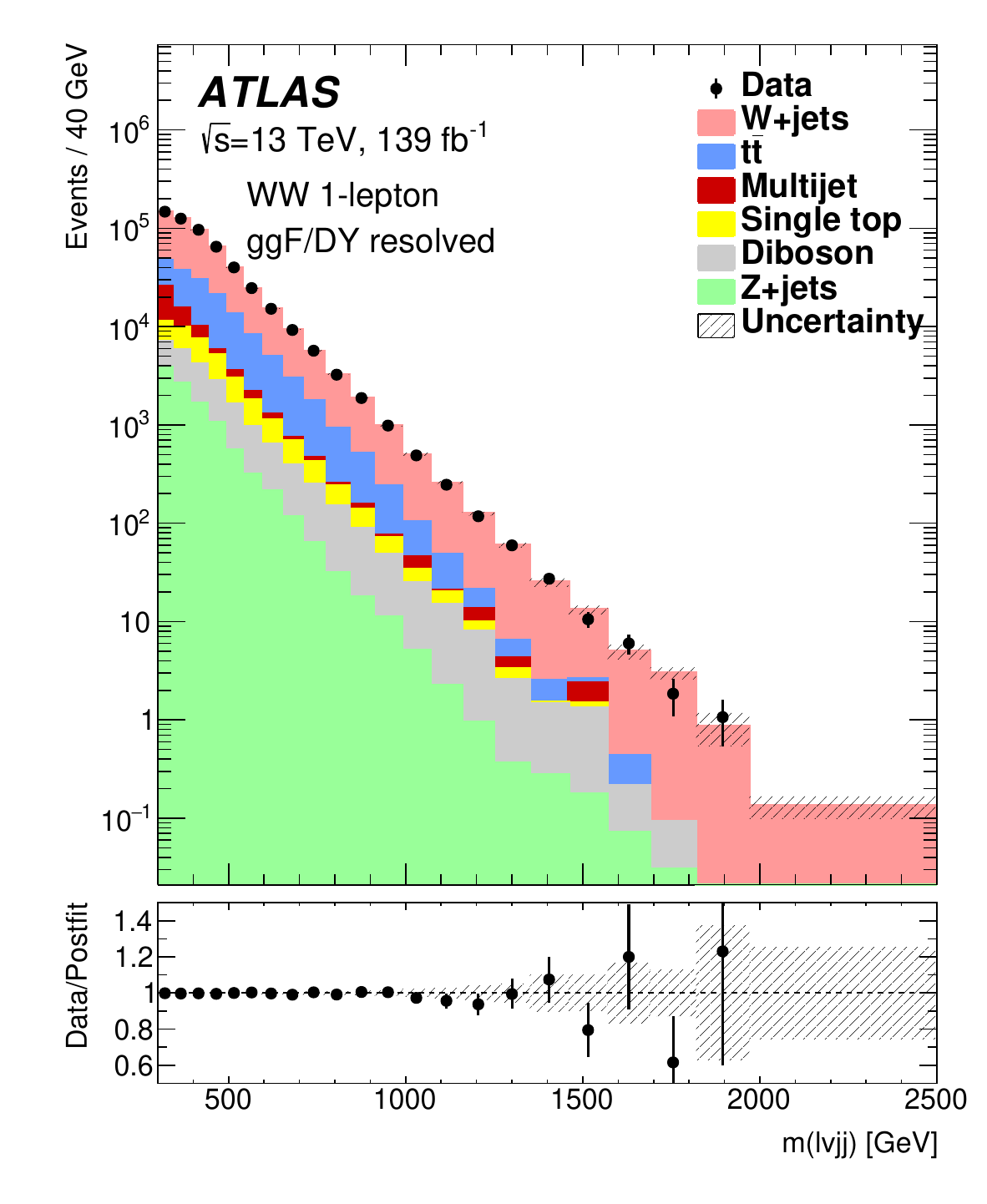}}
\caption{Comparisons of the observed data and the expected background distributions of $\mlvjj$ or $\mlvJ$ in the 6 $WW$ SRs of the \olep channel.
The background predictions are obtained through a background-only simultaneous fit to the 6 $WW$ and 15 $ZZ$ SRs and their respective $\Vjets$ and \ttbar CRs (see text).
The bottom panes show the ratios of the observed data to the background predictions.
The blue triangles indicate bins where the ratio is non-zero and outside the vertical range of the plot.
The hatched bands represent the uncertainties in the total background predictions, combining statistical and systematic contributions. }
\label{fig:mass1lep}
\end{figure}

\begin{figure}[!htbp]
\centering
\subfigure[VBF, merged HP]{\includegraphics[trim=0 8mm 0 8mm, width=0.32\textwidth]{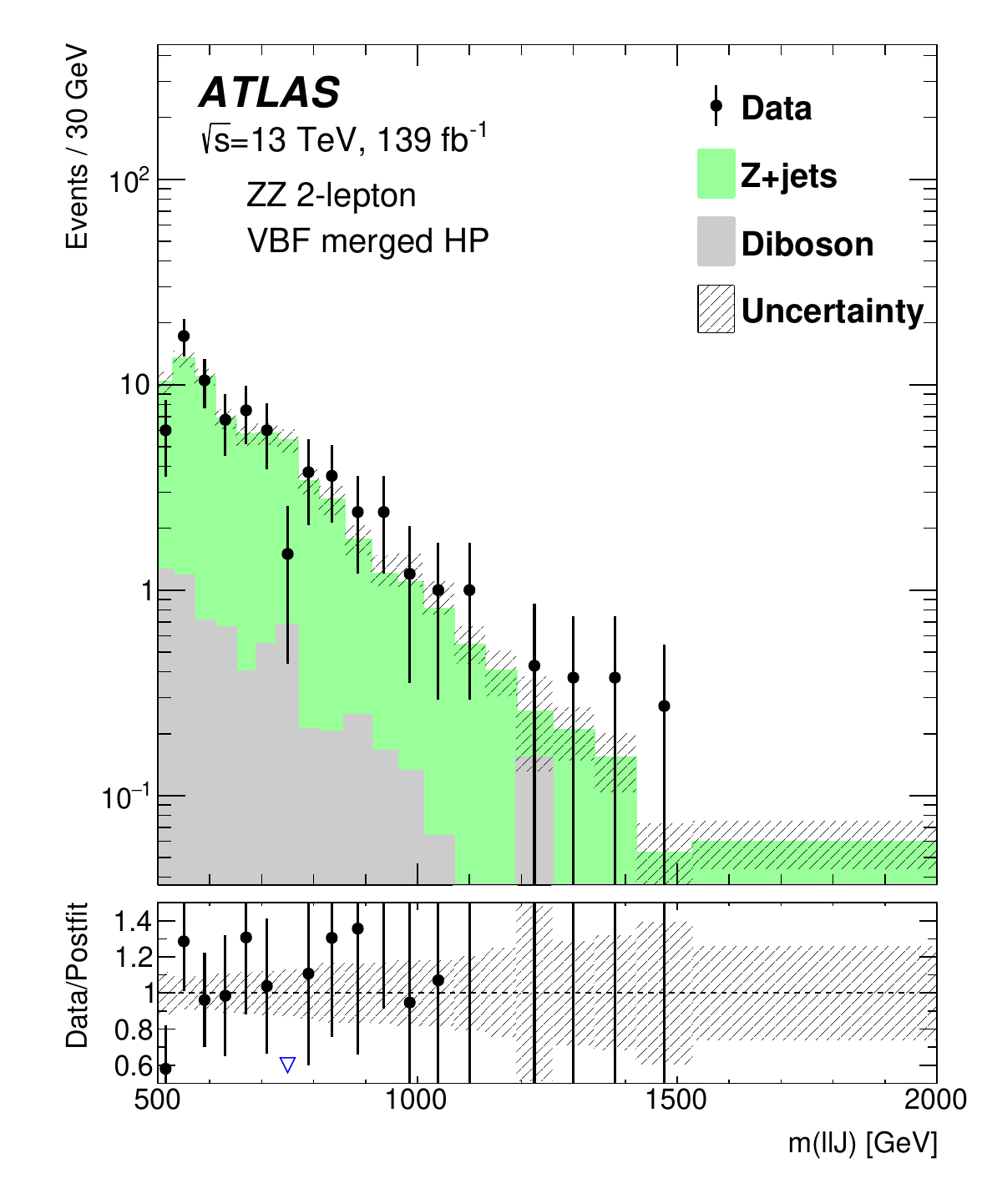}}
\subfigure[VBF, merged LP]{\includegraphics[trim=0 8mm 0 8mm, width=0.32\textwidth]{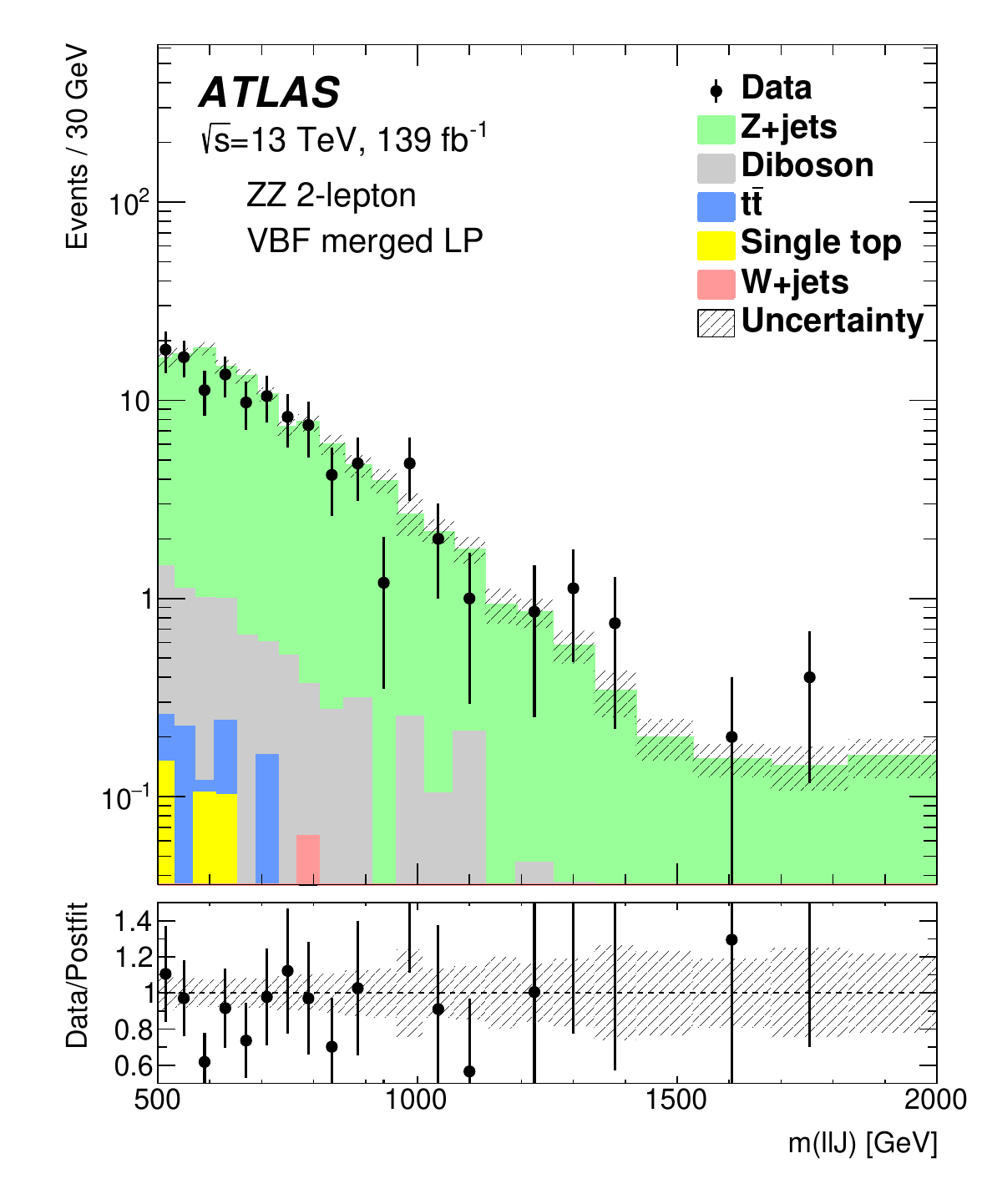}}
\subfigure[VBF, resolved] {\includegraphics[trim=0 8mm 0 8mm, width=0.32\textwidth]{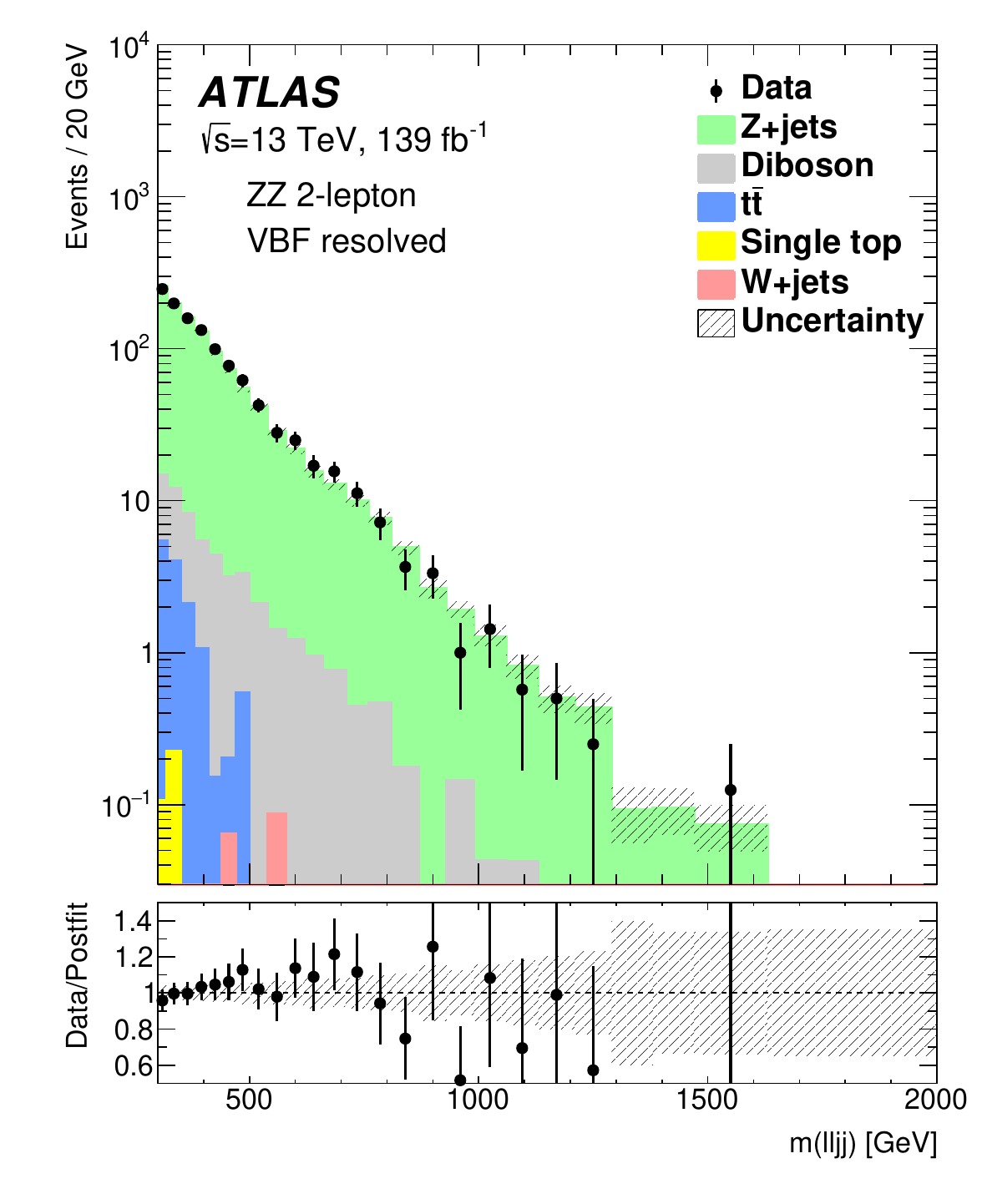}} \\
 
\subfigure[ggF/DY, merged HP tagged]  {\includegraphics[trim=0 8mm 0 8mm, width=0.32\textwidth]{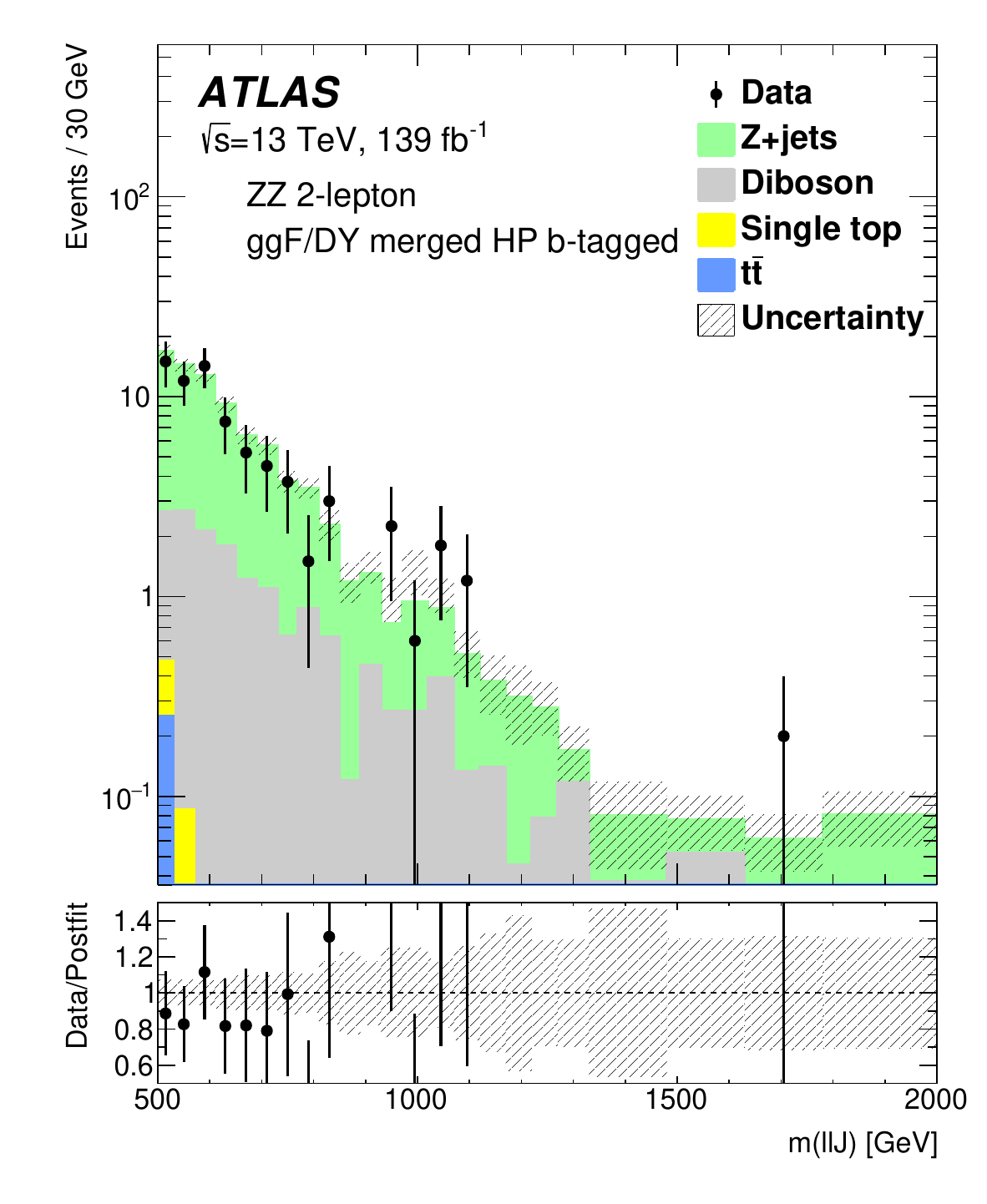}}
\subfigure[ggF/DY, merged HP untagged]{\includegraphics[trim=0 8mm 0 8mm, width=0.32\textwidth]{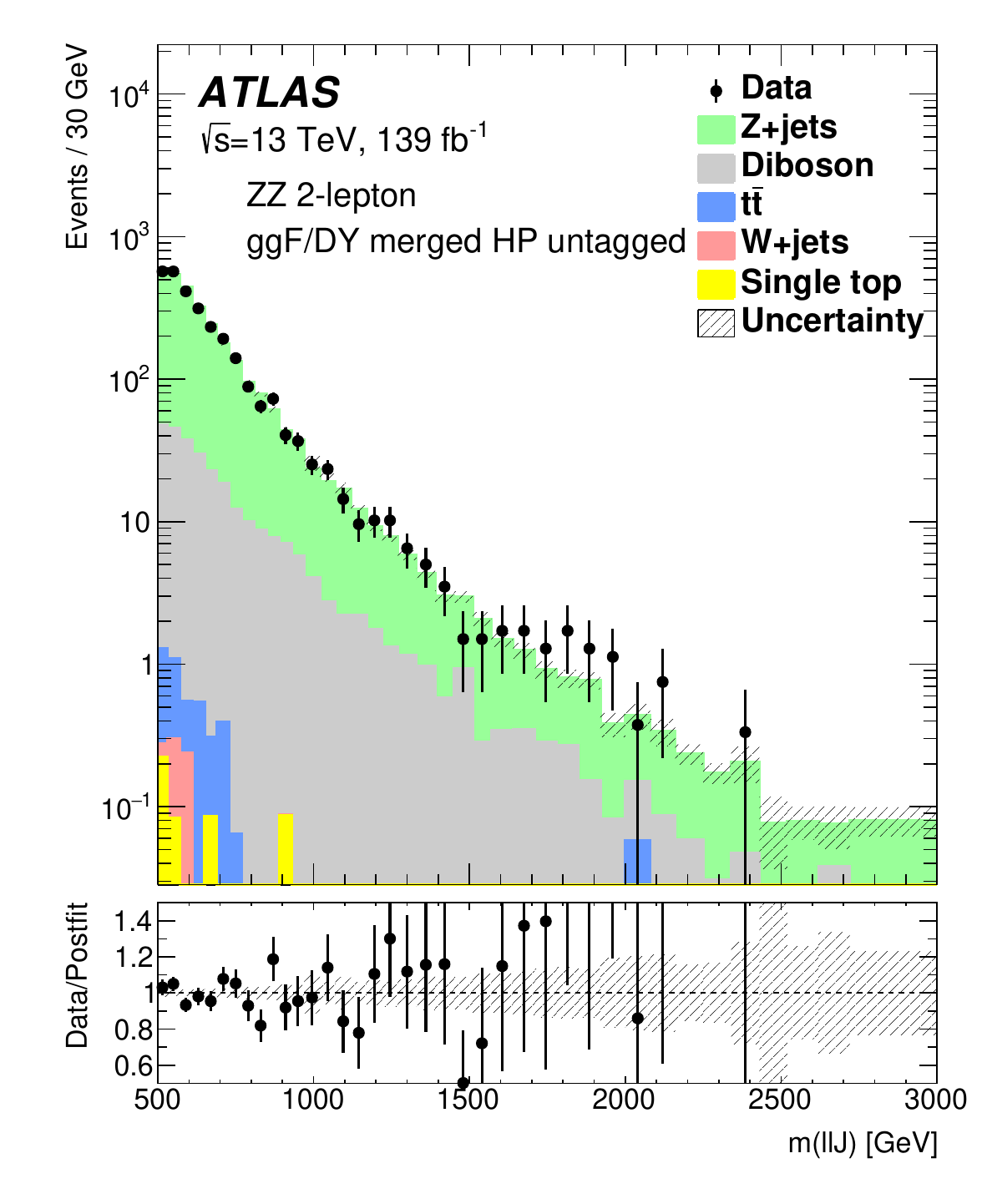}}
\subfigure[ggF/DY, merged LP tagged]  {\includegraphics[trim=0 8mm 0 8mm, width=0.32\textwidth]{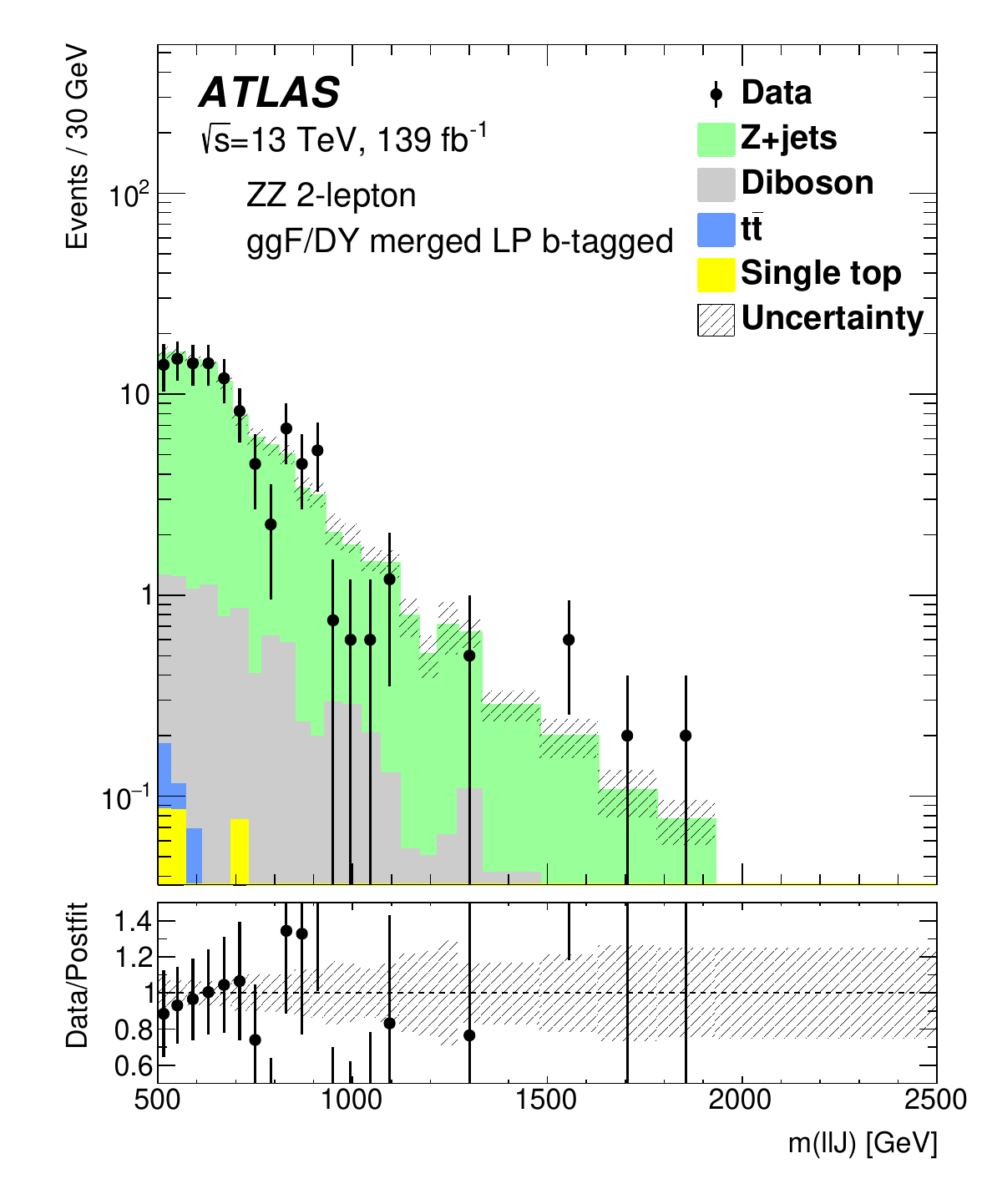}} \\
 
\subfigure[ggF/DY, merged LP untagged]{\includegraphics[trim=0 8mm 0 8mm, width=0.32\textwidth]{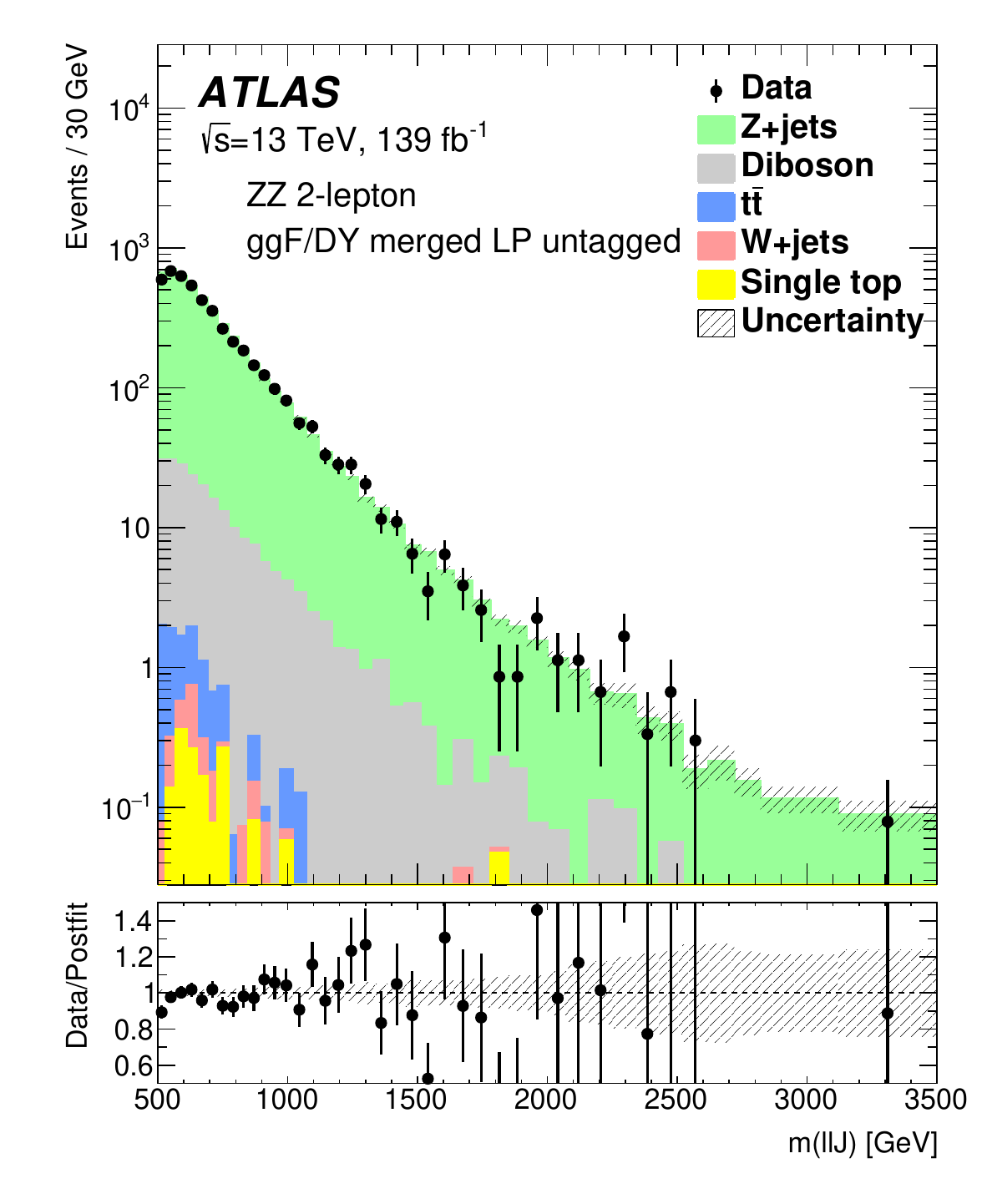}}
\subfigure[ggF/DY, resolved tagged]   {\includegraphics[trim=0 8mm 0 8mm, width=0.32\textwidth]{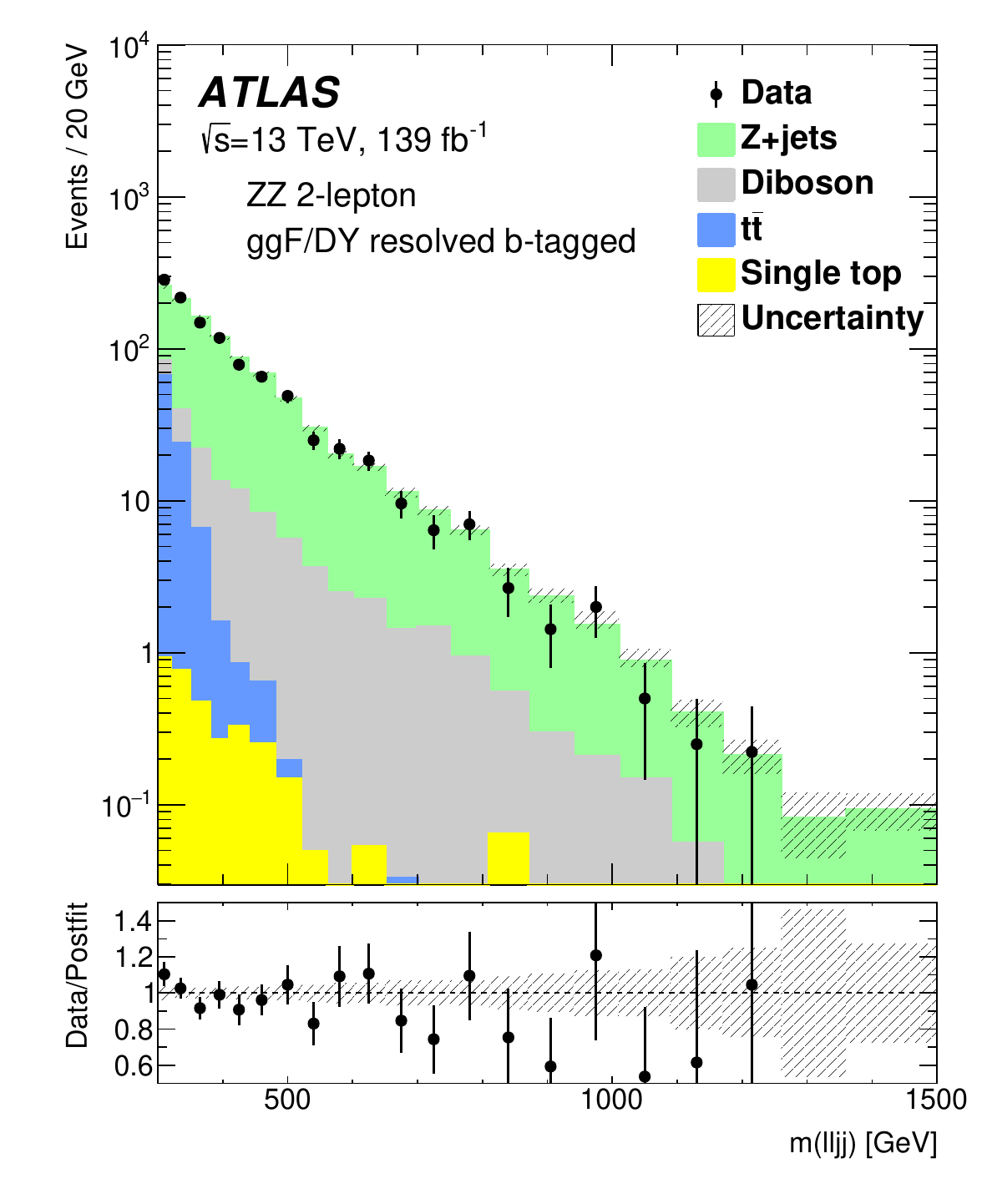}}
\subfigure[ggF/DY, resolved untagged] {\includegraphics[trim=0 8mm 0 8mm, width=0.32\textwidth]{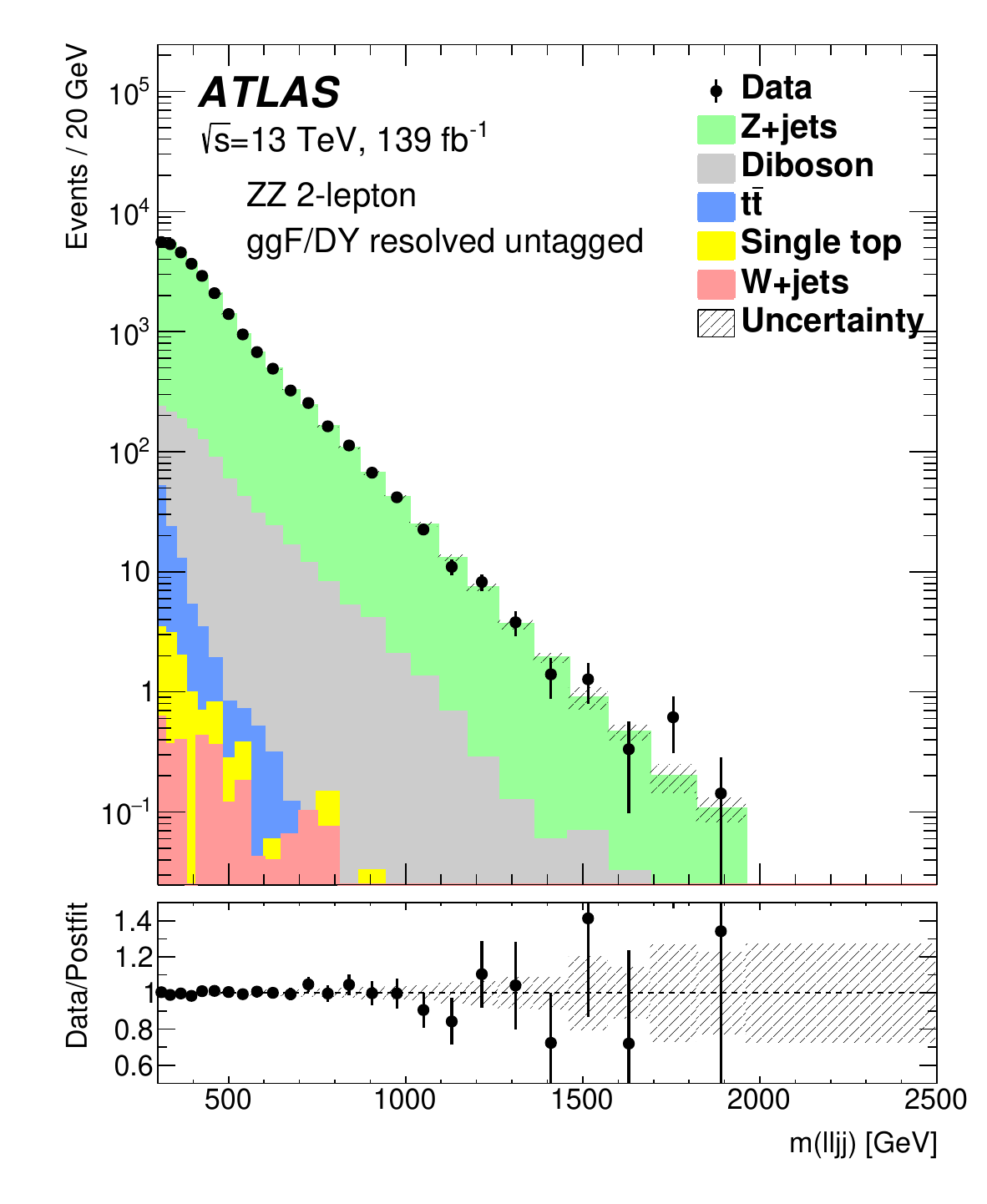}}
\caption{Comparisons of the observed data and the expected background distributions of $\mlljj$ or $\mllJ$  in the 9 $ZZ$ SRs of the \tlep channel.
The background predictions are obtained through a background-only simultaneous fit to the 6 $WW$ and 15 $ZZ$ SRs and their respective $\Vjets$ and \ttbar CRs (see text).
The bottom panes show the ratios of the observed data to the background predictions.
The blue triangles indicate bins where the ratio is non-zero and outside the vertical range of the plot.
The hatched bands represent the uncertainties in the total background predictions, combining statistical and systematic contributions. }
\label{fig:mass2lep}
\end{figure}

To test the compatibility of the data and the background expectations, the data are first fit to the background-only hypothesis for both the $WW+ZZ$ and $WZ$ fits.
Good agreement is found between the observed mass distributions and the estimated post-fit background contributions in all SRs.
As examples, the data are compared with the expected backgrounds from the $WW+ZZ$ fit in Figure~\ref{fig:mass0lep} for the \mT\ distributions of the \zlep channel, in Figure~\ref{fig:mass1lep} for the $\mlvjj/\mlvJ$ distributions of the \olep channel, and in Figure~\ref{fig:mass2lep} for the $\mlljj/\mllJ$ distributions of the \tlep channel.
The largest excess is seen at $\mT\sim 1.5$~\TeV\ in Figure~\ref{fig:mass0lep}(a).
This excess is estimated to have a local significance of $2.8$ standard deviations when modelled using RS radion production.
The differences between the pre- and post-fit background estimates are less than 10\% for a majority of the bins of the mass distributions. Larger differences, but comparable with the sizes of the statistical uncertainties, are observed for some bins with small numbers of events. Moreover the fits do not significantly alter the shapes of the background distributions for most of the signal regions. The largest change in the shape, up to 15\%, is seen in the mass distribution of the VBF merged HP signal region in the 1-lepton channel. The difference is traced primarily to the difference in the simulated $\Vjets$ samples produced with the \textsc{Sherpa} and \textsc{MadGraph5} programs.
Table~\ref{tab:WWZZ_SR_Yield} shows the post-fit estimated background event yields from different sources in all $WW$ and $ZZ$ SRs compared with the numbers of observed events in data. A similar level of agreement is obtained for the $WZ$ fit.

\begin{landscape}
 
\begin{table}[!hp]
\caption{\label{tab:WWZZ_SR_Yield} The expected background events with contributions from individual sources in 6 $WW$ and 15 $ZZ$ SRs compared with the data.
The backgrounds are estimated from a background-only simultaneous fit to all $WW$ and $ZZ$ SRs and their corresponding CRs.}
\centering
\resizebox{1.2\textwidth}{!}{
\begin{tabular}{c c r l
S[table-format=6.4, table-number-alignment=right]@{ $\pm$ }S[table-format=4.4, table-number-alignment=left]
S[table-format=5.1, table-number-alignment=right]@{ $\pm$ }S[table-format=4.1, table-number-alignment=left]
S[table-format=6.3, table-number-alignment=right]@{ $\pm$ }S[table-format=4.3, table-number-alignment=left]
S[table-format=5.1, table-number-alignment=right]@{ $\pm$ }S[table-format=4.1, table-number-alignment=left]
S[table-format=5.2, table-number-alignment=right]@{ $\pm$ }S[table-format=4.2, table-number-alignment=left]
S[table-format=5, table-number-alignment=right]@{ $\pm$ }S[table-format=4, table-number-alignment=left]
S[table-format=6, table-number-alignment=right]@{ $\pm$ }S[table-format=4, table-number-alignment=left]
r}
\toprule\midrule
\multirow{2}{*}{Channel} & $V\to qq$ & \multicolumn{2}{c}{Signal}  & \multicolumn{14}{c}{Background estimates} & \multirow{2}{*}{Data} \\
\cmidrule{5-18}
& recon. & \multicolumn{2}{c}{regions} & \multicolumn{2}{c}{\Wjets} & \multicolumn{2}{c}{\Zjets} & \multicolumn{2}{c}{\ttbar} & \multicolumn{2}{c}{Diboson} & \multicolumn{2}{c}{Single-$\mathrm{t}$} & \multicolumn{2}{c}{Multijet} & \multicolumn{2}{c}{Total} &                       \\
\hline
\multirow{8}{*}{0-lepton ($ZZ$)} 
\rule{0pt}{3ex}
& \multicolumn{17}{c}{VBF category} \\
\cmidrule{2-19}
& \multirow{2}{*}{Merged} & \multicolumn{2}{c}{HP} & 169&12 & 228&16 & 102&10 & 51&10 & 24&4 &\multicolumn{2}{c}{-}& 574&25 & 589 \\
&                         & \multicolumn{2}{c}{LP} & 370&23 & 411&20 &  75&8  & 30&4  & 21&4 &\multicolumn{2}{c}{-}& 906&33 & 936 \\
\cmidrule{2-19}
\rule{0pt}{3ex}
& \multicolumn{17}{c}{ggF/DY category} \\
\cmidrule{2-19}
& \multirow{4}{*}{Merged} & \multirow{2}{*}{HP} & Tag   &    133&14  &    270&40   &  437&31  &  100&10  &  45&7   & \multicolumn{2}{c}{-}&    982&60   &    978 \\
&                         &                     & Untag &   7600&400 & 14 300&600  & 6030&270 & 2300&180 & 840&110 & \multicolumn{2}{c}{-}& 31 100&800  & 31 074 \\
&                         & \multirow{2}{*}{LP} & Tag   &    259&28  &    560&50   &  342&24  &   67&7   &  43&7   & \multicolumn{2}{c}{-}&   1270&70   &   1277 \\
&                         &                     & Untag & 16 300&900 & 28 600&1100 & 5040&220 & 1760&150 & 600&80  & \multicolumn{2}{c}{-}& 52 400&1500 & 52 396 \\
\hline
\multirow{10}{*}{1-lepton ($WW$)}
\rule{0pt}{3ex}
& \multicolumn{17}{c}{VBF category} \\
\cmidrule{2-19}
& \multirow{2}{*}{Merged} & \multicolumn{2}{c}{HP} &    530&28  &  8.3&0.5 &  321&22  & 141&27 &  113&21  & \multicolumn{2}{c}{-} &   1110&50  &   1096 \\
&                         & \multicolumn{2}{c}{LP} &   1380&40  & 24.5&1.1 &  228&17  & 150&33 &   83&16  & \multicolumn{2}{c}{-} &   1870&60  &   1846 \\
&                Resolved & \multicolumn{2}{c}{}   & 11 360&190 &  530&10  & 4060&130 & 590&80 & 1070&210 &        960&110        & 18 570&340 & 18 530 \\
\cmidrule{2-19}
\rule{0pt}{3ex}
& \multicolumn{17}{c}{ggF/DY category} \\
\cmidrule{2-19}
& Merged   & \multicolumn{2}{c}{HP} &  24820&170  &   463&5  &  13890&220  &  4910&250  &  2800&400  & \multicolumn{2}{c}{-} &  46900&500  &  47330 \\
&          & \multicolumn{2}{c}{LP} &  60270&240  &  1095&8  &  11050&160  &  3950&210  &  1970&250  & \multicolumn{2}{c}{-} &  78300&400  &  78380 \\
& Resolved & \multicolumn{2}{c}{}   & 443500&1800 & 12480&40 & 126000&1500 & 16800&1200 & 21200&2800 &      27200&1400       & 647000&4000 & 645610 \\
\hline
\multirow{13}{*}{2-lepton ($ZZ$)} 
\rule{0pt}{3ex}
& \multicolumn{17}{c}{VBF category} \\
\cmidrule{2-19}
& \multirow{2}{*}{Merged} & \multicolumn{2}{c}{HP} & \multicolumn{2}{c}{0}   &   87&6  &  0.081&0.009 &  9.6&1.2 &  \multicolumn{2}{c}{0} &\multicolumn{2}{c}{-}&   97&6  &  101 \\
&                         & \multicolumn{2}{c}{LP} &      0.133&0.011        &  170&8  &   0.85&0.07  &  9.9&1.2 &     0.43&0.07          &\multicolumn{2}{c}{-}&  181&8  &  162 \\
&                Resolved & \multicolumn{2}{c}{}   &      0.272&0.012        & 1566&29 &   17.0&0.7   &   72&10  &     0.48&0.32          &\multicolumn{2}{c}{-}& 1656&31 & 1685 \\
\cmidrule{2-19}
\rule{0pt}{3ex}
& \multicolumn{17}{c}{ggF/DY category} \\
\cmidrule{2-19}
& \multirow{4}{*}{Merged}   & \multirow{2}{*}{HP} & Tag   &   0.0135&0.0043        &     85&6   &   0.283&0.035 & 21.1&2.3 &  0.34&0.05 &\multicolumn{2}{c}{-}&    107&7   &     94 \\
&                           &                     & Untag &   0.772 &0.010         &   3300&40  &    4.27&0.08  &  361&32  &  0.58&0.11 &\multicolumn{2}{c}{-}&   3670&50  &   3671 \\
&                           & \multirow{2}{*}{LP} & Tag   &   0.0135&0.0043        &    138&8   &   0.313&0.034 & 12.8&1.4 &  0.30&0.04 &\multicolumn{2}{c}{-}&    152&8   &    141 \\
&                           &                     & Untag &   2.341 &0.017         &   5920&50  &   10.16&0.16  &  278&26  &  2.03&0.29 &\multicolumn{2}{c}{-}&   6220&60  &   6095 \\
& \multirow{2}{*}{Resolved} & \multicolumn{2}{l}{Tag}     &  \multicolumn{2}{c}{-} &   1323&26  &     110&10    &  159&12  &   4.7&0.8  &\multicolumn{2}{c}{-}&   1600&30  &   1583 \\
&                           & \multicolumn{2}{l}{Untag}   &   4.681 &0.026         & 42 750&160 &   110.6&1.5   & 1800&100 &  13.4&2.0  &\multicolumn{2}{c}{-}& 44 650&190 & 44 604 \\
\midrule
\bottomrule
\end{tabular}}
\end{table}
 
\end{landscape}

\subsection{Limits on the production of heavy resonances}
 
Constraints on the production of heavy resonances are derived by repeating the fit to the signal-plus-background hypothesis for different signal models.
Upper limits on cross-sections are calculated with a modified frequentist method \cite{Read:2002hq}, also known as CL$_\mathrm{s}$, using the $\tilde{q}_{\mu}$ test statistic in the asymptotic approximation~\cite{Cowan:2010js} for resonance masses below 3~(1)~\TeV\ for ggF/DY (VBF) production.
For heavier resonances, the low number of events makes the approximation unreliable and the limits are obtained through pseudo-experiments instead.
The observed (expected) limits from pseudo-experiments are approximately 20--30\% (10--20\%) higher than those from the asymptotic calculations at the highest resonance masses in the search.

\subsubsection{Limits on the production of RS radions}
\begin{figure}[!htb]
\centering
\subfigure[]{\includegraphics[width=0.48\textwidth]{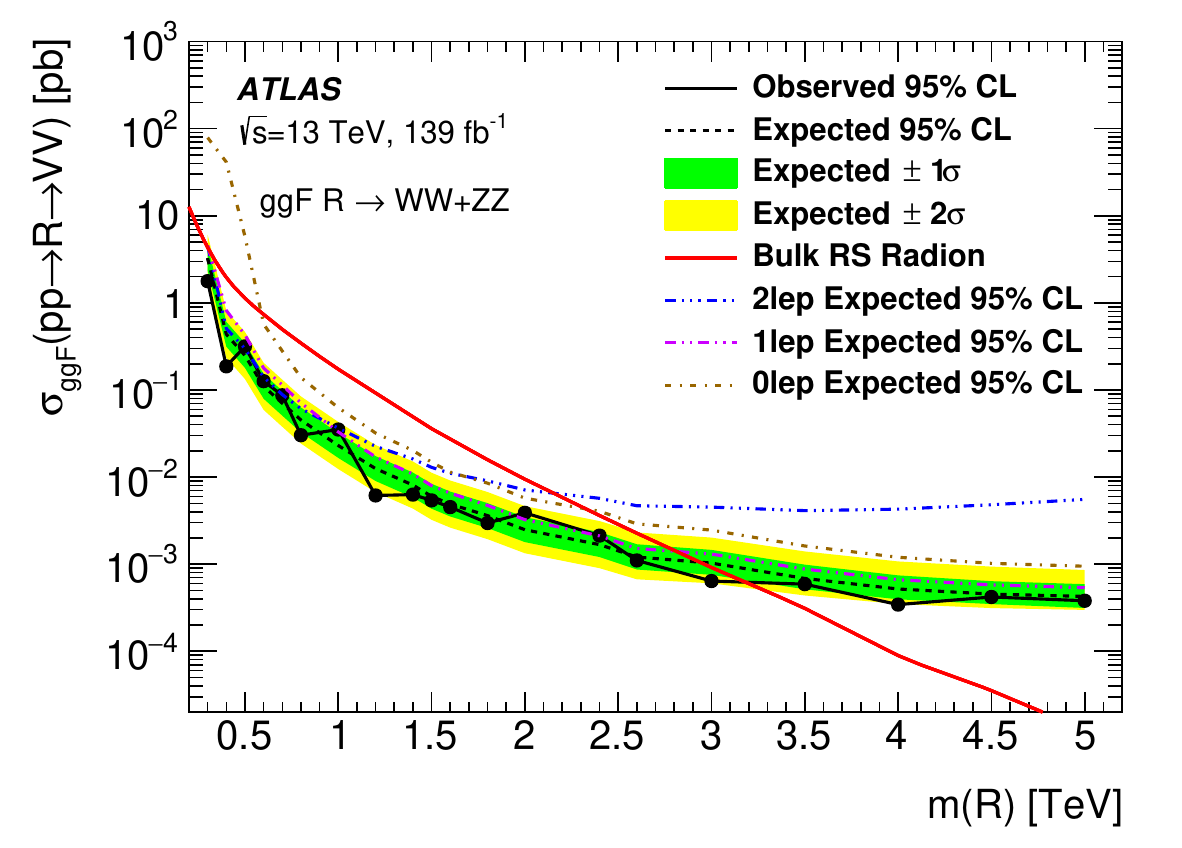}}
\subfigure[]{\includegraphics[width=0.48\textwidth]{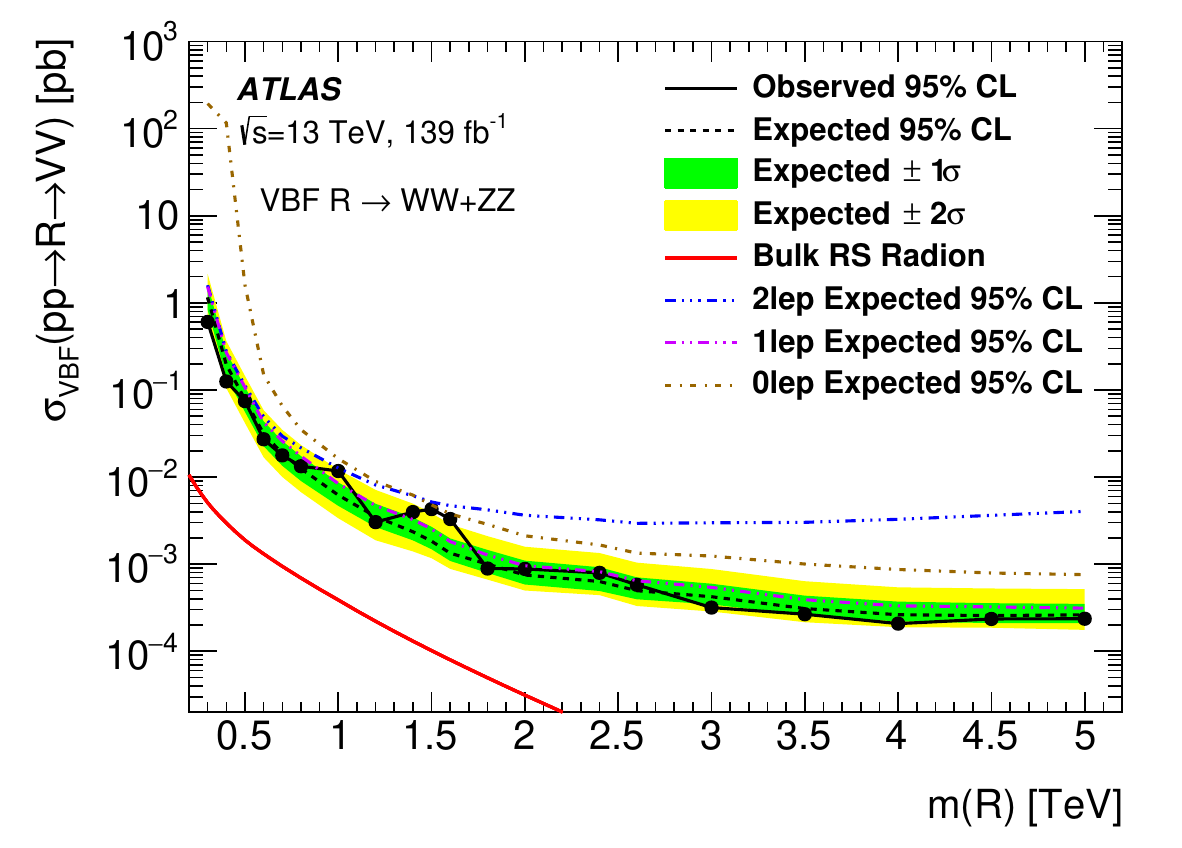}}
\caption{Observed (black solid curve) and expected (black dashed curve) 95\% CL upper limits on the (a) ggF and (b) VBF production cross-section of an RS radion at $\sqrt{s}=13$~\TeV\ in its diboson ($WW$ and $ZZ$) decay mode as functions of the RS radion mass, combining the $R\to WW$ and $R\to ZZ$ searches in the three leptonic channels.
The theoretical prediction for $\mathcal{B}(R\to WW)/\mathcal{B}(R\to ZZ)$ is assumed for the combination of the $WW$ and $ZZ$ decay modes.
The green (inner) and yellow (outer) bands represent $\pm 1\sigma$ and $\pm 2\sigma$ uncertainty in the expected limits.
Limits expected from individual leptonic channels (dot-dashed curves in blue, magenta, and brown) are also shown for comparison.
Limits are calculated in the asymptotic approximation below 3~(1)~\TeV\ and are obtained from pseudo-experiments above that for ggF (VBF) production.
Theoretical predictions (red solid curve) as functions of the RS radion mass are overlaid.}
\label{fig:LimitRad}
\end{figure}

Upper limits on the production cross-section of an RS radion in its $VV$ decay modes, $\sigma(pp\to R\to VV)$, are obtained by combining the $R\to WW$ and $R\to ZZ$ searches in the three leptonic final states.
The \olep channel is sensitive to the $R\to WW$ decay while the \zlep and \tlep channels are sensitive to the $R\to ZZ$ decay.
The limits are derived separately for the ggF and VBF processes through the $WW+ZZ$ fits for different RS radion mass hypotheses.
Figure~\ref{fig:LimitRad} shows the observed and expected upper limits at 95\% confidence level~(CL) as functions of its mass for both the ggF and VBF processes.
The observed limits for the VBF process are noticeably higher than the expected limits around an RS radion mass of 1.5~\TeV, reflecting the excess seen in the \mT\ distribution from the merged HP signal region of the \zlep channel.
The observed (expected) combined limit on $\sigma(pp\to R\to VV)$ varies from \RggFLowObs~(\RggFLowExp)~pb at 300~\GeV\ to \RggFHigObs~(\RggFHigExp)~fb at 5~\TeV\ for the ggF production process and from \RVBFLowObs~(\RVBFLowExp)~pb at 300~\GeV\ to \RVBFHigObs~(\RVBFHigExp)~fb at 5~\TeV\ for the VBF production process.
These observed (expected) upper limits exclude the ggF production of an RS radion with a mass below \RggFMasObs~(\RggFMasExp)~\TeV\ while no mass exclusion can be derived for the VBF production.
 
Except for masses below approximately 1~\TeV, the \olep channel dominates the combined search sensitivities for both the ggF and VBF processes.
This is not surprising as $\mathcal{B}(R\to WW)$ is $\approx 2\times\mathcal{B}(R\to ZZ)$ in the RS radion model.
The \tlep channel is the most sensitive at low masses, benefiting from its good mass resolution and small background contributions, and is the least sensitive at high masses because of its small expected signal yields.
The \zlep channel has the worst sensitivity at low masses and contributes non-negligibly at high masses.
 
For the RS radion search here as well as the HVT and RS graviton searches presented below, the resolved analysis is important for masses below 600~\GeV\ while the merged analysis dominates the search for higher masses.
The LP signal regions improve the cross-section sensitivities of the merged analysis by approximately 5\% for the entire mass range of the search.
The missing $X\to WV\to\lvqq,\tau\nu qq$ signal contribution in the \zlep channel as discussed in Section~\ref{sec:signal}, if included, would lower the expected cross-section upper limits by up to 4\% at 2~\TeV\ and up to 10\% at 5~\TeV.

\subsubsection{Limits on the production of HVT resonances}
\begin{figure}[!htb]
\centering
\subfigure[]{\includegraphics[width=0.48\textwidth]{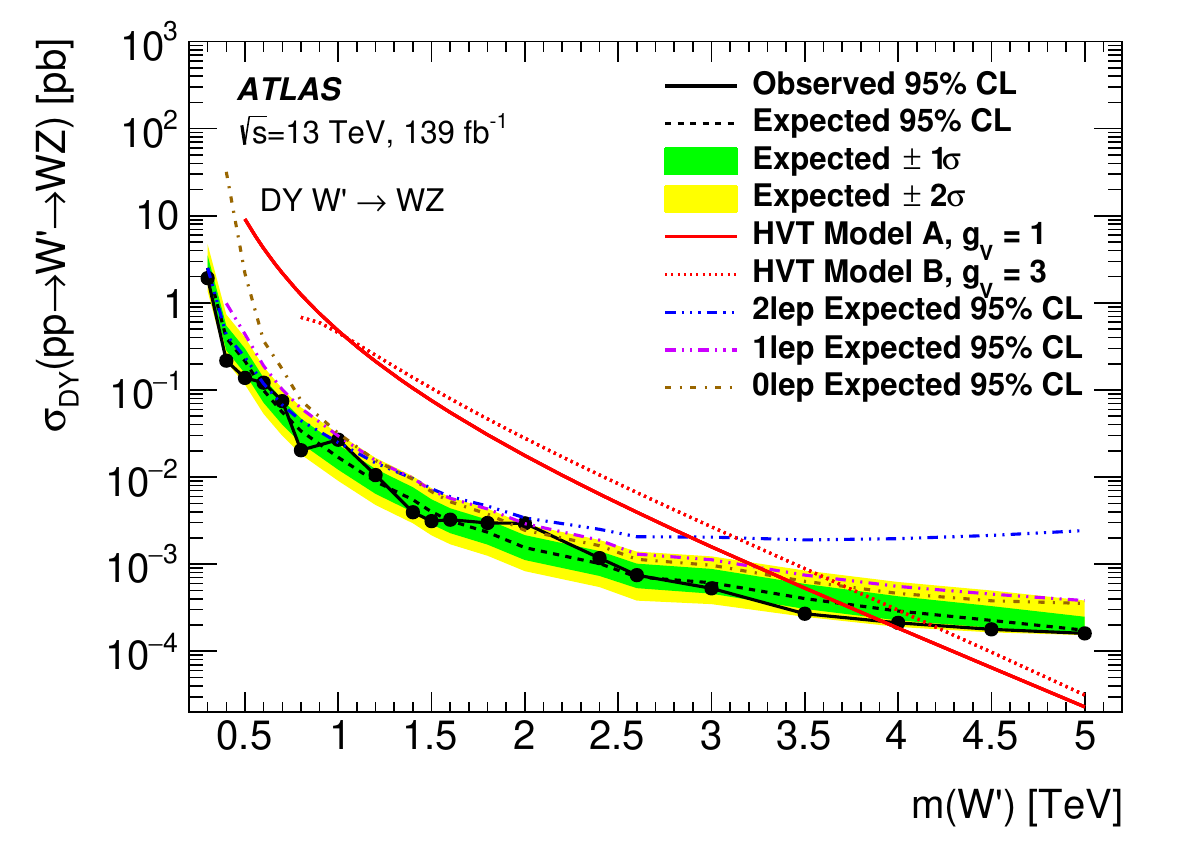}}
\subfigure[]{\includegraphics[width=0.48\textwidth]{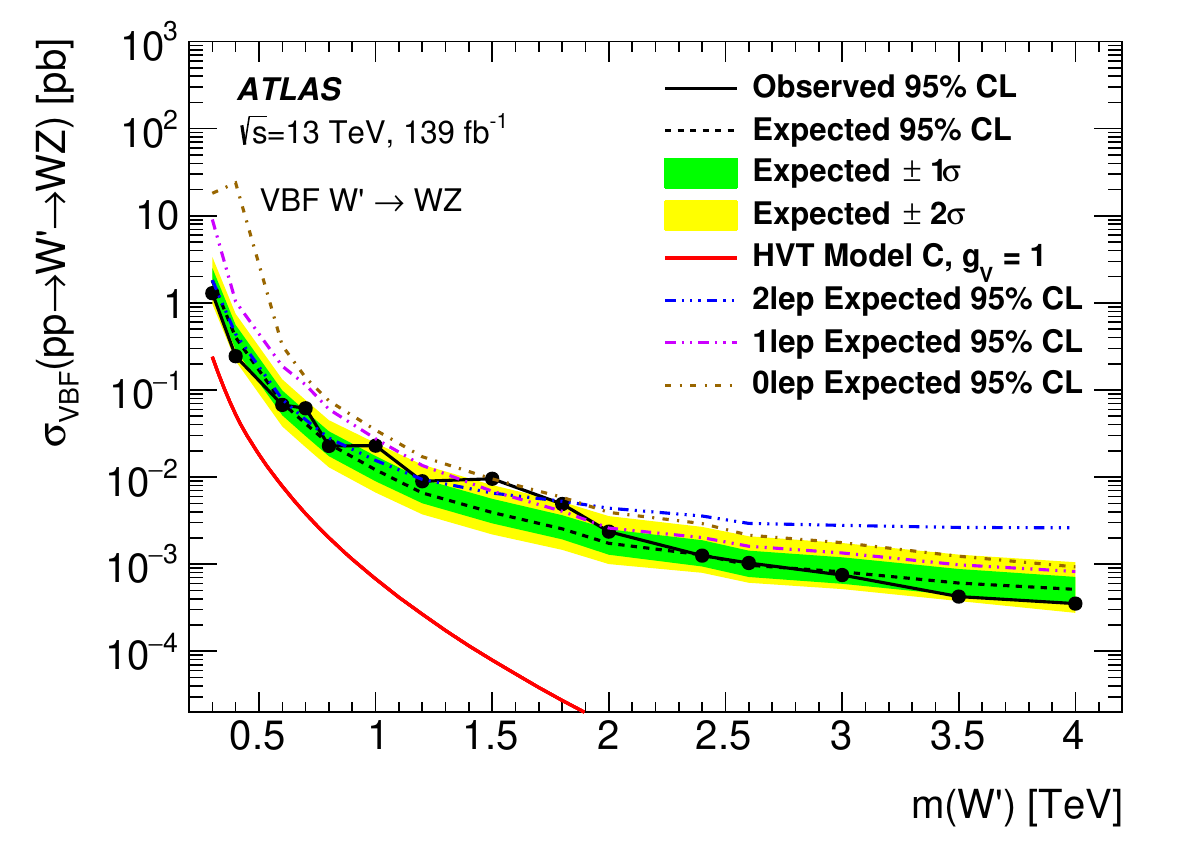}}
\caption{Observed (black solid curve) and expected (black dashed curve) 95\% CL upper limits on the (a) DY and (b) VBF production cross-section of an HVT $W'$ boson at $\sqrt{s}=13$~\TeV\ in the $WZ$ decay mode as functions of its mass, combining searches in the three leptonic channels.
The green (inner) and yellow (outer) bands represent $\pm 1\sigma$ and $\pm 2\sigma$ uncertainty in the expected limits.
Limits expected from individual leptonic channels (dot-dashed curves in blue, magenta, and brown) are also shown for comparison.
Limits are calculated in the asymptotic approximation below 3~(1)~\TeV\ and are obtained from pseudo-experiments above that for DY (VBF) production.
Theoretical predictions as functions of the $W'$ boson mass are overlaid in (a) for Model A (red solid curve) and Model B (red dotted curve) and in (b) for Model C (red solid curve). }
\label{fig:LimitHVTW}
\end{figure}

\begin{figure}[!htb]
\centering
\subfigure[]{\includegraphics[width=0.48\textwidth]{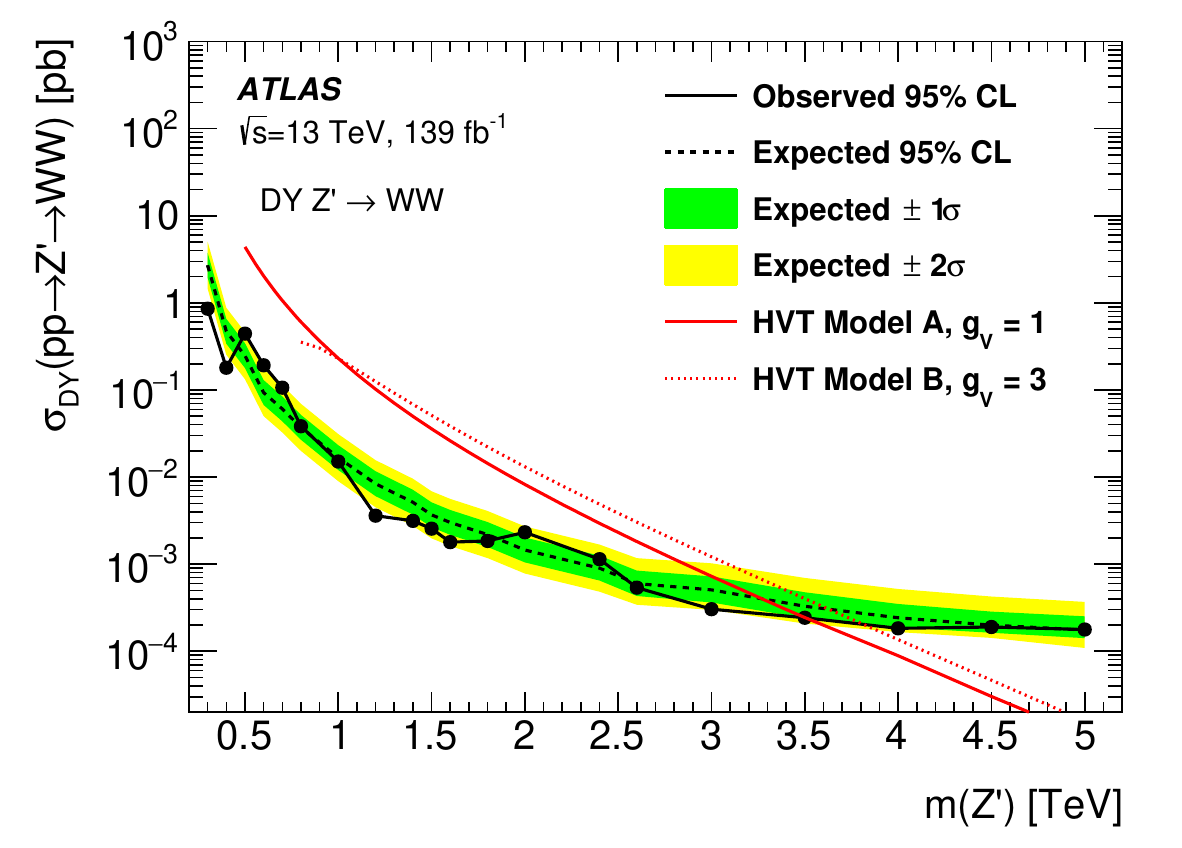}}
\subfigure[]{\includegraphics[width=0.48\textwidth]{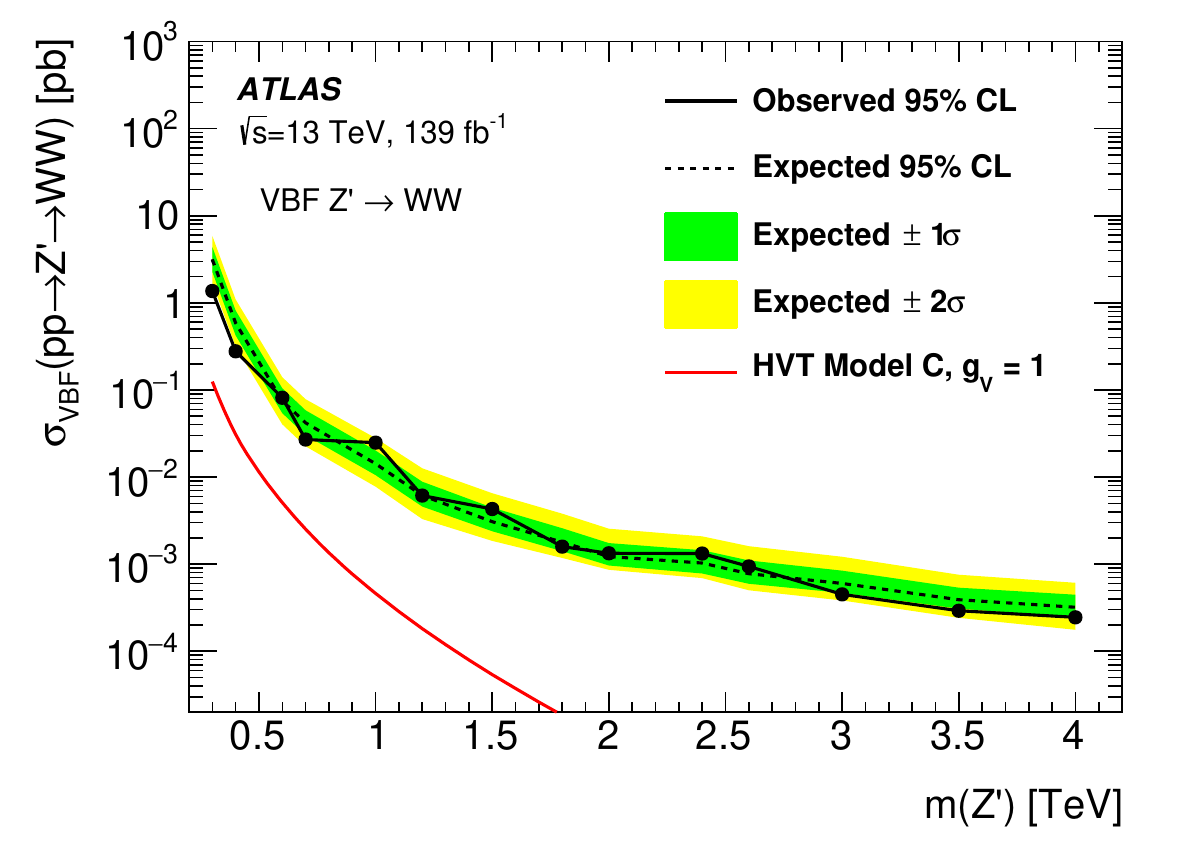}}
\caption{Observed (black solid curve) and expected (black dashed curve) 95\% CL upper limits on the (a) DY and (b) VBF production cross-section of an HVT $Z'$ boson at $\sqrt{s}=13$~\TeV\ in the $WW$ decay mode as functions of its mass from the search in the \olep channel.
The green (inner) and yellow (outer) bands represent $\pm 1\sigma$ and $\pm 2\sigma$ uncertainty in the expected limits.
Limits are calculated in the asymptotic approximation below 3~(1)~\TeV\ and are obtained from pseudo-experiments above that for DY (VBF) production.
Theoretical predictions as functions of the $Z'$ boson mass are overlaid in (a) for Model A (red solid curve) and Model B (red dotted curve) and in (b) for Model C (red solid curve). }
\label{fig:LimitHVTZ}
\end{figure}

Upper limits on the production cross-sections of HVT $W'$ and $Z'$ bosons in their $WZ$ and $WW$ decays are obtained through the $WZ$ and $WW+ZZ$ fits, respectively.
All leptonic channels contribute to the $W'\to WZ$ search while only the \olep channel contributes to the $Z'\to WW$ search.
Limits as functions of resonance masses are shown in Figures~\ref{fig:LimitHVTW} and~\ref{fig:LimitHVTZ} for $W'\to WZ$ and $Z'\to WW$, respectively, for both DY and VBF processes.
The theoretical predictions of the HVT Model A, Model B, and Model C are also shown for comparison.
The observed (expected) $\sigma(pp\to W'\to WZ)$ limit ranges from \WpDYLowObs~(\WpDYLowExp)~pb at 300~\GeV\ to \WpDYHigObs~(\WpDYHigExp)~fb at 5~\TeV\ for DY production and from \WpVBFLowObs~(\WpVBFLowExp)~pb at 300~\GeV\ to \WpVBFHigObs~(\WpVBFHigExp)~fb at 4~\TeV\ for VBF production.
These observed (expected) limits exclude an HVT $W'$ boson produced in the DY process lighter than \WpModelAObs~(\WpModelAExp)~\TeV\ for Model A and \WpModelBObs~(\WpModelBExp)~\TeV\ for Model B, but fail to exclude any mass region in the VBF process for the benchmark Model C.
For both production processes, the \tlep channel is the most sensitive for masses up to $\sim 1.5$~\TeV.
At high masses, the sensitivity is dominated by the 0-lepton and 1-lepton channels.

Only the \olep channel contributes to the $Z'\to WW$ search.
The observed~(expected) $\sigma(pp\to Z'\to WW)$ limit ranges from \ZpDYLowObs~(\ZpDYLowExp)~pb  at 300~\GeV\ to \ZpDYHigObs~(\ZpDYHigExp)~fb at 5~\TeV\ for the DY process and from \ZpVBFLowObs~(\ZpVBFLowExp)~pb at 300~\GeV\ to \ZpVBFHigObs~(\ZpVBFHigExp)~fb at 4~\TeV\ for the VBF process.
These limits exclude an HVT $Z'$ boson lighter than \ZpModelAObs~(\ZpModelAExp)~\TeV\ for Model A and \ZpModelBObs~(\ZpModelBExp)~\TeV\ for Model~B in the DY process.

\subsubsection{Limits on the production of RS gravitons}
\begin{figure}[!htb]
\centering
\subfigure[]{\includegraphics[width=0.48\textwidth]{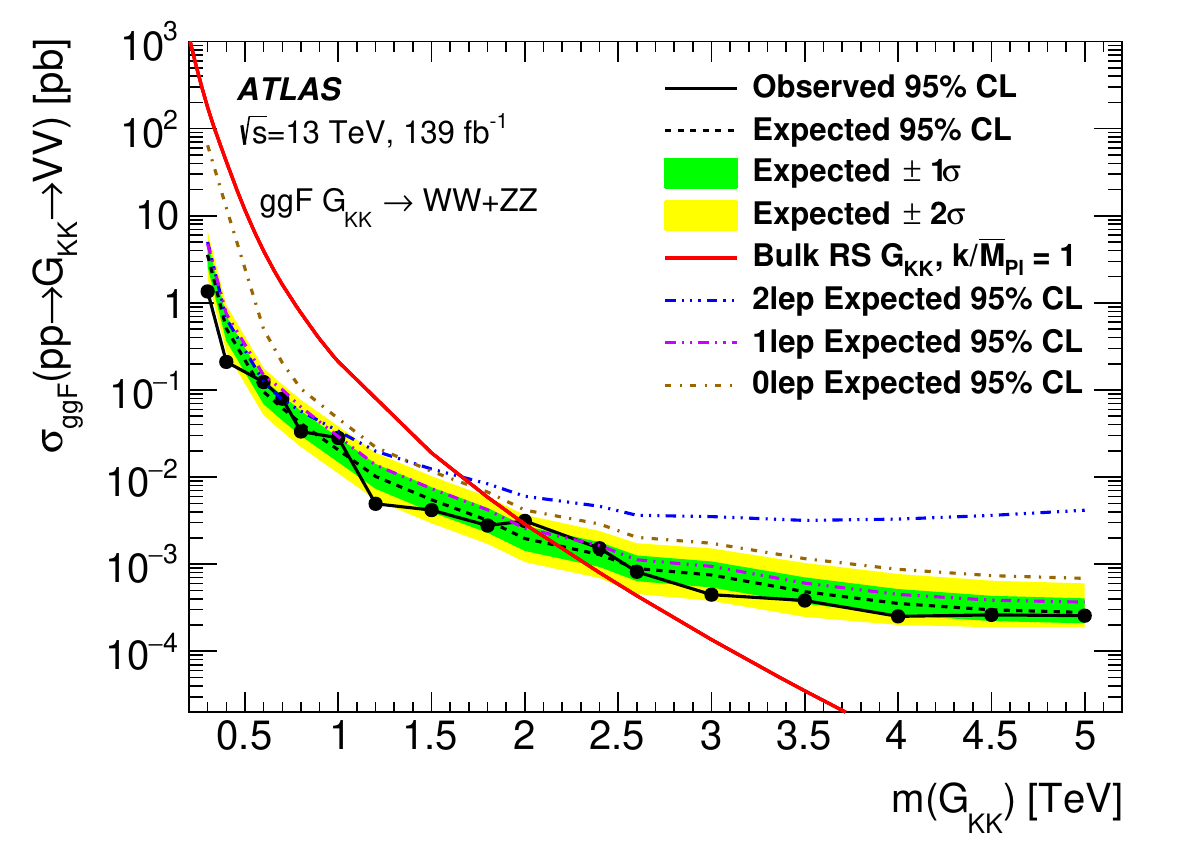}}
\subfigure[]{\includegraphics[width=0.48\textwidth]{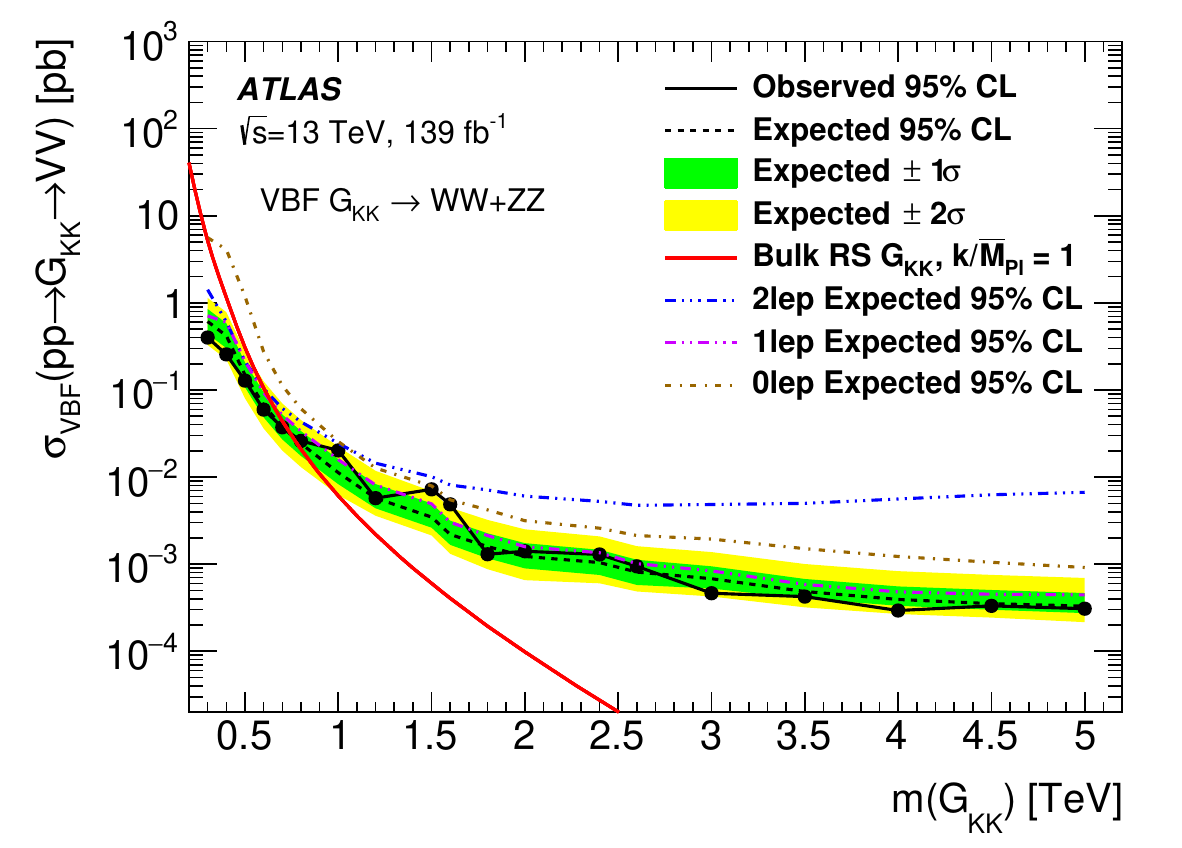}}
\caption{Observed (black solid curve) and expected (black dashed curve) 95\% CL upper limits on the (a) ggF and (b) VBF production cross-section of an RS graviton at $\sqrt{s}=13$~\TeV\ in its diboson ($WW$ and $ZZ$) decay mode as functions of its mass, combining the searches for the $\Gkk\to WW$ and $\Gkk\to ZZ$ decays in the three leptonic channels.
The theoretical prediction for $\mathcal{B}(\Gkk\to WW)/\mathcal{B}(\Gkk\to ZZ)$ is assumed for the  combination of the $WW$ and $ZZ$ decay modes.
The green (inner) and yellow (outer) bands represent $\pm 1\sigma$ and $\pm 2\sigma$ uncertainty in the expected limits.
Limits expected from individual leptonic channels (dashed curves in blue, magenta, and brown) are also shown for comparison.
Limits are calculated in the asymptotic approximation below 3~(1)~\TeV\ and are obtained from pseudo-experiments above that for ggF (VBF) production.
Theoretical predictions (red solid curves) for the chosen model as functions of the RS graviton mass are overlaid.}
\label{fig:LimitGkk}
\end{figure}

Upper limits on the production cross-section of an RS graviton in its $VV$ decays, $\sigma(pp\to \Gkk\to VV)$, are obtained following the same procedure used to derive the RS radion limits.
The observed and expected upper limits as functions of its mass for both the ggF and VBF processes are shown in Figure~\ref{fig:LimitGkk}.
The observed (expected) limit varies from \GkkggFLowObs~(\GkkggFLowExp)~pb at 300~\GeV\ to \GkkggFHigObs~(\GkkggFHigExp)~fb at 5~\TeV\ for the ggF production process and from \GkkVBFLowObs~(\GkkVBFLowExp)~pb at 300~\GeV\ to \GkkVBFHigObs~(\GkkVBFHigExp)~fb at 5~\TeV\ for the VBF production process.
Compared with theory cross-sections, with the benchmark parameters, the observed (expected) upper limits at 95\%~CL exclude the production of an RS graviton  lighter than \GkkggFMasObs~(\GkkggFMasExp)~\TeV\ in the ggF process and lighter than \GkkVBFMasObs~(\GkkVBFMasExp)~\TeV\ in the VBF process.
 
Similar to the RS radion case, the \olep channel contributes dominantly to the combined search sensitivities at high masses while the \tlep channel is slightly more sensitive than the \olep channel for masses below $\sim 1$~\TeV.
The \zlep channel is the least sensitive at low masses, but provides significant contributions at high masses.

\subsection{Comparisons of the limits}

Table~\ref{tab:MassLimits} summarises the observed and expected 95\% CL lower limits on the masses of the resonances in the benchmark models studied in this paper. These mass limits and the cross-section upper limits presented above are significantly more stringent than those published previously from similar searches.
The cross-section upper limits are a factor of three or more lower than those of the search using the same data set, but in the hadronic $VV\to JJ$ final state~\cite{HDBS-2018-31}.
Compared to the searches with the 36.1~\ifb\ data set in the same leptonic final states~\cite{EXOT-2016-28,EXOT-2016-29}, an improvement of a factor of three or more in the cross-section upper limits is also obtained for most of the searches at the highest masses. The observed lower mass limits of this search for the HVT $W'$ and $Z'$ as well as for the RS graviton are also similar to those from the combinations of searches in the fully-leptonic, semi-leptonic, and fully-hadronic final states from smaller datasets of $\sim 36$~\ifb~\cite{EXOT-2017-31,CMS-B2G-18-006}.

\begin{table}[!htb]
\caption{Observed (expected) 95\% CL lower limits on the mass, in \TeV, of different resonances in the benchmark models studied. The symbol ``$-$'' means no limit is set.}
\label{tab:MassLimits}\vspace*{-0.2cm}
\begin{center}
\begin{tabular}{cc|c|ccc|cc} \hline\hline
& Production     &  \multirow{2}{*}{RS radion}  &   \multicolumn{3}{c|}{HVT}   &  \multirow{2}{*}{RS graviton} & \\ \cline{4-6}
& process       &                           &        & $W'$    &  $Z'$    &                          \\ \hline
& \multirow{2}{*}{ggF/DY} & \multirow{2}{*}{$\RggFMasObs\;(\RggFMasExp)$}  & Model A & $\WpModelAObs\;(\WpModelAExp)$ & $\ZpModelAObs\;(\ZpModelAExp)$ & \multirow{2}{*}{$\GkkggFMasObs\;(\GkkggFMasExp)$} \\
&                        &                                                & Model B & $\WpModelBObs\;(\WpModelBExp)$ & $\ZpModelBObs\;(\ZpModelBExp)$ &                                                   \\ \hline
& VBF                    &  $-$                                           & Model C & $-$                            & $-$                            & $\GkkVBFMasObs\;(\GkkVBFMasExp)$  \\ \hline\hline
 
\end{tabular}
\end{center}
\end{table}

% End of text imported from the .//tex/results.tex input file
 
% The next lines are included from the .//tex/summary.tex input file
\section{Summary}
\label{sec:summary}
 
Searches for the production of heavy diboson resonances are performed using the proton--proton collision data with an integrated luminosity of \intLumi~\fbi\ at $\sqrt{s}=13$~\TeV. The data were recorded by the ATLAS experiment between 2015 and 2018 at the LHC. The $WW$, $ZZ$ and $WZ$ decay modes of the heavy resonances in the \zlep, \olep and \tlep final states are considered. The data are found to be in good agreement with  background expectations. Upper limits on the production of heavy resonances in the mass range 300--5000~\GeV\ through gluon--gluon fusion, Drell--Yan or vector-boson fusion processes are derived for Standard Model extensions with an additional neutral scalar, a heavy vector triplet, or warped extra dimensions.
 
Combining the $WW$ and $ZZ$ decay modes, the observed 95\% confidence-level upper limit on $\sigma(pp\to X\to VV)$ for the ggF~(VBF) process ranges from \RggFLowObs~(\RVBFLowObs)~pb at 300~\GeV\ to \RggFHigObs~(\RVBFHigObs)~fb at 5~\TeV\ for an RS radion and from \GkkggFLowObs~(\GkkVBFLowObs)~pb at 300~\GeV\ to \GkkggFHigObs~(\GkkVBFHigObs)~fb at 5~\TeV\ for an RS graviton. These observed limits set lower mass limits of \RggFMasObs~\TeV\ for the ggF production of an RS radion, and \GkkggFMasObs~(\GkkVBFMasObs)~\TeV\ for the ggF~(VBF) production of an RS graviton.

For the production of $W'$ and $Z'$ bosons in the HVT framework, the observed upper limit on $\sigma(pp\to W'\to WZ)$ varies from \WpDYLowObs~pb at 300~\GeV\ to \WpDYHigObs~fb at 5~\TeV\ for DY production and from \WpVBFLowObs~pb at 300~\GeV\ to \WpVBFHigObs~fb at 4~\TeV\ for VBF production. Similarly, the limits on $\sigma(pp\to Z'\to WW)$ are observed to vary from \ZpDYLowObs~pb at 300~\GeV\ to \ZpDYHigObs~fb at 5~\TeV\ for DY production and from \ZpVBFLowObs~pb at 300~\GeV\ to \ZpVBFHigObs~fb at 4~\TeV\ for VBF production. In the benchmark Model A~(Model B), these cross-section upper limits exclude the ggF production of a $W'$ boson with $m(W')<\WpModelAObs\ (\WpModelBObs)$~\TeV\ and a $Z'$ boson with $m(Z')<\ZpModelAObs\ (\ZpModelBObs)$~\TeV.
% End of text imported from the .//tex/summary.tex input file

\section*{Acknowledgements}
 
% The next lines are included from the .//acknowledgements/Acknowledgements.tex input file

We thank CERN for the very successful operation of the LHC, as well as the
support staff from our institutions without whom ATLAS could not be
operated efficiently.
 
We acknowledge the support of ANPCyT, Argentina; YerPhI, Armenia; ARC, Australia; BMWFW and FWF, Austria; ANAS, Azerbaijan; SSTC, Belarus; CNPq and FAPESP, Brazil; NSERC, NRC and CFI, Canada; CERN; ANID, Chile; CAS, MOST and NSFC, China; COLCIENCIAS, Colombia; MSMT CR, MPO CR and VSC CR, Czech Republic; DNRF and DNSRC, Denmark; IN2P3-CNRS and CEA-DRF/IRFU, France; SRNSFG, Georgia; BMBF, HGF and MPG, Germany; GSRT, Greece; RGC and Hong Kong SAR, China; ISF and Benoziyo Center, Israel; INFN, Italy; MEXT and JSPS, Japan; CNRST, Morocco; NWO, Netherlands; RCN, Norway; MNiSW and NCN, Poland; FCT, Portugal; MNE/IFA, Romania; MES of Russia and NRC KI, Russia Federation; JINR; MESTD, Serbia; MSSR, Slovakia; ARRS and MIZ\v{S}, Slovenia; DST/NRF, South Africa; MICINN, Spain; SRC and Wallenberg Foundation, Sweden; SERI, SNSF and Cantons of Bern and Geneva, Switzerland; MOST, Taiwan; TAEK, Turkey; STFC, United Kingdom; DOE and NSF, United States of America. In addition, individual groups and members have received support from BCKDF, CANARIE, Compute Canada and CRC, Canada; ERC, ERDF, Horizon 2020, Marie Sk{\l}odowska-Curie Actions and COST, European Union; Investissements d'Avenir Labex, Investissements d'Avenir Idex and ANR, France; DFG and AvH Foundation, Germany; Herakleitos, Thales and Aristeia programmes co-financed by EU-ESF and the Greek NSRF, Greece; BSF-NSF and GIF, Israel; La Caixa Banking Foundation, CERCA Programme Generalitat de Catalunya and PROMETEO and GenT Programmes Generalitat Valenciana, Spain; G\"{o}ran Gustafssons Stiftelse, Sweden; The Royal Society and Leverhulme Trust, United Kingdom.
 
The crucial computing support from all WLCG partners is acknowledged gratefully, in particular from CERN, the ATLAS Tier-1 facilities at TRIUMF (Canada), NDGF (Denmark, Norway, Sweden), CC-IN2P3 (France), KIT/GridKA (Germany), INFN-CNAF (Italy), NL-T1 (Netherlands), PIC (Spain), ASGC (Taiwan), RAL (UK) and BNL (USA), the Tier-2 facilities worldwide and large non-WLCG resource providers. Major contributors of computing resources are listed in Ref.~\cite{ATL-SOFT-PUB-2020-001}.
 
% End of text imported from the .//acknowledgements/Acknowledgements.tex input file
 
\clearpage

\printbibliography
 
\clearpage
% ATLAS Collaboration author list
% Reference date of HDBS-2018-10 is 2020-02-05
% Author list last updated on date 12-OCT-20
% Data extracted on 12-Oct-2020 for paper reference HDBS-2018-10
% at 10:53am
 
\begin{flushleft}
\hypersetup{urlcolor=black}
{\Large The ATLAS Collaboration}

\bigskip

\AtlasOrcid[0000-0002-6665-4934]{G.~Aad}$^\textrm{\scriptsize 102}$,    
\AtlasOrcid[0000-0002-5888-2734]{B.~Abbott}$^\textrm{\scriptsize 128}$,    
\AtlasOrcid{D.C.~Abbott}$^\textrm{\scriptsize 103}$,    
\AtlasOrcid[0000-0002-2788-3822]{A.~Abed~Abud}$^\textrm{\scriptsize 36}$,    
\AtlasOrcid[0000-0002-1002-1652]{K.~Abeling}$^\textrm{\scriptsize 53}$,    
\AtlasOrcid[0000-0002-2987-4006]{D.K.~Abhayasinghe}$^\textrm{\scriptsize 94}$,    
\AtlasOrcid[0000-0002-8496-9294]{S.H.~Abidi}$^\textrm{\scriptsize 166}$,    
\AtlasOrcid[0000-0002-8279-9324]{O.S.~AbouZeid}$^\textrm{\scriptsize 40}$,    
\AtlasOrcid{N.L.~Abraham}$^\textrm{\scriptsize 155}$,    
\AtlasOrcid[0000-0001-5329-6640]{H.~Abramowicz}$^\textrm{\scriptsize 160}$
% ...
\bigskip
\\

% ...
$^{36}$CERN, Geneva; Switzerland.\\
% ...
$^{40}$Niels Bohr Institute, University of Copenhagen, Copenhagen; Denmark.\\
% ...
$^{53}$II. Physikalisches Institut, Georg-August-Universit\"{a}t G\"ottingen, G\"ottingen; Germany.\\
% ...
$^{94}$Department of Physics, Royal Holloway University of London, Egham; United Kingdom.\\
% ...
$^{102}$CPPM, Aix-Marseille Universit\'e, CNRS/IN2P3, Marseille; France.\\
$^{103}$Department of Physics, University of Massachusetts, Amherst MA; United States of America.\\
% ...
$^{128}$Homer L. Dodge Department of Physics and Astronomy, University of Oklahoma, Norman OK; United States of America.\\
% ...
$^{155}$Department of Physics and Astronomy, University of Sussex, Brighton; United Kingdom.\\
% ...
$^{160}$Raymond and Beverly Sackler School of Physics and Astronomy, Tel Aviv University, Tel Aviv; Israel.\\
% ...
$^{166}$Department of Physics, University of Toronto, Toronto ON; Canada.\\
% ...

\end{flushleft}

% Created with Glance <Atlas.Glance@cern.ch>

\end{document}